\documentclass[sn-mathphys-num]{sn-jnl}

\usepackage{graphicx}%
\usepackage{multirow}%
\usepackage{amsmath,amssymb,amsfonts}%
\usepackage{amsthm}%
\usepackage{mathrsfs}%
\usepackage[title]{appendix}%
\usepackage{xcolor}%
\usepackage{textcomp}%
\usepackage{manyfoot}%
\usepackage{booktabs}%
\usepackage{algorithm}%
\usepackage{algorithmicx}%
\usepackage{algpseudocode}%
\usepackage{listings}%
\usepackage{float}
\usepackage{dcolumn}
\usepackage{bm}
\usepackage{enumitem}
\usepackage{tikz}
\usetikzlibrary{arrows.meta}
\usepackage{subcaption}

\theoremstyle{thmstyleone}%
\theoremstyle{thmstyletwo}%

\theoremstyle{thmstylethree}%

\begin{document}

\title[Article Title]{Non-Radial Free Geodesics. I. In Spatially Flat FLRW Spacetime}


\author[1]{\fnm{Omar} \sur{Nemoul}}\email{omar.nemoul@umc.edu.dz}

\author[1]{\fnm{Hichem} \sur{Guergouri}}\email{hichem.guergouri@univ-bejaia.dz}

\author[1,2]{\fnm{Jamal} \sur{Mimouni}}\email{jamalmimouni@umc.edu.dz}

\affil*[1]{\orgdiv{Research Unit in Scientific Mediation}, \orgname{CERIST}, \orgaddress{\city{Constantine}, \postcode{25016}, \country{Algeria}}}

\affil[2]{\orgdiv{Laboratoire de Physique Mathematique et Subatomique (LPMS)}, \orgname{University of Constantine 1}, \orgaddress{\city{Constantine}, \postcode{25017}, \country{Algeria}}}


\abstract{This paper provides an analytical examination of non-radial geodesics within the context of the spatially flat Friedmann–Lemaître–Robertson–Walker (FLRW) spacetime. Using the symmetry properties of the system, two constants of motion related to this dynamical system are derived. This treatment enables us to explicitly compute the radial and angular peculiar velocities, as well as the evolution of the comoving radial distance and angle over time. The study introduces three distinct methods to characterize geodesic solutions, and additionally examines the radial geodesic limit, the null geodesic limit, and the comoving geodesic limit. Additionally, the paper investigates both non-radial and radial free geodesics within the Lambda Cold Dark Matter ($\Lambda$CDM) model, presenting detailed results and discussions. Lastly, we explore the total angle swept by a free particle from the initial singularity to infinity as measured by a comoving observer, offering insights into the dynamics of free particles in FLRW spacetime.}




\maketitle

\section{Introduction}\label{sec1}
Experiments investigating the Cosmic Microwave Background (CMB), such as the Cosmic Background Explorer (COBE)~\cite{COBE}, Wilkinson Microwave Anisotropy Probe (WMAP)~\cite{WMAP}, and the Planck satellite~\cite{Planck}, along with studies examining the universe's accelerating expansion rate~\cite{Perlmutter,Riess}, have provided significant support for the $\Lambda$CDM model~\cite{Peebles}. These extensive studies have consistently shown that the universe is dominated by a mysterious form of repulsive gravity known as dark energy causing its accelerating expansion. Moreover, on a large scale, the observable universe exhibits spatial homogeneity, isotropy, and geometric flatness. These features are fundamental in validating the Friedmann-Lemaître-Robertson-Walker (FLRW) metric as the standard model for describing our spacetime~\cite{Friedmann,Lemaitre,Robertson,Walker}.

The study of geodesics in FLRW spacetime is crucial for several reasons, primarily because it provides deep insights into the geometry and dynamics of the universe.  Such studies are fundamental for analyzing the path of light and other particles as they move through the universe. This can be vital for interpreting astronomical observations, such as the CMB, gravitational lensing~\cite{Schneider}, the dynamical evolution of galaxies~\cite{Binney}, and modeling the large-scale universe. Furthermore, analyzing these paths also allows for the testing and refinement of cosmological models, helping to constrain parameters and explore theories like dark energy~\cite{Carroll} and general relativity modifications~\cite{Saridakis}. It would be ideal if the study included the gravitational influence exerted by nearby masses; however, considering free geodesics could serve as a zero-th approximation in scenarios characterized by relatively weak gravitational fields. Research on free geodesic motions in FLRW spacetimes, particularly for radially freely-falling test particles in the absence of non-gravitational forces, has been rigorously investigated in Refs.~\cite{Baumann,Whiting,Davis,Gron,Barnes,Kerachian,Vachon,Cotaescu,Omar}. While several studies~\cite{Whiting,Gron,Barnes,Kerachian,Vachon} have mainly focused on solving geodesic equations, other research efforts~\cite{Cotaescu,Omar} used the symmetry of the system to construct the exact solution.

Since our 3D space is assumed to be isotropic and homogeneous, it follows that radial and non-radial geodesics are equivalent, up to a reference choice (change of origin). However, this work aims to investigate non-radial geodesics by determining both the radial $\chi(t)$ and angular $\phi(t)$ of a freely-falling test particle relative to a comoving observer, providing its radial and angular (transverse) motion across the observer's sky.

First, we begin by offering an in-depth review of what has been already done in radial $\chi$- motion problem through a direct approach based on the Euler-Lagrange equation~\cite{Euler,Lagrange} to determine the timelike geodesics for radial motion. We will derive a constant of motion $A$ related to the conjugate radial variable without solving tedious geodesic equations—a task that typically requires the explicit computation of Christoffel symbols and the resolution of two independent differential equations. This approach is commonly presented in nearly all introductory texts on general relativity (e.g., See~\cite{Hartle}). Subsequently, we present in this review two different methods for describing the physical radial solutions. The first parameterization, denoted by $(\chi_i,A)$, includes the initial comoving radial distance $\chi_i$ at an initial time $t_i$ and the constant of motion $A$. The second parameterization, represented by $(\chi_i,v_i)$, uses both the comoving radial distance $\chi_i$ and the peculiar velocity $v_i$ at an initial time $t_i$.

Moving on, we address the general motion problem which includes both radial $\chi$ and angular $\phi$ movements. Addressing this problem by directly solving either the Euler-Lagrange equations or the geodesic equations with respect to unknown sets of functions $(\chi,\phi)$ and $(t,\chi,\phi)$, respectively, is an extremely challenging task. The difficulty primarily arises from the complex nature of the resulting coupled partial differential equations system. To overcome this obstacle, the following strategy exploiting the inherent symmetry of the system, will yields two constants of motion. The first constant $B$ arises from the system's isotropic symmetry concerning the angular variable $\phi$, while the second one is derived by recognizing that the magnitude of peculiar velocity is a 3D scalar vector. Using this strategy, we can extend its known expression from radial to non-radial motion, thereby deriving the second constant of motion $A$, which now corresponds to both radial and angular peculiar velocity components. This approach significantly simplifies the process, making it possible to avoid directly solving both the Euler-Lagrange and geodesic equations. Moreover, we present three different methods for characterizing the physical solutions $(\chi(t),\phi(t))$. The first parameterization, signified by $(\chi_i,\phi_i,A,B)$, uses both the comoving radial distance $\chi_i$, and the initial angle $\phi_i$ at a chosen initial time $t_i$, and the two constants of motion $A$ and $B$. The second parameterization, indicated by $(\chi_i,\phi_i,v_{\chi,i},v_{\phi,i})$, uses the initial peculiar velocity components $v_{\chi,i}$ and $v_{\phi,i}$ as substitutes for the constants of motion $A$ and $B$. In contrast, the third parameterization, represented by $(\chi_i,\phi_i,v_{\text{pec},i},\psi_i)$, adopts polar coordinates $(v_{\text{pec},i},\psi_i)$ for expressing the initial peculiar velocity instead of $(v_{\chi,i},v_{\phi,i})$, where $v_{\text{pec},i}$ stands for the magnitude of the initial peculiar velocity, while $\psi_i$ is the deviation angle of the 3D initial vector $\vec v_{\text{pec},i}$ from the initial radial axis. Following this, we will prove that the non-radial geodesics successfully meet the radial geodesic limit, the null geodesic limit, and the comoving geodesic limit. Furthermore, we will provide a geometrical interpretation based on the fact that our geodesics are represented by straight-line trajectories in the comoving frame. This can be proven by deriving the geometrical formula for a non-radial straight line, starting from the resulting non-radial solution. Subsequently, our discussion will cover the Killing vectors related to the symmetries of the dynamical system, aiming to determine their precise expressions.

Finally, we will apply our results to the currently accepted cosmological model ($\Lambda$CDM), using the recent Planck results~\cite{Planck}. These results will be visually represented through a series of graphs and animations in both comoving and physical frames. Additionally, we introduce the concept of the total angle to quantify the entire angular journey a free particle can undertake from the Big Bang singularity at $t=0$ to an indefinitely extended time. We will analyze its characteristics within the $\Lambda$CDM model, comparing it to its counterpart in a static (non-expanding) flat space. In Appendices~\ref{A4} and~\ref{A5}, we rigorously demonstrate that the non-radial solutions indeed satisfy both the Euler-Lagrange and the geodesic equations. Throughout this paper, we use Greek letters $\mu,\nu$, etc., with values $0$,$1$,$2$, and $3$, to represent spacetime indices. Spatial indices will be denoted by Latin letters $i,j$, etc., with values $1$,$2$, and $3$. Our analysis adopts a metric signature of $(+,-,-,-)$ in a spacetime coordinate system defined by cosmic time $t$ and comoving spatial coordinates $x^i$. $(x,y,z)$ and $(\chi,\theta,\phi)$ represent the comoving Cartesian and spherical coordinate systems, respectively. Additionally, we employ a system of units where the speed of light is set to $c=1$.
\section{General Geodesics in spatially flat FLRW Spacetime}\label{sec2}

\subsection{For general coordinates system}\label{sec2.1}
Before addressing general geodesics for free motion in flat FLRW spacetime, it is crucial to consider the problem within a general coordinate system $(t, x^i)$. The FLRW spacetime serves as a fundamental mathematical framework for understanding the large-scale structure and evolution of the universe. This model assumes that the universe is, on average, both homogeneous (similar at all points) and isotropic (appearing the same in all directions) on large cosmic scales. It provides a powerful tool for comprehending the expansion of the universe and its defining characteristics. The FLRW spacetime metric in a general coordinate system $(t, x^i)$ can be expressed as follows
\begin{equation}
ds^2 = g_{\mu\nu} dx^\mu dx^\nu = dt^2 - a^2(t) \gamma_{ij}(\vec{x}) dx^i dx^j,
\label{eq1}
\end{equation}
where $a(t)$ is the scale factor that represents the expansion of the 3D space, and $\gamma_{ij}(\vec{x})$ is the spatially homogeneous and isotropic 3D metric in the comoving coordinate system $\vec{x}=(x^i)$. Geodesics trace the shortest path in spacetime and their trajectories can be determined by minimizing the spacetime interval $\Delta s$. Consequently, we formulate an action principle as
\begin{align}
S[x^i(t)] &= -m \int ds = -m \int dt \sqrt{1 - a^2(t) \gamma_{ij}(\vec{x}) \dot{x}^i \dot{x}^j},
\label{eq2}
\end{align}
where the Lagrangian takes the form
\begin{equation}
L(x^i(t), \dot{x}^i(t), t) = -m \sqrt{1 - a^2(t) \gamma_{ij}(\vec{x}) \dot{x}^i \dot{x}^j}.
\label{eq3}
\end{equation}
To ensure that the $L$ is expressed in energy units and to obtain the correct nonrelativistic limit consistent with Newton's law, we have multiplied by the factor “$-m$”. From Eq.~\eqref{eq1}, we readily obtain
\begin{equation}
\frac{dt}{ds} = \frac{1}{\sqrt{1 - a^2(t) \gamma_{ij}(\vec{x}) \dot{x}^i \dot{x}^j}}.
\label{eq4}
\end{equation}
Applying the Euler-Lagrange equations for the three spatial coordinates $x^i(t)$ yields three differential equations
\begin{equation}
\frac{d}{dt} \frac{\partial L}{\partial \dot{x}^i} = \frac{\partial L}{\partial x^i}.
\label{eq5}
\end{equation}
This yields
\begin{align}
&\frac{d}{dt} \left[ \frac{a^2(t) \gamma_{ij}(\vec{x}) \dot{x}^j}{\sqrt{1 - a^2(t) \gamma_{i'j'}(\vec{x}) \dot{x}^{i'} \dot{x}^{j'}}} \right] = \frac{a^2(t) \dot{x}^k \dot{x}^l}{2\sqrt{1 - a^2(t) \gamma_{i'j'}(\vec{x}) \dot{x}^{i'} \dot{x}^{j'}}} \frac{\partial \gamma_{kl}(\vec{x})}{\partial x^i}.
\label{eq6}
\end{align}
These equations, alongside the expression for $\frac{dt}{ds}$ as indicated in Eq.~\eqref{eq4}, are equivalent to the system of four geodesic equations
\begin{equation}
\frac{d^2 x^\mu}{ds^2} + \Gamma_{\alpha\beta}^\mu \frac{dx^\alpha}{ds} \frac{dx^\beta}{ds} = 0,
\label{eq7}
\end{equation}
where $\Gamma_{\alpha\beta}^\mu$ are the Christoffel symbols. Solving these equations to find general geodesics is a complex task. Hence, for simplicity, much research effort has focused on radial geodesics.

\subsection{For comoving spherical coordinates}\label{sec2.2}
In the comoving spherical coordinate system, where $x^i = (\chi, \theta, \phi)$, the spacetime interval in Eq.~\eqref{eq1} will have the following form
\begin{equation}
ds^2 = dt^2 - a^2(t)\left[d\chi^2 + \chi^2(d\theta^2 + \sin^2\theta d\phi^2)\right],
\label{eq8}
\end{equation}
where $\chi\in[0,+\infty)$, $\theta\in[0,\pi]$, and $\phi\in(-\pi,\pi]$. Consequently, the action in this context is expressed as
\begin{equation}
S[\chi(t), \theta(t), \phi(t)] = -m \int ds = -m \int dt \sqrt{1 - a^2(t)\left(\dot{\chi}^2 + \chi^2 \dot{\theta}^2 + \chi^2 \sin^2\theta \dot{\phi}^2\right)}
\label{eq9}
\end{equation}
where the Lagrangian is given by
\begin{equation}
L(\chi(t), \dot{\chi}(t), \theta(t), \dot{\theta}(t), \dot\phi(t),t) = -m\sqrt{1 - a^2(t)\left(\dot{\chi}^2 + \chi^2 \dot{\theta}^2 + \chi^2 \sin^2\theta \dot{\phi}^2\right)}
\label{eq10}
\end{equation}

From Eq.~\eqref{eq8}, the differentiation of cosmic time relative to the spacetime interval (proper time) can be straightforwardly derived as
\begin{equation}
\frac{dt}{ds} = \frac{1}{\sqrt{1 - a^2(t)\left(\dot{\chi}^2 + \chi^2 \dot{\theta}^2 + \chi^2\sin^2\theta \dot{\phi}^2\right)}}.
\label{eq11}
\end{equation}
Applying the Euler-Lagrange equations yields three differential equations for the three comoving coordinates 
\begin{subequations}
$(\chi(t), \theta(t), \phi(t))$ as
\begin{align}
\frac{d}{dt} \frac{\partial L}{\partial \dot{\chi}} &= \frac{\partial L}{\partial \chi}, \label{eq12a}\\
\frac{d}{dt} \frac{\partial L}{\partial \dot{\theta}} &= \frac{\partial L}{\partial \theta}, \label{eq12b}\\
\frac{d}{dt} \frac{\partial L}{\partial \dot{\phi}} &= \frac{\partial L}{\partial \phi}.\label{eq12c}
\end{align}
\end{subequations}
This yields
\begin{subequations}
\begin{align}
&\frac{d}{dt} \left[\frac{a^2(t) \dot{\chi}}{\sqrt{1 - a^2(t)\left(\dot{\chi}^2 + \chi^2 \dot{\theta}^2 + \chi^2 \sin^2\theta \dot{\phi}^2\right)}}\right] = \frac{a^2(t)\chi\left(\dot{\theta}^2 + \sin^2\theta \dot{\phi}^2\right)}{\sqrt{1 - a^2(t)\left(\dot{\chi}^2 + \chi^2 \dot{\theta}^2 + \chi^2 \sin^2\theta \dot{\phi}^2\right)}}, \label{eq13a}\\
&\frac{d}{dt} \left[\frac{a^2(t) \chi^2 \dot{\theta}}{\sqrt{1 - a^2(t)\left(\dot{\chi}^2 + \chi^2 \dot{\theta}^2 + \chi^2 \sin^2\theta \dot{\phi}^2\right)}}\right] = \frac{a^2(t) \chi^2 \sin\theta \cos\theta \dot{\phi}^2}{\sqrt{1 - a^2(t)\left(\dot{\chi}^2 + \chi^2 \dot{\theta}^2 + \chi^2 \sin^2\theta \dot{\phi}^2\right)}}, \label{eq13b}\\
&\frac{d}{dt} \left[\frac{a^2(t) \chi^2 \sin^2\theta \dot{\phi}}{\sqrt{1 - a^2(t)\left(\dot{\chi}^2 + \chi^2 \dot{\theta}^2 + \chi^2 \sin^2\theta \dot{\phi}^2\right)}}\right] = 0.\label{eq13c}
\end{align}
\end{subequations}
It can be shown that the set of these three differential equations, when combined with the expression for $\frac{dt}{ds}$ as outlined in Eq.~\eqref{eq11}, are equivalent to the system of four geodesic equations Eq.~\eqref{eq7} for the coordinates system $x^\mu = (t, \chi, \theta, \phi)$, which are
\begin{subequations}
\begin{align}
&\frac{d^2 t}{ds^2} + a(t) \dot{a}(t)\left[\left(\frac{d\chi}{ds}\right)^2 + \chi^2 \left(\frac{d\theta}{ds}\right)^2 + \chi^2 \sin^2\theta \left(\frac{d\phi}{ds}\right)^2\right] = 0, \label{eq14a}\\
&\frac{d^2 \chi}{ds^2} + 2 \frac{\dot{a}(t)}{a(t)} \frac{dt}{ds}\frac{d\chi}{ds} - \chi\left(\frac{d\theta}{ds}\right)^2 - \chi \sin^2\theta \left(\frac{d\phi}{ds}\right)^2 = 0, \label{eq14b}\\
&\frac{d^2 \theta}{ds^2} + 2 \frac{\dot{a}(t)}{a(t)} \frac{dt}{ds}\frac{d\theta}{ds} + \frac{2}{\chi} \frac{d\chi}{ds}\frac{d\theta}{ds} - \sin\theta \cos\theta \left(\frac{d\phi}{ds}\right)^2 = 0, \label{eq14c}\\
&\frac{d^2 \phi}{ds^2} + 2 \frac{\dot{a}(t)}{a(t)} \frac{dt}{ds}\frac{d\phi}{ds} + \frac{2}{\chi} \frac{d\chi}{ds}\frac{d\phi}{ds} + \frac{2 \cos\theta}{\sin\theta} \frac{d\theta}{ds}\frac{d\phi}{ds} = 0.\label{eq14d}
\end{align}
\label{eq14}
\end{subequations}
These equations are obtained after calculating the non-zero Christoffel symbol components (See Appendix~\ref{A1}) for the metric element given by Eq.~\eqref{eq8}, and then substituting them in the geodesic equations~\eqref{eq7}. As previously mentioned, solving these differential equations to find general geodesics presents a complex task. Hence, to simplify the analysis and without losing generality, our focus in the following section will be on radial geodesics.

\section{Radial Geodesic Motion in FLRW}\label{sec3}
In a homogeneous and isotropic universe, where no directions or locations are favored, it is possible to transform any general geodesic curve into one with a purely radial spatial component through a suitable choice of local coordinates. Thus, without losing generality, we confine our attention to radial geodesics. Within the comoving spherical coordinates system $x^i=(\chi,\theta,\phi)$, we consider both $\theta$ and $\phi$ to be constants, while the comoving radial coordinate $\chi=\chi(t)$ varies with time. These assumptions simplify the spacetime interval Eq.~\eqref{eq8} into the following form
\begin{equation}
ds^2 = dt^2 - a^2(t) d\chi^2.
\label{eq15}
\end{equation}
The action becomes in this case as
\begin{equation}
S[\chi(t)] = -m \int ds = -m \int dt \sqrt{1 - a^2(t) \dot{\chi}^2},
\label{eq16}
\end{equation}
with its Lagrangian 
\begin{equation}
L(\dot{\chi}(t),t) = -m \sqrt{1 - a^2(t) \dot{\chi}^2}.
\label{eq17}
\end{equation}
One can simply obtain from Eq.~\eqref{eq15} the differentiation of the cosmic time with respect to the spacetime interval (proper time) as
\begin{equation}
\frac{dt}{ds} = \frac{1}{\sqrt{1 - a^2(t) \dot{\chi}^2}}.
\label{eq18}
\end{equation}
\subsection{Determining the constant of motion}\label{sec3.1}
It is evident that this Lagrangian is symmetric under any time-independent translation $\epsilon$ of the comoving radial variable $\chi$, as the Lagrangian does not depend on $\chi$
\begin{equation}
\chi \longrightarrow \chi' = \chi + \epsilon.
\label{eq19}
\end{equation}
According to Emmy Noether's theorem~\cite{Noether}, this symmetry implies a conservation law, represented by a conserved quantity (constant of motion) that remains invariant over time. This constant of motion is derived from the Euler-Lagrange equation~\eqref{eq12a}:
\begin{equation}
\frac{d}{dt} \left[ \frac{a^2(t) \dot{\chi}}{\sqrt{1 - a^2(t) \dot{\chi}^2}} \right] = 0.
\label{eq20}
\end{equation}
Thus, we obtain
\begin{equation}
\frac{a^2(t) \dot{\chi}}{\sqrt{1 - a^2(t) \dot{\chi}^2}} = A.
\label{eq21}
\end{equation}
Here, $A$ is an arbitrary real number representing the initial radial comoving velocity of the particle. Inverting Eq.~\eqref{eq21} straightforwardly yields the radial comoving velocity $\dot{\chi}$ as
\begin{equation}
\dot{\chi}(t) = \frac{1}{a(t)} \frac{A}{\sqrt{a^2(t) + A^2}}.
\label{eq22}
\end{equation}
In this case, the peculiar velocity $v_{\text{pec}}$ has only a radial component $v_\chi$ which can be written as
\begin{equation}
v_{\text{pec}}(t) = v_\chi(t) = a(t) \dot{\chi}(t) = \frac{A}{\sqrt{a^2(t) + A^2}}.
\label{eq23}
\end{equation}
Eq.~\eqref{eq22} is consistent with the result obtained in Ref.~\cite{Davis,Gron,Barnes,Kerachian,Vachon,Cotaescu,Omar}. Due to the homogeneous and isotropic nature of our spatial space, this formula applies to all geodesic motions, including non-radial geodesics as well, making Eq.~\eqref{eq23} a general formula for calculating the magnitude of the peculiar velocity $v_{\text{pec}}$ for general geodesics. This observation shall be used later in our discussion to address the problem of non-radial geodesics. As $\theta$ and $\phi$ are constants, the remaining Euler-Lagrange equations~\eqref{eq12b} and~\eqref{eq12c} vanish identically. It can be shown that the differential equation~\eqref{eq20}, when combined with the expression for $\frac{dt}{ds}$ as indicated in Eq.~\eqref{eq18}, are equivalent to the two geodesic equations
\begin{subequations}
\begin{align}
&\frac{d^2 t}{ds^2} + a(t) \dot{a}(t) \left(\frac{d\chi}{ds}\right)^2 = 0, \label{eq24a}\\
&\frac{d^2 \chi}{ds^2} + 2 \frac{\dot{a}(t)}{a(t)} \frac{dt}{ds} \frac{d\chi}{ds} = 0.\label{eq24b}
\end{align}
\end{subequations}
These two geodesic equations~\eqref{eq24a} and~\eqref{eq24b} are induced from Eqs.~\eqref{eq14a} and~\eqref{eq14b}, respectively, under the consideration of radial motion. Meanwhile, the remaining equations~\eqref{eq14c} and~\eqref{eq14d} are identically zero in this context. Furthermore, it can be demonstrated that the radial velocity solution~\eqref{eq22}, along with the expression for $\frac{dt}{ds}$ as provided in Eq.~\eqref{eq18}, satisfy the geodesic equations~\eqref{eq14a} and~\eqref{eq14b}.
\subsection{The initial condition for peculiar velocity}\label{sec3.2}
From Eq.~\eqref{eq23}, it is evident that to determine the peculiar velocity of the particle, the constant of motion $A$ must be fixed. However, instead of relying on $A$ to constrain the motion, a more practical method involves using a measured quantity, such as the initial peculiar velocity $v_i = v_{\text{pec}}(t_i)$ at an arbitrary chosen initial time $t_i$. This can be achieved by substituting $t = t_i$ into Eq.~\eqref{eq21}, resulting in the following expression
\begin{equation}
A = \frac{a_i^2 \dot{\chi}_i}{\sqrt{1 - a_i^2 \dot{\chi}_i^2}} = \frac{a_i v_i}{\sqrt{1 - v_i^2}},\label{eq25}
\end{equation}
where we use the notation $a(t_i) = a_i$ and $\dot{\chi}(t_i) = \dot{\chi}_i$. By replacing this expression of $A$ into Eqs.~\eqref{eq22} and~\eqref{eq23}, we obtain
\begin{equation}
\dot{\chi}(t) = \frac{a_i v_i}{a(t) \sqrt{a^2(t)(1 - v_i^2) + a_i^2 v_i^2}},
\label{eq26}
\end{equation}
and
\begin{equation}
v_{\text{pec}}(t) = a(t) \dot{\chi}(t) = \frac{a_i v_i}{\sqrt{a^2(t)(1 - v_i^2) + a_i^2 v_i^2}}.
\label{eq27}
\end{equation}
This gives the radial comoving velocity $\dot{\chi}$ and its corresponding radial peculiar velocity $v_{\text{pec}}$ at any time $t$, expressed in terms of the peculiar velocity $v_i$ at an arbitrary initial time $t_i$.

\subsection{Geodesics Parametrization for Radial Motion}\label{sec3.3}
Geodesic motion is given by integrating over the radial comoving velocity, we have
\begin{equation}
\chi(t) = \left| \int_{t_i}^{t} \dot\chi(t) \, dt + \chi_i \right|.
\label{eq28}
\end{equation}
To ensure the comoving radial distance $\chi$ remains positive at all times, we place the right-hand side of the equation within an absolute value. Building on previous work, we outline two methods to characterize a specific geodesic solution for our test particle, as detailed in the following:
\begin{itemize}[leftmargin=*]
\item \textbf{$(\chi_i, A)$ initial conditions:}
Using Eqs.~\eqref{eq28} and~\eqref{eq22}, we express the comoving radial distance $\chi$ of a radial geodesic at any given time $t$, under the specific initial conditions $(\chi_i, A)$ as follows
\begin{align}
\chi(t; \chi_i, A) = \left| \int_{t_i}^{t} \frac{1}{a(t')} \frac{A}{\sqrt{a^2(t') + A^2}} \, dt' + \chi_i \right|.
\label{eq29}
\end{align}
\item  \textbf{$(\chi_i, v_i)$ initial conditions:}
Using Eqs.~\eqref{eq28} and~\eqref{eq26}, we can formulate the radial comoving distance $\chi$ of a radial geodesic at any given time $t$, under the specific initial conditions $(\chi_i, v_i)$, in the following manner
\begin{align}
\chi(t; \chi_i, v_i) = \left| \int_{t_i}^{t} \frac{a_i v_i}{a(t') \sqrt{a^2(t')(1 - v_i^2) + a_i^2 v_i^2}} \, dt' + \chi_i \right|.
\label{eq30}
\end{align}
\end{itemize}
We are free to select the initial time $t_i$ for fixing both the initial comoving distance and initial peculiar velocity. However, Eq.~\eqref{eq30} has limitations when choosing $t_i = 0$. In this scenario, $a_i = 0$ and $v_i = \pm 1$ leading to an undefined integral term, $\frac{0}{0}$. This issue arises because the peculiar velocity of all freely-falling particles at the Big Bang singularity $t = 0$  equals the speed of light, thus making it unsuitable for determining the initial radial peculiar velocity. To address this problem, one can choose the present time $t_0$ as the reference time to fix the peculiar velocity $v_0 = v_{\text{pec}}(t_0)$, while still considering the Big Bang singularity time $t = 0$ for fixing the initial comoving distance $\chi_i = \chi(0)$. Consequently, in this case, the solution is parameterized by the initial conditions $(\chi_i, v_0)$. We implemented this method in the recent paper~\cite{Omar}, where we investigate the behaviors of all geodesics using this parametrization.
\subsection{Null Geodesics limit for radial motion}\label{sec3.4}
In Ref.~\cite{Omar}, we show the correspondence between the zero-mass limit $m = 0$, $A = \pm\infty$, and $v_i = \pm 1$. Now, applying these limits to Eqs.~\eqref{eq39} and~\eqref{eq30}, we obtain the corresponding null geodesics
\begin{equation}
\chi(t) = \left| \pm \int_{t_i}^{t} \frac{dt'}{a(t')} + \chi_i \right|.
\label{eq31}
\end{equation}
\subsection{Comoving geodesics limit for radial motion}\label{sec3.5}
If we set $A = 0$ in Eq.~\eqref{eq29} or $v_i = 0$ in Eq.~\eqref{eq30}, the particle maintains a constant comoving distance
\begin{equation}
\chi(t)  = \chi_i.
\label{eq32}
\end{equation}
\section{Non-Radial Geodesic Motion in flat FLRW Spacetime}\label{sec4}
Now, we explore the timelike geodesics for general free motion, including non-radial trajectories, within the spatially flat FLRW spacetime. Without loss of generality, we restrict the motion to the $(xy)$ plane by setting $\theta = \frac{\pi}{2}$. This constraint simplifies the spacetime interval~\eqref{eq8} into the following form
\begin{equation}
ds^2 = dt^2 - a^2(t)(d\chi^2 + \chi^2 d\phi^2),
\label{eq33}
\end{equation}
where $\phi$ is the angle measured from the origin relative to the $x$-axis. The action in this case is as follows
\begin{align}
S[\chi(t), \phi(t)] = -m \int ds = -m \int dt \sqrt{1 - a^2(t) \dot{\chi}^2 - a^2(t) \chi^2 \dot{\phi}^2},
\label{eq34}
\end{align}
where the Lagrangian is given by
\begin{align}
L(\chi(t), \dot{\chi}(t), \dot{\phi}(t),t) = -m \sqrt{1 - a^2(t) \dot{\chi}^2 - a^2(t) \chi^2 \dot{\phi}^2}.
\label{eq35}
\end{align}
From Eq.~\eqref{eq33}, the differentiation of cosmic time with respect to the spacetime interval (proper time) can be easily derived in the following form
\begin{equation}
\frac{dt}{ds} = \frac{1}{\sqrt{1 - a^2(t) \dot{\chi}^2 - a^2(t) \chi^2 \dot{\phi}^2} }.
\label{eq36}
\end{equation}
Applying the Euler-Lagrange equations for $(\chi(t), \phi(t))$, Eqs.~\eqref{eq12a} and~\eqref{eq12c}, it yields
\begin{subequations}
\begin{align}
&\frac{d}{dt} \left[ \frac{a^2(t) \dot{\chi}(t)}{\sqrt{1 - a^2(t) \dot{\chi}^2 - a^2(t) \chi^2 \dot{\phi}^2}} \right] = \frac{a^2(t) \chi(t)\dot{\phi}^2(t)}{\sqrt{1 - a^2(t) \dot{\chi}^2 - a^2(t) \chi^2 \dot{\phi}^2}}, \label{eq37a}\\
&\frac{d}{dt} \left[ \frac{a^2(t) \chi^2(t) \dot{\phi}(t)}{\sqrt{1 - a^2(t) \dot{\chi}^2 - a^2(t) \chi^2 \dot{\phi}^2}} \right] = 0,\label{eq37b}
\end{align}
\end{subequations}
where the Euler-Lagrange equation~\eqref{eq12b} for the variable $\theta$ vanishes identically since we have set $\theta$ as a constant. It can be shown that the set of these two differential equations, when combined with the expression for $\frac{dt}{ds}$ as given in Eq.~\eqref{eq36}, are equivalent to the system of the following three geodesic equations
\begin{subequations}
\begin{align}
&\frac{d^2 t}{ds^2} + a(t) \dot{a}(t)\left[\left(\frac{d\chi}{ds}\right)^2 + \chi^2 \left(\frac{d\phi}{ds}\right)^2 \right] = 0,\label{eq38a}\\
&\frac{d^2 \chi}{ds^2} + 2 \frac{\dot{a}(t)}{a(t)} \frac{dt}{ds} \frac{d\chi}{ds} - \chi \left(\frac{d\phi}{ds}\right)^2 = 0, \label{eq38b}\\
&\frac{d^2 \phi}{ds^2} + 2 \frac{\dot{a}(t)}{a(t)} \frac{dt}{ds} \frac{d\phi}{ds} + \frac{2}{\chi} \frac{d\chi}{ds} \frac{d\phi}{ds} = 0.\label{eq38c}
\end{align}
\end{subequations}
These three geodesic equations~\eqref{eq38a},~\eqref{eq38b}, and~\eqref{eq38c} are induced from Eqs.~\eqref{eq14a},~\eqref{eq14b}, and~\eqref{eq14d}, respectively, for motion confined to the $(xy)$ plane. In this context, the various terms of the left-handed side of the remaining equation~\eqref{eq14c} vanishes identically. However, since both the Euler-Lagrange equations and geodesic equations are sets of partial differential equations involving the coupled functions $(\chi, \phi)$, directly solving these equations to determine free geodesics is a highly complex and challenging task. To address this effectively, we undertake two steps to derive the solutions, identifying two constants of motion that characterize the comoving velocities $(\dot{\chi}, \dot{\phi})$. The first constant of motion can be readily obtained from the expression~\eqref{eq37b}, while the second one will be determined by using the constant of motion for radial geodesic motion~\eqref{eq21}, taking into consideration the general formula for the peculiar velocity. Let us now delve into these two steps in more detail.

\subsection{Determining the Constant of Motions}\label{sec4.1}
It is evident that the Lagrangian~\eqref{eq35} exhibits a symmetry under any time-independent rotation $\delta$ of the angular variable $\phi$, as it is independent of $\phi$
\begin{equation}
\phi \longrightarrow \phi' = \phi + \delta.
\label{eq39}
\end{equation}
According to Noether's theorem~\cite{Noether}, this symmetry in the dynamical system implies the existence of a conservation law represented by a conserved quantity (constant of motion) that remains invariant over time. This constant of motion can be directly derived from the Euler-Lagrange equation~\eqref{eq37b}, which yields
\begin{equation}
\frac{a^2(t) \chi^2(t) \dot{\phi}(t)}{\sqrt{1 - a^2(t) \dot{\chi}^2(t) - a^2(t) \chi^2(t) \dot{\phi}^2(t)}} = B,
\label{eq40}
\end{equation}
where $B$ is an arbitrary real number related to both the initial angular and radial velocities of the particle. It is important to note that the sign of $B$ determines the sign of $\dot{\phi}$. Therefore, if $B > 0$, this indicates an increasing angle $\phi$ for the free particle's trajectory, as measured from the origin point relative to the $x$-axis, while if $B < 0$ the angle decreases. In the case where $B = 0$, the particle remains at a fixed angle, implying purely radial motion. Consequently, $\phi$ must be a monotonic function, either increasing or decreasing, depending on the sign of $B$. Inverting Eq.~\eqref{eq40} allows us to directly derive the square of the angular peculiar velocity $v_\phi^2$ as
\begin{equation}
v_\phi^2(t) = a^2(t) \chi^2 \dot{\phi}^2 = \frac{B^2 (1 - a^2(t) \dot{\chi}^2)}{a^2(t) \chi^2 + B^2}.
\label{eq41}
\end{equation}
We still need to determine the second constant of motion. Using the fact that the peculiar velocity magnitude $v_\text{pec}(t)$ is a 3D scalar (invariant under spatial coordinate transformations), which is obvious from its spatial covariant form
\begin{equation}
v^2_\text{pec}(t)=a^2(t)\gamma_{ij}(\vec{x}) \dot x^i(t) \dot x^j(t).
\label{eq42}
\end{equation}
In our homogeneous and isotropic 3D space, any general geodesic curve can be transformed to have a purely radial spatial part through an appropriate choice of local spatial coordinates. This allows us to generalize the 3D scalar peculiar velocity expression~\eqref{eq23} for radial motion to encompass all general geodesic motions. Consequently, we can write the following equation
\begin{align}
v_{\text{pec}}^2(t) &= v_\chi^2(t) + v_\phi^2(t) \nonumber\\&= a^2(t) \dot{\chi}^2(t) + a^2(t) \chi^2(t) \dot{\phi}^2(t) \nonumber\\&= \frac{A^2}{a^2(t) + A^2}.
\label{eq43}
\end{align}
The real number $A$ is now related to both the initial condition for radial and angular peculiar velocities. It is worth mentioning that the sign of $A$ is redundant for our physical motion, i.e., does not affect the physical nature of the motion where both “$A$” and “$-A$” represent the same physics. If $A=0$, it follows that $v_\text{pec}(t)=0$, resulting in both $v_\chi(t)$ and $v_\phi(t)$ being zero. This scenario corresponds to comoving geodesics. Accordingly, $A=0\Rightarrow B=0$, although the converse does not necessarily hold true. If $A=+\infty$, it leads to $v_\text{pec}(t)=1$ and one can show from Eq.~\eqref{eq40} that $A=+\infty\Leftrightarrow B=+\infty$. We obtain a system of two equations~\eqref{eq40} and~\eqref{eq43}, involving the comoving velocity variables $(\dot{\chi}, \dot{\phi})$. Solving these two equations for $(\dot{\chi}, \dot{\phi})$ gives
\begin{subequations}
\label{eq44}
\begin{align}
\dot{\chi}(t) &= \text{sgn}(\dot{\chi}(t)) \frac{1}{a(t)} \sqrt{\frac{A^2 - \frac{B^2}{\chi^2(t)}}{a^2(t) + A^2}}, \label{eq44a}\\
\dot{\phi}(t) &= \frac{\frac{B}{\chi^2(t)}}{a(t) \sqrt{a^2(t) + A^2}},\label{eq44b}
\end{align}
\end{subequations}
and the peculiar velocity components $(v_\chi, v_\phi)$ are
\begin{subequations}
\label{eq45}
\begin{align}
v_\chi(t) &= \text{sgn}(v_\chi(t)) \sqrt{\frac{A^2 - \frac{B^2}{\chi^2(t)}}{a^2(t) + A^2}}, \label{eq45a}\\
v_\phi(t) &= \frac{\frac{B}{\chi(t)}}{\sqrt{a^2(t) + A^2}}.\label{eq45b}
\end{align}
\end{subequations}
Indeed, for $B = 0$, it becomes evident that $\dot{\phi} = v_\phi = 0$. In this situation, both $\dot{\chi}$ and $v_\chi$ precisely match the expressions for radial free geodesics as described in Eqs.~\eqref{eq22} and~\eqref{eq23}, respectively. This limit effectively reduces the motion to a purely radial trajectory, completely eliminating any angular component. Furthermore, we observe that both radial and angular peculiar velocities decrease over time, approaching zero as the scale factor goes to infinity. In our previous work (see Ref.~\cite{Omar}), we discussed how all free geodesics in FLRW spacetime converge with the Hubble flow when we consider a cosmological constant $\Lambda$ model with an equation of state parameter $w_\Lambda=-1$. This satisfies convergent the condition $w_d < -\frac{1}{3}$ for the dominant cosmological component as time progresses towards infinity, as detailed in Ref.~\cite{Barnes}.\\
\subsection{The initial Conditions for Peculiar Velocity Components}\label{sec4.2}
From Eqs.~\eqref{eq45a} and~\eqref{eq45b}, it is clear that in order to determine the peculiar velocity components $v_\chi(t)$ and $v_\phi(t)$ of the particle, the constants of motion $A$ and $B$ must be fixed. As performed in the previous section, we shall express $A$ and $B$ in terms of the initial peculiar velocities $v_{\chi, i} = v_\chi(t_i)$ and $v_{\phi, i} = v_\phi(t_i)$, which are the initial radial and angular peculiar velocities at a chosen initial time $t_i$, respectively. To achieve this, one would substitute $t = t_i$ into Eqs.~\eqref{eq45a} and~\eqref{eq45b} and then invert these equations to solve for $A$ and $B$, which gives
\begin{subequations}
\begin{align}
A &= \text{sgn}(A) a_i \sqrt{\frac{v_{\chi, i}^2 + v_{\phi, i}^2}{1 - v_{\chi, i}^2 - v_{\phi, i}^2}}, \label{eq46a}\\
B &= \frac{a_i \chi_i v_{\phi, i}}{\sqrt{1 - v_{\chi, i}^2 - v_{\phi, i}^2}}.\label{eq46b}
\end{align}
\end{subequations}
By substituting the derived expressions for $A$ and $B$ back into Eqs.~\eqref{eq44} and~\eqref{eq45}, we obtain
\begin{subequations}
\begin{align}
\dot{\chi}(t) &= \text{sgn}(\dot{\chi}(t)) \frac{a_i}{a(t)} \sqrt{\frac{v_{\chi, i}^2 + v_{\phi, i}^2 - \frac{\chi_i^2 v_{\phi, i}^2}{\chi^2(t)}}{a^2(t)(1 - v_{\chi, i}^2 - v_{\phi, i}^2) + a_i^2(v_{\chi, i}^2 + v_{\phi, i}^2)}},\label{eq47a}\\
\dot{\phi}(t) &= \frac{a_i}{a(t)}  \frac{\frac{\chi_i v_{\phi, i}}{\chi^2(t)}}{\sqrt{a^2(t)(1 - v_{\chi, i}^2 - v_{\phi, i}^2) + a_i^2(v_{\chi, i}^2 + v_{\phi, i}^2)}},\label{eq47b}
\end{align}
\end{subequations}
and
\begin{subequations}
\begin{align}
v_\chi(t) &= a(t) \dot{\chi}(t) = \text{sgn}(v_\chi(t)) a_i \sqrt{\frac{v_{\chi, i}^2 + v_{\phi, i}^2 - \frac{\chi_i^2 v_{\phi, i}^2}{\chi^2(t)}}{a^2(t)(1 - v_{\chi, i}^2 - v_{\phi, i}^2) + a_i^2(v_{\chi, i}^2 + v_{\phi, i}^2)}},\label{eq48a}\\
v_\phi(t) &= a(t)\chi(t) \dot{\phi}(t) =  \frac{\frac{a_i \chi_i v_{\phi, i}}{\chi(t)}}{\sqrt{a^2(t)(1 - v_{\chi, i}^2 - v_{\phi, i}^2) + a_i^2(v_{\chi, i}^2 + v_{\phi, i}^2)}},\label{eq48b}
\end{align}
\end{subequations}
where $0\leq v_{\chi, i}^2 + v_{\phi, i}^2 \leq 1$. These expressions determine the comoving velocity components $(\dot{\chi}, \dot{\phi})$, and their corresponding peculiar velocities $(v_\chi, v_\phi)$ at any time $t$, in terms of the initial peculiar velocities $(v_{\chi, i}, v_{\phi, i})$ at an initial time $t_i$. In addition, we can use the polar coordinates $(v_{\text{pec},i},\psi_i)$ instead of $(v_{\chi,i},v_{\phi,i})$ to parameterize the comoving and peculiar velocity components. Here, $v_{\text{pec},i}$ represents the magnitude of $\vec v_{\text{pec},i}$ and $\psi_i$ indicates the deviation angle of $\vec v_{\text{pec},i}$ from the initial radial axis. It is important to note that this initial radial axis at $t=t_i$ itself is oriented at an angle $\phi_i$ relative to the $x$-axis, which makes the initial launch angle to be $\phi_i+\psi_i$. This polar coordinate system offers a more effective and practical approach for determining the comoving and peculiar velocity components as
\begin{subequations}
\begin{align}
\dot{\chi}(t) &= \text{sgn}(\dot{\chi}(t)) \frac{a_iv_{\text{pec}, i}}{a(t)} \sqrt{\frac{1 - \frac{\chi_i^2}{\chi^2(t)}\sin^2{\psi_i}}{a^2(t)(1 - v_{\text{pec}, i}^2) + a_i^2v_{\text{pec}, i}^2}},\label{eq49a}\\
\dot{\phi}(t) &= \frac{a_iv_{\text{pec}, i}}{a(t)}  \frac{\frac{\chi_i}{\chi^2(t)}\sin{\psi_i}}{\sqrt{a^2(t)(1 - v_{\text{pec}, i}^2) + a_i^2v_{\text{pec}, i}^2}},\label{eq49b}
\end{align}
\end{subequations}
and
\begin{subequations}
\begin{align}
v_\chi(t) &= a(t) \dot{\chi}(t) =  \text{sgn}(v_\chi(t)) a_iv_{\text{pec}, i} \sqrt{\frac{1 - \frac{\chi_i^2}{\chi^2(t)}\sin^2{\psi_i}}{a^2(t)(1 - v_{\text{pec}, i}^2) + a_i^2v_{\text{pec}, i}^2}},\label{eq50a}\\
v_\phi(t) &= a(t)\chi(t)  \dot{\phi}(t) = \frac{ \frac{\chi_i}{\chi(t)}a_iv_{\text{pec}, i}\sin{\psi_i}}{\sqrt{a^2(t)(1 - v_{\text{pec}, i}^2) + a_i^2v_{\text{pec}, i}^2}},\label{eq50b}
\end{align}
\end{subequations}
where
\begin{subequations}
\begin{align}
v_{\text{pec}, i}&=\sqrt{v_{\chi, i}^2 + v_{\phi, i}^2},\label{eq51a}\\
\psi_i&=\arctan{\left(\frac{v_{\phi, i}}{v_{\chi, i}}\right)},\label{eq51b}
\end{align}
\end{subequations}
are constrained to $0\leq v_{\text{pec}, i}\leq1$ and $-\pi<\psi_i\leq\pi$. Notice that, at $t=t_i$, inward radial motion corresponds to $\psi_i=0$, while outward radial motion is related to $\psi_i=\pi$.
\subsection{Geodesics Parametrization for General Motion}\label{sec4.3}
Geodesic motion can be determined by integrating Eqs.~\eqref{eq44a} and~\eqref{eq44b}, but this treatment does not easily apply to the radial $\chi$ equation. This issue is thoroughly explored in Appendix~\ref{A2} and a detailed analysis provided. Now, building upon previous work, we establish three distinct methods for characterizing a specific geodesic solution for our freely-falling test particle, as described below
\begin{itemize}[leftmargin=*]
    \item ($\chi_i, \phi_i, A, B,\text{sgn}(\dot\chi_i)$) \textbf{initial conditions:}
\end{itemize}
Using Eqs.~\eqref{eq44a} and~\eqref{eq44b}, we write the comoving radial distance $\chi$ (as detailed in Appendix~\ref{A2}) and the angle $\phi$ for a free test-particle at any given time $t$, under the specific initial conditions $(\chi_i, \phi_i, A, B)$ and $\text{sgn}(\dot\chi_i)$ as
\begin{subequations}
\label{eq52}
\begin{align}
&\chi\left(t; \chi_i, A, B,\text{sgn}(\dot\chi_i)\right) =\sqrt{\left( \int_{t_i}^t \frac{|A| \, dt'}{a(t') \sqrt{a^2(t') + A^2}} + \text{sgn}(\dot{\chi}_i) \sqrt{\chi_i^2 - \frac{B^2}{A^2}} \right)^2 + \frac{B^2}{A^2}},\label{eq52a}\\
&\phi\left(t; \chi_i, \phi_i, A, B,\text{sgn}(\dot\chi_i)\right) = \int_{t_i}^{t} \frac{ \frac{B}{\chi^2\left(t'; \chi_i, A, B,\text{sgn}(\dot\chi_i)\right)}}{a(t') \sqrt{a^2(t') + A^2}} \, dt' + \phi_i,\label{eq52b}
\end{align}
\end{subequations}
with the initial radial distance $\chi_i\geq\left|B/A\right|$, the initial angle $-\pi<\phi_i\leq\pi$, $A=0\Rightarrow B=0$ and the sign of the initial comoving radial veclocity $\text{sgn}(\dot\chi_i)=\pm$ or $0$.
\begin{itemize}[leftmargin=*]
\item \textbf{($\chi_i,\phi_i, v_{\chi, i}, v_{\phi, i}$) initial conditions:}
\end{itemize}
Using Eqs.~\eqref{eq47a} and~\eqref{eq47b}, we can express the radial comoving distance $\chi$ and the angle $\phi$ of a free test-particle at any given time $t$, under the specific initial conditions $(\chi_i, \phi_i, v_{\chi, i}, v_{\phi, i})$ as
\begin{subequations}
\label{eq53}
\begin{align}
\begin{split}
&\chi(t; \chi_i, v_{\chi, i}, v_{\phi, i}) = \biggl[\biggl( \int_{t_i}^{t} \frac{a_i}{a(t')} \sqrt{\frac{v_{\chi, i}^2 + v_{\phi, i}^2}{a^2(t')(1 - v_{\chi, i}^2 - v_{\phi, i}^2) + a_i^2(v_{\chi, i}^2 + v_{\phi, i}^2)}} \, dt' \\& \ \ \ \ \ \ \ \ \ \ \ \ \ \ \ \ \ \ \ \ \ \ \ \ \ +\frac{\chi_i v_{\chi, i}}{\sqrt{v_{\chi, i}^2 + v_{\phi, i}^2}} \biggr)^2+\frac{\chi_i^2 v_{\phi, i}^2}{v_{\chi, i}^2 + v_{\phi, i}^2}\biggr]^{1/2}
,\label{eq53a}
\end{split}
\\
&\phi(t; \chi_i, \phi_i,v_{\chi, i}, v_{\phi, i}) = \int_{t_i}^{t} \frac{a_i}{a(t')} \frac{\frac{\chi_i v_{\phi, i}}{\chi^2(t'; \chi_i, v_{\chi, i}, v_{\phi, i})}}{\sqrt{a^2(t')(1 - v_{\chi, i}^2 - v_{\phi, i}^2) + a_i^2(v_{\chi, i}^2 + v_{\phi, i}^2)}} \, dt' + \phi_i,\label{eq53b}
\end{align}
\end{subequations}
with $\chi_i\geq0$, $-\pi<\phi_i\leq\pi$, and $0\leq v_{\chi, i}^2 + v_{\phi, i}^2 \leq 1$.
\begin{itemize}[leftmargin=*]
\item ($\chi_i,\phi_i, v_{\text{pec},i},\psi_i$) \textbf{initial conditions:}
\end{itemize}
Using Eqs.~\eqref{eq49a} and~\eqref{eq49b}, we can express the radial comoving distance $\chi$ and the angle $\phi$ of a free test-particle at any given time $t$, under the specific initial conditions $(\chi_i,\phi_i, v_{\text{pec}, i},\psi_i)$ as
\begin{subequations}
\label{eq54}
\begin{align}
\begin{split}
&\chi(t; \chi_i, v_{\text{pec}, i},\psi_i) = \biggl[\biggl( \int_{t_i}^{t} \frac{a_i}{a(t')} \frac{v_{\text{pec}, i}}{\sqrt{a^2(t')(1 - v_{\text{pec}, i}^2) + a_i^2v_{\text{pec}, i}^2}} \, dt' \\
& \ \ \ \ \ \ \ \ \ \ \ \ \ \ \ \ \ \ \ \ \ \ \ \ \ \ \ \ \ + \chi_i\cos\psi_i \biggr)^2 + \chi^2_i\sin^2\psi_i\biggr]^{1/2}, \label{eq54a}
\end{split}
\\
&\phi(t; \chi_i,\phi_i, v_{\text{pec}, i},\psi_i) = \int_{t_i}^{t} \frac{a_i}{a(t')} \frac{\frac{\chi_iv_{\text{pec}, i}\sin{\psi_i}}{\chi^2(t';\chi_i, v_{\text{pec}, i},\psi_i)}}{\sqrt{a^2(t')(1 - v_{\text{pec}, i}^2) + a_i^2v_{\text{pec}, i}^2}} \, dt' + \phi_i,\label{eq54b}
\end{align}
\end{subequations}
with $\chi_i\geq0$, $-\pi<\phi_i\leq\pi$, $0\leq v_{\text{pec}, i}\leq1$, and $-\pi<\psi_i\leq\pi$.
\\

In the first parametrization~\eqref{eq52}, once $A$ and $B$ are fixed, the range of the initial comoving radial distance $\chi_i$ must be greater than or equal to the potential closest approaching distance $|B/A|$ . However, in the case of the second~\eqref{eq53} and third~\eqref{eq54} formulations, the initial comoving distance $\chi_i$ is not related to any other initial parameters. This independence makes these latter two approaches more practical for describing the non-radial solutions.
\subsection{The Closest Approaching Distance}\label{sec4.4}
From the $\dot{\chi}$ formula in Eq.~\eqref{eq44a}, including the term $\sqrt{A^2 - \frac{B^2}{\chi^2(t)}}$, it necessarily follows that $\chi(t) \geq \left|\frac{B}{A}\right|$ at any time $t$. This is consistent with the result obtained in Eq.~\eqref{eq52a} indicating that for any non-zero $A$ and real $B$, the closest approaching distance, denoted by $\chi_*$, to which the free particle can approach the center in the comoving frame at a specific moment called $t_*$. The following relations define the radial distance $\chi_*$ and angle $\phi_*$ of this closest approaching point under different parameterizations

\begin{subequations}
\begin{flalign}
\chi_*= \chi(t_*) &=\left|\frac{B}{A}\right|
= \frac{\chi_i |v_{\phi, i}|}{\sqrt{v_{\chi, i}^2 + v_{\phi, i}^2}}
=\chi_i|\sin{\psi_i}|,\label{eq55a}\\
\phi_*=\phi(t_*) &= \phi_i-\text{sgn}(B) \text{sgn}(\dot{\chi}_i) \arccos\left( \frac{\left|\frac{B}{A}\right|}{\chi_i} \right)\nonumber\\
&=\phi_i -\text{sgn}(v_{\phi,i}) \text{sgn}(v_{\chi,i}) \arccos\left( \frac{|v_{\phi,i}|}{\sqrt{v_{\chi,i}^2 + v_{\phi,i}^2}} \right)\nonumber\\
&=\text{sgn}(\psi_i (\pi - \psi_i)) \left( |\psi_i| - \frac{\pi}{2} \right) + \phi_i.\label{eq55b}&&
\end{flalign}
\end{subequations}
The angle $\phi_*$, corresponding to the point of closest approach, is derived from Eq.~\eqref{eq66} which will be discussed later and matches the previously proven Eq.~\eqref{eq52b} already established. The peculiar velocity components at this specific time $t_*$ are given by
\begin{subequations}
\begin{flalign}
\dot{\chi}_* =\dot{\chi}(t_*)&= 0, \label{eq56a}\\
\dot{\phi}_* =\dot{\phi}(t_*)&= \frac{A^2}{a_* B \sqrt{a_*^2 + A^2}} \nonumber\\
&=\frac{\frac{a_i (v_{\chi, i}^2 + v_{\phi, i}^2)}{a_* \chi_i v_{\phi, i}}}{\sqrt{a_*^2 (1 - v_{\chi, i}^2 - v_{\phi, i}^2) + a_i^2 (v_{\chi, i}^2 + v_{\phi, i}^2)}}\nonumber\\
&=\frac{\frac{a_i v_{\text{pec},i}}{a_* \chi_i \sin{\psi_i}}}{\sqrt{a_*^2 (1 - v_{\text{pec}, i}^2) + a_i^2 v_{\text{pec}, i}^2}}.&&\label{eq56b}
\end{flalign}
\end{subequations}
In this context, an asterisk (*) is used to denote values at time $t_*$, indicating the moment that particle reaches its closest distance $\chi_*$ to the center. The condition for determining $t_*$ in terms of the initial condition $(\chi_i, A, B)$
\begin{equation}
\int_{t_i}^{t_*} \frac{|A|dt'}{a(t') \sqrt{a^2(t') + A^2}}  + \text{sgn}(\dot\chi_i)\sqrt{\chi_i^2 - \frac{B^2}{A^2}} = 0.
\label{eq57}
\end{equation}
It is important to note that this point of closest approach defined by the coordinates in Eqs.~\eqref{eq55a} and~\eqref{eq55b}, is observed only in the comoving frame. However, in the physical frame, a different point of minimum approach might exist at time $t_\text{min}$, determined by the following conditions
\begin{subequations}
\begin{align}
&\dot\chi_\text{phy}(t_\text{min})=\frac{d}{dt}\left(a(t)\chi(t)\right)\biggl|_{t=t_\text{min}}=0,\\
&\ddot\chi_\text{phy}(t_\text{min})
\label{eq58}=\frac{d^2}{dt^2}\left(a(t)\chi(t)\right)\biggl|_{t=t_\text{min}}>0.
\end{align}
\end{subequations}
\subsection{Geometrical Interpretation for Non-Radial Solution}\label{sec4.5}
Now, let's focus on the geometrical representation of the geodesic trajectory in the comoving frame. Clearly, free geodesics in the comoving frame are straight lines with an inclination angle $\phi_i+\psi_i$ (the initial launch angle), a claim substantiated in Appendix~\ref{A3}. Without loss of generality, Figure~\ref{fig1} depicts such a general geodesic trajectory with in the $(xy)$ plane relative to a comoving reference frame with the origin point $O$. From the expression for radial distance~\eqref{eq52a}, we can see a precise match with the right-angled triangle $O M_t M_*$ in Figure~\ref{fig1}, having sides $R(t)$, $\chi_*$, and the hypotenuse $\chi(t)$. They are given as follows
\begin{figure}[H]
\centering
\begin{tikzpicture}
\draw (2,2) -- (10,10); 
\draw[dashed] (6,2) -- (4,4); 
\draw[dotted] (4,4) -- (2.5,5.5);
\draw[dashed] (6,2) -- (26/3,10); 
\draw[dotted] (8,8) -- (6.5,9.5);
\draw[dotted] (6,6) -- (5.3,6.7);
\draw[dashed] (6,2) -- (6,6); 
\fill (6,2) circle (2pt);
\fill (4,4) circle (2pt);
\fill (8,8) circle (2pt);
\fill (6,6) circle (2pt);
\node at (7.2, 4.8) {$\chi_i$};
\node at (3.9, 5.9) {$R(t)$};
\node[rotate=45] at (6, 8) {$\int_{t_i}^t \frac{|A| \, dt'}{a(t') \sqrt{a^2(t') + A^2}}$};
\node[rotate=90] at (6.2, 4.2) {$\chi(t)$};
\node[rotate=45] at (4, 7.6) {$\sqrt{\chi_i^2-\frac{B^2}{A^2}}$};
\node[rotate=-45] at (4.65, 2.85) {$\chi_*=\left|\frac{B}{A}\right|$};
\node[rotate=45] at (9.4,9.7) {\scriptsize trajectory};
\node at (8.65, 7.9) {\scriptsize $M_i$ at $t_i$};
\node at (4.4, 4) {\scriptsize $M_*$};
\node at (6.6,5.9) {\scriptsize $M_t$ at $t$};
\node at (7,2.9) {$\phi(t)$};
\node at (6.6,2.4) {$\phi_i$};
\node at (7.7,8.45) {$\psi_i$};
\node at (7.2,7.8) {$\vec v_{\text{pec},i}$};
\node at (9.9,1.7) {\large $x$};
\draw[{<[length=4mm, width=2.3mm]}-{>[length=4mm, width=2.3mm]}] (2.5,5.5) -- (6.5,9.5);
\draw[{<[length=4mm, width=2.3mm]}-{>[length=4mm, width=2.3mm]}] (3.3,4.7) -- (5.3,6.7);
\draw[{<[length=4mm, width=2.3mm]}-{>[length=4mm, width=2.3mm]}] (5.3,6.7) -- (7.3,8.7);
\node at (6, 1.7) {$O$};
\coordinate (arcCenter) at (6,2);
\draw[->] (arcCenter) ++(0:1) arc (0:90:1);
\draw[->] (arcCenter) ++(0:0.5) arc (0:71:0.5);
\draw[-{>[length=2mm, width=2mm]}] (3,2) -- (10,2);
\coordinate (arcCenter) at (8,8);
\draw[->] (arcCenter) ++(71:0.3) arc (71:225:0.3);
\draw[-{Latex[length=4mm, width=2.3mm]}, line width=1.2pt] (8,8) -- (7,7);
\end{tikzpicture}
\caption{Visualizing diagram of the geodesic trajectory in the comoving frame.}
\label{fig1}
\end{figure}
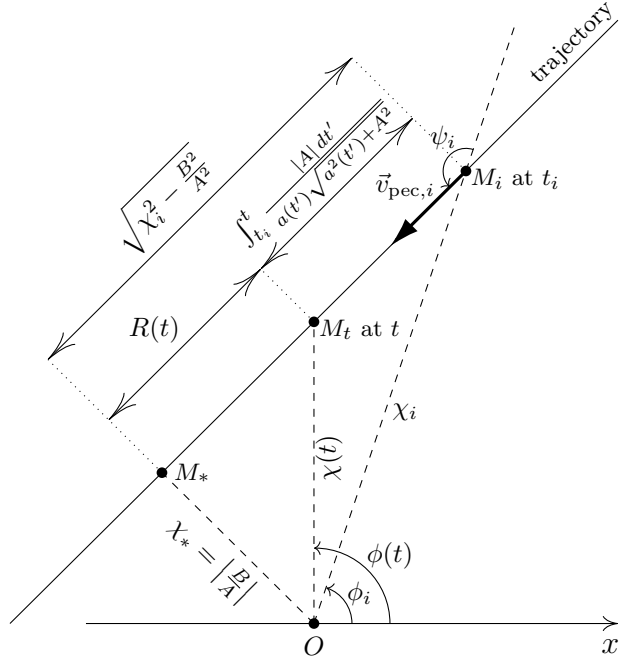
\begin{subequations}
  \begin{align}
&R(t)=M_* M_t =\left| \int_{t_i}^{t} \frac{\left| A \right| dt'}{a(t') \sqrt{a^2(t') + A^2}} + \text{sgn}(\dot{\chi}_i) \sqrt{\chi_i^2 - \frac{B^2}{A^2}} \right|\label{eq59a}&\\
&\chi_*=OM_*= \left| \frac{B}{A} \right|\label{eq59b}&\\
&\chi(t)=OM_t\label{eq59c}&
\end{align}
\end{subequations}
The Pythagorean theorem is clearly satisfied by Eq.~\eqref{eq52a} within the comoving frame's context.
\begin{equation}
\chi(t) = \sqrt{R^2(t) + \chi_*^2}.
\label{eq60}
\end{equation}
In conclusion, this analysis, together with Fig.\ref{fig1}, clarifies that the radial distance $\chi(t)$ solution in Eq.\eqref{eq52a} for general geodesics beginning at the initial comoving distance $\chi_i$ from point $O$, can be formulated in terms of the radial distance $R(t)$ for radial geodesics starting at $\sqrt{\chi_i^2 - \frac{B^2}{A^2}}$ relative to the comoving point $M_*$. Therefore, the expression of $R(t)$ can be matched with the solution~\eqref{eq29} for Radial Motion (RM) as following
\begin{equation}
R(t)=\chi_{\text{RM}}(t;\sqrt{\chi_i^2 - \frac{B^2}{A^2}},A),
\label{eq61}
\end{equation}
where we have used the redundancy of $\text{sgn}(A)$ to align it with $\text{sgn}(\dot{\chi}_i)$. In our analysis, the measurement of $\phi(t)$ remains consistent across both comoving and physical frames. As a result, $\phi(t)$ explicitly depends only on the radial distance $\chi$ and not on the scale factor $a(t)$. Accordingly, we will derive the trajectory equation geometrically, establishing the relationship between $\phi$ and $\chi$. From Fig.~\ref{fig1}, one can express $\phi(t)$ in terms of the following oriented angles
\begin{align}
\phi(t)=&\measuredangle M_iOM_t +\phi_i\nonumber\\
=&\measuredangle M_iOM_* + \measuredangle M_*OM_t + \phi_i\nonumber\\
=&\text{sgn}\left(\measuredangle M_iOM_*\right)\left|\measuredangle M_iOM_*\right| +\text{sgn}\left(\measuredangle M_*OM_t\right)\left|\measuredangle M_*OM_t\right| + \phi_i.
\label{eq62}
\end{align}
It can be shown that the sign of the oriented angles $\measuredangle M_iOM_*$ and $\measuredangle M_*OM_t$ are given by
\begin{subequations}
\begin{align}
\text{sgn}\left(\measuredangle M_iOM_*\right)&=-\text{sgn}(B) \text{sgn}(\dot\chi_i),\label{eq62a}\\
\text{sgn}\left(\measuredangle M_*OM_t\right)&=\text{sgn}(B) \text{sgn}(\dot{\chi}(t)),\label{eq62b}
\end{align}
\end{subequations}
while the absolute angles $\left|\measuredangle M_iOM_*\right|$ and $\left|\measuredangle M_*OM_t\right|$ can be straightforwardly derived in terms of the arctan from Figure~\ref{fig1} as follows
\begin{subequations}
\begin{align}
\left|\measuredangle M_iOM_*\right|&=\arctan{\left(\frac{\sqrt{\chi_i^2 - \frac{B^2}{A^2}}}{\left|\frac{B}{A}\right|}\right)},\label{eq64a}\\
\left|\measuredangle M_*OM_t\right|&=\arctan{\left(\frac{\sqrt{\chi^2(t) - \frac{B^2}{A^2}}}{\left|\frac{B}{A}\right|}\right)}.\label{eq64b}
\end{align}
\end{subequations}
Incorporating all these elements, we obtain the following result
\begin{flalign}
\phi(t) = \text{sgn}(B)\Biggl[&\text{sgn}(\dot{\chi}(t)) \arctan{\left(\frac{\sqrt{\chi^2(t) - \frac{B^2}{A^2}}}{\left|\frac{B}{A}\right|}\right)}
-\text{sgn}(\dot{\chi}_i) \arctan{\left(\frac{\sqrt{\chi_i^2 - \frac{B^2}{A^2}}}{\left|\frac{B}{A}\right|}\right)}\Biggr]+\phi_i.
\label{eq65}&&
\end{flalign}
Alternatively, using the “arccos" function, the angle can be expressed as
\begin{flalign}
\phi(t) = \text{sgn}(B) \left[\text{sgn}(\dot{\chi}(t)) \arccos\left(\frac{\left|\frac{B}{A}\right|}{\chi(t)}\right)-\text{sgn}(\dot{\chi}_i) \arccos\left(\frac{\left|\frac{B}{A}\right|}{\chi_i}\right)\right] +\phi_i.
\label{eq66}&&
\end{flalign}
In terms of
$(v_{\chi,i},v_{\phi,i})$
\begin{flalign}
\phi(t) = \text{sgn}(v_{\phi,i}) \Biggl[ &\text{sgn}(\dot{\chi}(t)) \arccos\left(\frac{\frac{\chi_i}{\chi(t)}|v_{\phi,i}|}{\sqrt{v_{\chi,i}^2 + v_{\phi,i}^2}} \right) 
- \text{sgn}(v_{\chi,i}) \arccos\left( \frac{|v_{\phi,i}|}{\sqrt{v_{\chi,i}^2 + v_{\phi,i}^2}} \right) \Biggr] + \phi_i.
\label{eq67}&&
\end{flalign}
In terms of $(v_{\text{pec},i},\psi_i)$
\begin{flalign}
\phi(t) = \text{sgn}(\psi_i (\pi - \psi_i)) \left[ \text{sgn}(\dot{\chi}(t)) \arccos\left(\frac{\chi_i |\sin(\psi_i)|}{\chi(t)}\right) + |\psi_i| - \frac{\pi}{2} \right] + \phi_i,
\label{eq68}&&
\end{flalign}
where we have used the sign relations
\begin{subequations}
\begin{align}
&\text{sgn}(\dot{\chi}_i) = \text{sgn}(v_{\chi,i}) = \text{sgn}\left(\frac{\pi}{2} - |\psi_i|\right),\label{eq69a}\\
&\text{sgn}(B) = \text{sgn}(v_{\phi,i}) = \text{sgn}(\psi_i (\pi - \psi_i)).
\label{eq69b}
\end{align}
\end{subequations}
The trajectory equation can also be expressed in terms of $\chi(t)$ just by inverting Eq.~\eqref{eq68}, or directly by using the sines law for the $O M_i M_t$ triangle as illustrated in Fig.~\ref{fig1}, one can write
\begin{equation}
\chi(t)=\chi_i\left|\frac{\sin{(\psi_i)}}{\sin{(\phi(t)-\phi_i-\psi_i)}}\right|.
\label{eq70}
\end{equation}
In our solution detailed in Eq.~\eqref{eq52b}, it appears that $\phi(t)$ explicitly depends on the scale factor $a(t)$. If the geodesic trajectory represents a straight line, then the explicit relation to the scale factor $a(t)$ should not appear in the expression for $\phi(t)$. However, we know that the trajectory indeed follows a straight line, and thus the angle $\phi(t)$ can somehow be expressed as Eq.~\eqref{eq66} without including the scale factor $a(t)$ in its expression. This angle treatment is fully addressed in Appendix~\ref{A3}, by starting from the geodesic solution Eq.~\eqref{eq52b}, leading to the same geometrical formula as given in Eq.~\eqref{eq66}, thereby concluding that the non-radial geodesics are indeed straight lines in the comoving frame.
\subsection{Radial Geodesics limit}\label{sec4.6}
The radial geodesics limit corresponds to applying the limits $B = 0$ to Eqs.~\eqref{eq52}, or $v_{\phi,i} = 0$ to Eqs.~\eqref{eq53}, or $\psi_i=0,\pi$ to Eqs.~\eqref{eq54}. We retrieve Eqs.~\eqref{eq29} and~\eqref{eq30} for radial geodesics, with $\phi(t) = \phi_i$ remaining constant. This consistency check ensures that the new radial-angular solution appropriately satisfies the correct radial limit.
\subsection{Null geodesics limit}\label{sec4.7}
The null geodesics limit is characterized by the condition $A=B=+\infty$ in Eqs.~\eqref{eq52}, or $v_{\chi, i}^2 + v_{\phi, i}^2 = 1$ in Eqs.~\eqref{eq53}, or $v_{\text{pec},i}=1$ in Eqs.~\eqref{eq54}. Applying this limit we obtain the corresponding null geodesics
\begin{subequations}
\begin{align}
&\chi(t) = \sqrt{\left( \int_{t_i}^{t} \frac{dt'}{a(t')} + \chi_i \cos{\psi_i} \right)^2 + \chi_i^2 \sin^2{\psi_i}},\label{eq71a}\\
&\phi(t) = \int_{t_i}^{t} \frac{1}{a(t')} \frac{\chi_i \sin{\psi_i}}{\chi^2(t')} dt'+ \phi_i.\label{eq71b}
\end{align}
\end{subequations}
\subsection{Comoving geodesics limit}\label{sec4.8}
If we set $A= 0\Rightarrow B=0$ in Eqs.~\eqref{eq52}, or $v_\chi =v_\phi = 0$ in Eqs.~\eqref{eq53} in Eqs.~\eqref{eq53}, or $v_{\text{pec},i}=0$ in Eqs.~\eqref{eq54}, the particle remains at a constant comoving distance.
\begin{subequations}
\begin{align}
&\chi(t) = \chi_i,\label{eq72a}\\
&\phi(t) = \phi_i.\label{eq72b}
\end{align}
\end{subequations}
Indeed, in this case, the zero-limit of peculiar velocities in both the radial and angular components implies that the particle lies with the Hubble flow.
\subsection{Symmetry and Killing vectors}\label{sec4.9}
The comoving 4-momentum can be explicitly written as follows
\begin{align}
P^\mu = m \frac{dx^\mu}{ds}  &= \frac{m}{\sqrt{1 - v_{\text{pec}}^2(t)}} (1, \dot{\chi}, 0, \dot{\phi})= m \left( \sqrt{1 + \frac{A^2}{a^2(t)}}, \frac{\text{sgn}(\dot{\chi}(t))}{a^2(t)} \sqrt{A^2 - \frac{B^2}{\chi^2(t)}}, 0, \frac{B}{a^2(t) \chi^2(t)} \right).\label{eq73}
\end{align}
On the other hand, the conjugate momentum variables $(P_\chi,P_\theta, P_\phi)$, corresponding to the comoving coordinates $(\chi,\theta, \phi)$, can be computed from the Lagrangian~\eqref{eq35} as
\begin{subequations}
\label{eq74}
\begin{align}
& P_\chi = \frac{\partial L}{\partial \dot{\chi}} = \frac{ma^2(t) \dot{\chi}}{\sqrt{1 - v_{\text{pec}}^2}} = \text{sgn}(\dot{\chi})m \sqrt{A^2 - \frac{B^2}{\chi^2}}, \label{eq74a}& \\
& P_\theta = \frac{\partial L}{\partial \dot{\theta}} = 0, &\label{eq74b} \\
& P_\phi = \frac{\partial L}{\partial \dot{\phi}} = \frac{ma^2(t) \chi^2(t) \dot{\phi}(t)}{\sqrt{1 - v_{\text{pec}}^2(t)}} = mB.\label{eq74c} 
\end{align}
\end{subequations}
The energy $E=P_t$ can be obtained by performing a Legendre transform of the Lagrangian $L$ as follows
\begin{align}
P_t = P_\chi \dot{\chi} + P_\phi \dot{\phi} - L = \frac{m}{\sqrt{1 - v_{\text{pec}}^2}} = m \sqrt{1 + \frac{A^2}{a^2(t)}}.
\label{eq75}
\end{align}
Consistency between the comoving 4-momentum $P^\mu$ and the conjugate momentum $P_\mu$ can be shown by using the metric tensor $g^{\mu\nu}$ and $g_{\mu\nu}$ defined by the spacetime interval~\eqref{eq33} to raise and lower indices, respectively. Now, the energy-momentum relation (mass-shell condition) can be verified
\begin{equation}
g_{\mu\nu}P^\mu P^\nu=E^2-p^2=m^2,
\label{eq76}
\end{equation}
where the physical 3D momentum $p$ is defined as
\begin{equation}
p(t)=a(t)\sqrt{\gamma_{ij}P^i P^j}.
\label{eq77}
\end{equation}
In cosmology, it is well-known that the magnitude of the comoving 3D conjugate momentum is a conserved quantity for free motion. Consequently, we can infer that the constant of motion $A$ is expressed as follows
\begin{equation}
A^2=\frac{a^4(t)\gamma_{ij}P^i P^j}{m^2}=\frac{\gamma^{ij}P_i P_j}{m^2}=\frac{a^2(t)p^2(t)}{m^2}.
\label{eq78}
\end{equation}
Now, as previously mentioned, the constant of motion $B$ is related to the rotational symmetry described in Eq.~\eqref{eq39} corresponding to a Killing vector $\xi_B=\partial_\phi$. However, the symmetry associated with the constant of motion $A$ remains undetermined. Using Noether's theorem~\cite{Noether}, we can express the conserved quantities associated with a spatial symmetry $\delta x^i$ of this dynamical system as follows
\begin{equation}
\frac{\partial L}{\partial \dot{x}^i} \delta x^i = P_i \delta x^i.
\label{eq79}
\end{equation}
where $\delta x^i$ represents an infinitesimal symmetry transformation of the Lagrangian~\eqref{eq35}. To associate the constant of motion $A^2$ given in Eq.~\eqref{eq78} with such a symmetry, $\delta x^i$ must assume the form
\begin{equation}
\delta x^i =\varepsilon\xi_A^i= \frac{\varepsilon}{m} \gamma^{ij}P_j,
\label{eq80}
\end{equation}
for an infinitesimally small parameter $\varepsilon$ and $\xi_A$ represents the Killing vector for this symmetry. Therefore, the specific form of the spatial time-dependent infinitesimal transformation is explicitly given by
\begin{subequations}
\begin{align}
\chi &\longrightarrow \chi' = \chi + \frac{\varepsilon}{m} P_{\chi},\label{eq81a}\\
\phi &\longrightarrow \phi' = \phi + \frac{\varepsilon}{m} \frac{P_{\phi}}{\chi^2},\label{eq81b}
\end{align}
\end{subequations}
where the conjugate momenta $P_\chi$ and $P_\phi$ are expressed in Eqs.~\eqref{eq74a} and~\eqref{eq74c}, respectively, and satisfy the Euler-Lagrange equations~\eqref{eq37a} and~\eqref{eq37b} as follows
\begin{subequations}
\begin{align}
\dot P_\chi&=\frac{\dot\phi}{\chi}P_\phi,\label{eq82a}\\
\dot P_\phi&=-2\frac{\dot\chi}{\chi^3}P_\phi.\label{eq82b}
\end{align}
\end{subequations}
According to Noether's theorem, the transformations given in Eqs.~\eqref{eq81a} and~\eqref{eq81b} lead to the constant of motion $A^2$ in Eq.~\eqref{eq79}. It can be checked that this transformation indeed represents a symmetry of the Lagrangian~\eqref{eq35}. The Killing vector $\xi_A$ related to this symmetry can be expressed as
\begin{align}
\xi_A &=\frac{1}{m} P_\chi \partial_\chi + \frac{1}{m} \frac{P_\phi}{\chi^2} \partial_\phi = \pm \sqrt{A^2 - \frac{B^2}{\chi^2}} \partial_\chi + \frac{B}{\chi^2} \partial_\phi
\label{eq83}
\end{align}
Two points should be highlighted:
\begin{itemize}[leftmargin=*]
\item Eqs.~\eqref{eq81a} and~\eqref{eq81b} precisely satisfies the expected radial limit described by the translation~\eqref{eq19}, as it results $P_\chi=\text{constant}$ and $P_\phi=0$. Similarly, by setting $B=0$ in Eq.~\eqref{eq83}, we get the Killing vector corresponding to radial translation symmetry $\xi_A=\partial_\chi$.
\item The Killing vectors $\xi_A$ and $\xi_B$ related to the constants of motion $A$ and $B$, respectively, indeed satisfy the Killing equation (isometry condition) 
\begin{align}
\nabla_\mu \xi_\nu + \nabla_\nu \xi_\mu = 0
\end{align}
with $\nabla_\mu$ is the covariant derivative given by the connection~\eqref{eqA3}.
\end{itemize}
\section{Application to $\Lambda$CDM Model}\label{sec5}
The $\Lambda$CDM is a scientific framework for comprehending the universe, including its evolution, structure, and composition. It is based on observations such as the distribution of galaxies and the CMB. By comparing theoretical predictions with actual observations, the model's accuracy is assessed, leading to its continuous refinement. This model states that the universe is composed of basic components, each with a dimensionless density parameter at the present time of about (according to Planck Collaboration 2018).
\begin{align*}
\text{Radiation:} & \quad \Omega_r \approx 9 \times 10^{-5} \\
\text{Matter:} & \quad \Omega_m \approx 0.315 \\
\text{Cosmological constant:} & \quad \Omega_\Lambda \approx 0.685
\end{align*}
and it is based on the Friedmann-Lemaître equation
\begin{equation}
  \frac{\dot{a}(t)}{a(t)} = H(t) = H_0 \sqrt{\frac{\Omega_r}{a^4(t)} + \frac{\Omega_m}{a^3(t)} +  \Omega_\Lambda}
  \label{eq85}.
\end{equation}
$H_0$ is the Hubble parameter today (according to Plank 2018): $H_0 \approx 67.4 \, \left(\frac{\text{km} \cdot \text{s}^{-1}}{\text{Mpc}}\right)$.
We have considered that our 3D space is flat, with the curvature density parameter $\Omega_k=0$. If we neglect the radiation density parameter $\Omega_r\approx0$, it becomes straightforward to analytically solve the Friedman-Lemaître equation~\eqref{eq85} for the scale factor $a(t)$ as follows
\begin{equation}
a(t) = \left(\frac{\Omega_m}{\Omega_\Lambda}\right)^{1/3} \sinh^{2/3}\left(\frac{3}{2} \sqrt{\Omega_\Lambda}H_0t\right)
\label{eq86}.
\end{equation}
\subsection{Free Geodesics in $\Lambda$CDM model}\label{sec5a}
In this section, we use thee radial-angular geodesic solution~\eqref{eq54} determined by the initial conditions $(\chi_i,\phi_i,v_{\text{pec},i},\psi_i)$ to study freely-falling particles in FLRW spacetime within the context of the $\Lambda$CDM model. For this investigation, we set the initial time $t_i$ for the conditions at the current time, $t_i = 13.79$ Gyr. In what follows, we adopt the units of distance and time in billions of light-years (Gly) and billions of years (Gyr), respectively, so that $H_0^{-1} \approx 14.51$ Gyr. We position ourselves as a comoving observer at the origin $\chi = 0$ disregarding any peculiar velocities attributable to the Milky Way Galaxy, the Solar System, or Earth's own motion. We shall compute the integrals and generate the corresponding curves numerically. We now aim to create a series of illustrative graphs to visualize the dynamics of freely-falling particles in FLRW spacetime, based on specific initial conditions. The graphs we plan to generate include:
\begin{enumerate}
    \item The comoving radial distance $\chi$ plotted over time $t$.
    \item The trajectories of geodesics in the comoving $(xy)$ plane.
        \item The physical radial distance $\chi_\text{phy}$ plotted over time $t$.
    \item The trajectories of geodesics in the physical $(x_{\text{phy}},y_{\text{phy}})$ plane.
    \item The variation of the angle $\phi$ over time $t$.
\end{enumerate}
To construct these graphs, we use the established geodesic solution for the radial distance and angle relations, as detailed in~\eqref{eq54a} and~\eqref{eq54b}, respectively, along with the expression of the scale factor provided in Eq.~\eqref{eq86}. The physical radial distance is determined by multiplying the comoving radial distance by the scale factor $a(t)$, as follows
\begin{equation}
\chi_\text{phy}(t;\chi_i,v_{\text{pec},i},\psi_i)=a(t)\chi(t;\chi_i,v_{\text{pec},i},\psi_i).
\label{eq87}
\end{equation}
The geodesic trajectories will be traced in both the comoving and physical frames using their respective parametric relations. For the comoving trajectories
\begin{equation}
\left\{
\begin{aligned}
x(t) &= \chi(t)\cos{\phi(t)}\\
y(t) &= \chi(t)\sin{\phi(t)}
\end{aligned}
\right.\label{eq88},
\end{equation}
for the physical trajectories
\begin{equation}
\left\{
\begin{aligned}
x_\text{phy}(t) &= a(t)\chi(t)\cos{\phi(t)}\\
y_\text{phy}(t) &= a(t)\chi(t)\sin{\phi(t)}
\end{aligned}
\right.\label{eq89}.
\end{equation}
In addition to these graphical representations, we will augment our analysis of each geodesic trajectory with their corresponding animations for both the comoving and physical frames. Please refer to the Supplemental Material, for a dynamic visualization of the motion of freely-falling particles in our spacetime over time.

\begin{figure*}[p]
\centering
\begin{subfigure}{0.48\textwidth}
  \centering
\includegraphics[width=0.8\linewidth]{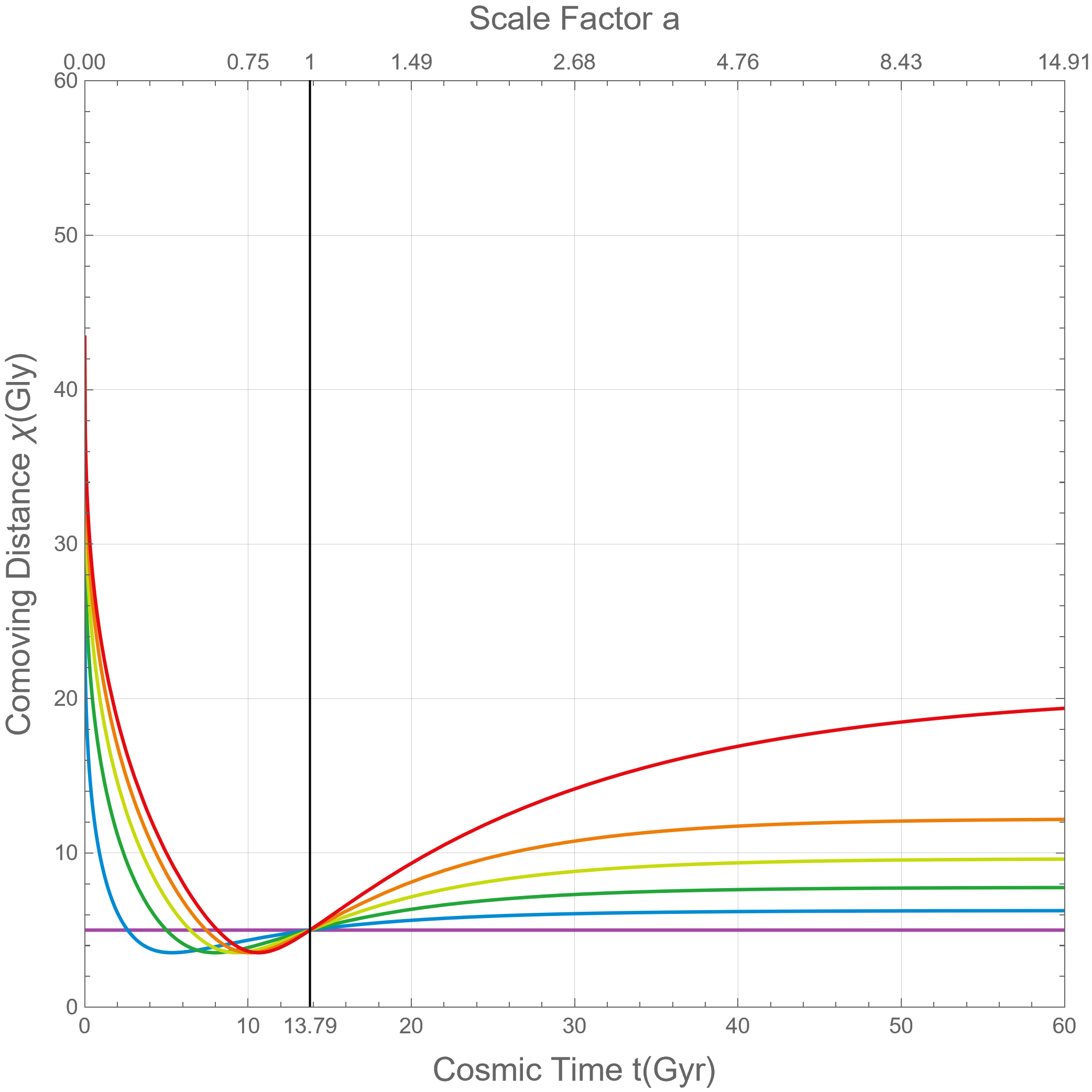}
  \caption{}
  \label{fig2a}
\end{subfigure}
\hfill
\begin{subfigure}{0.48\textwidth}
  \centering
\includegraphics[width=0.8\linewidth]{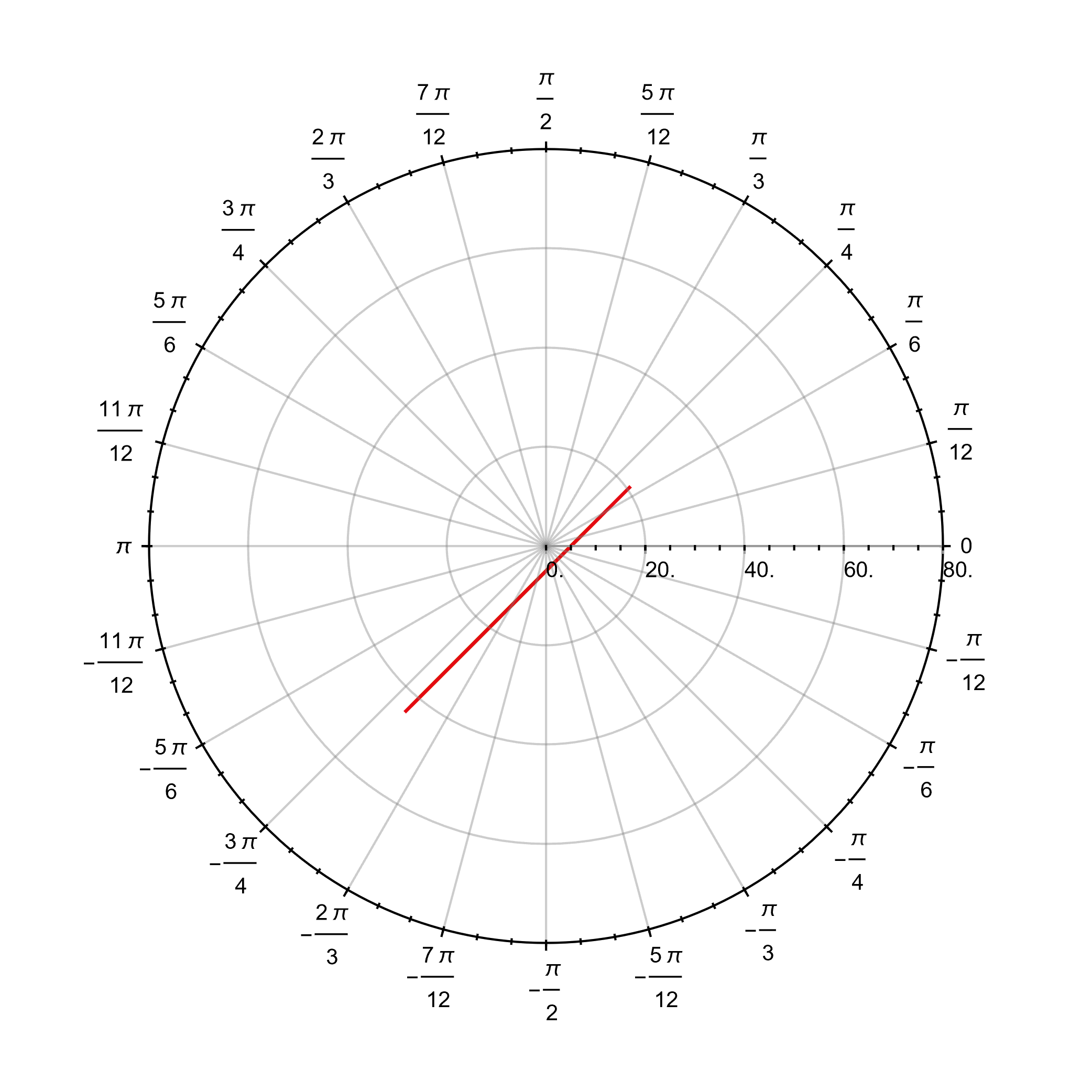}
  \caption{See Online Resource at~\cite{OnlineResource1} to view animation.}
  \label{fig2b}
\end{subfigure}
\begin{subfigure}{0.48\textwidth}
  \centering
\includegraphics[width=0.8\linewidth]{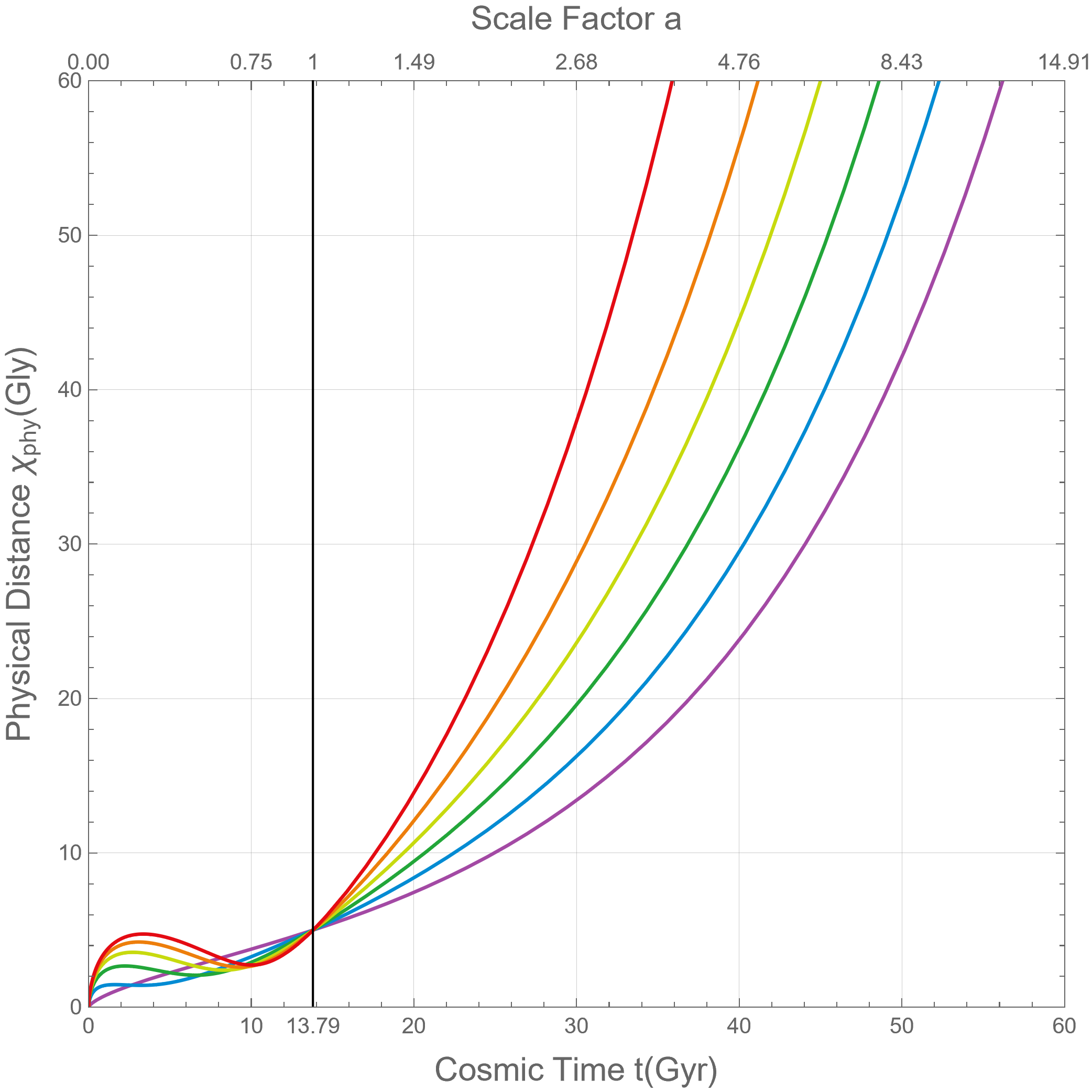}
  \caption{}
  \label{fig2c}
\end{subfigure}
\hfill
\begin{subfigure}{0.48\textwidth}
  \centering
\includegraphics[width=0.8\linewidth]{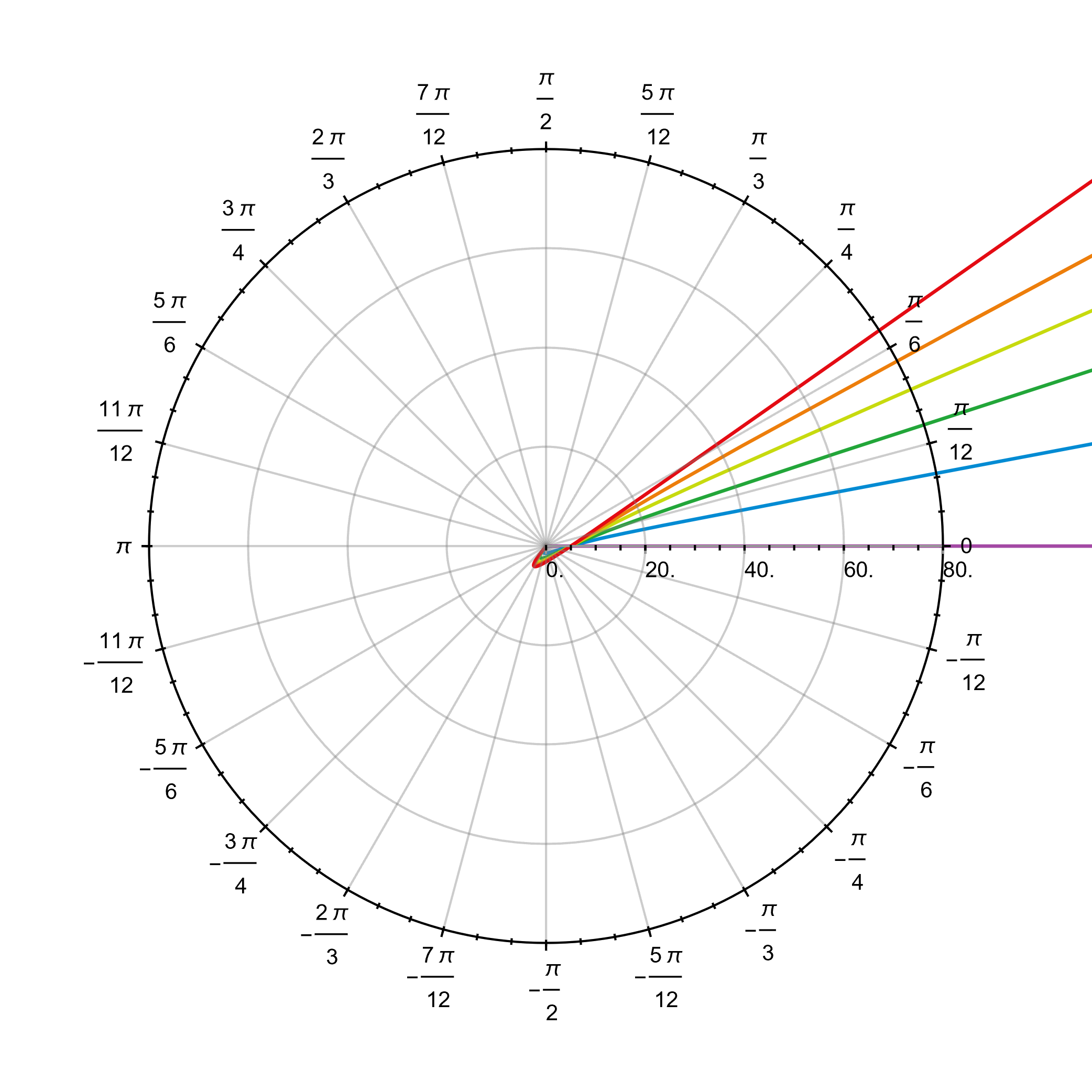}
  \caption{See Online Resource at~\cite{OnlineResource2} to view animation.}
  \label{fig2d}
\end{subfigure}
\begin{subfigure}{0.48\textwidth}
  \centering
\includegraphics[width=0.8\linewidth]{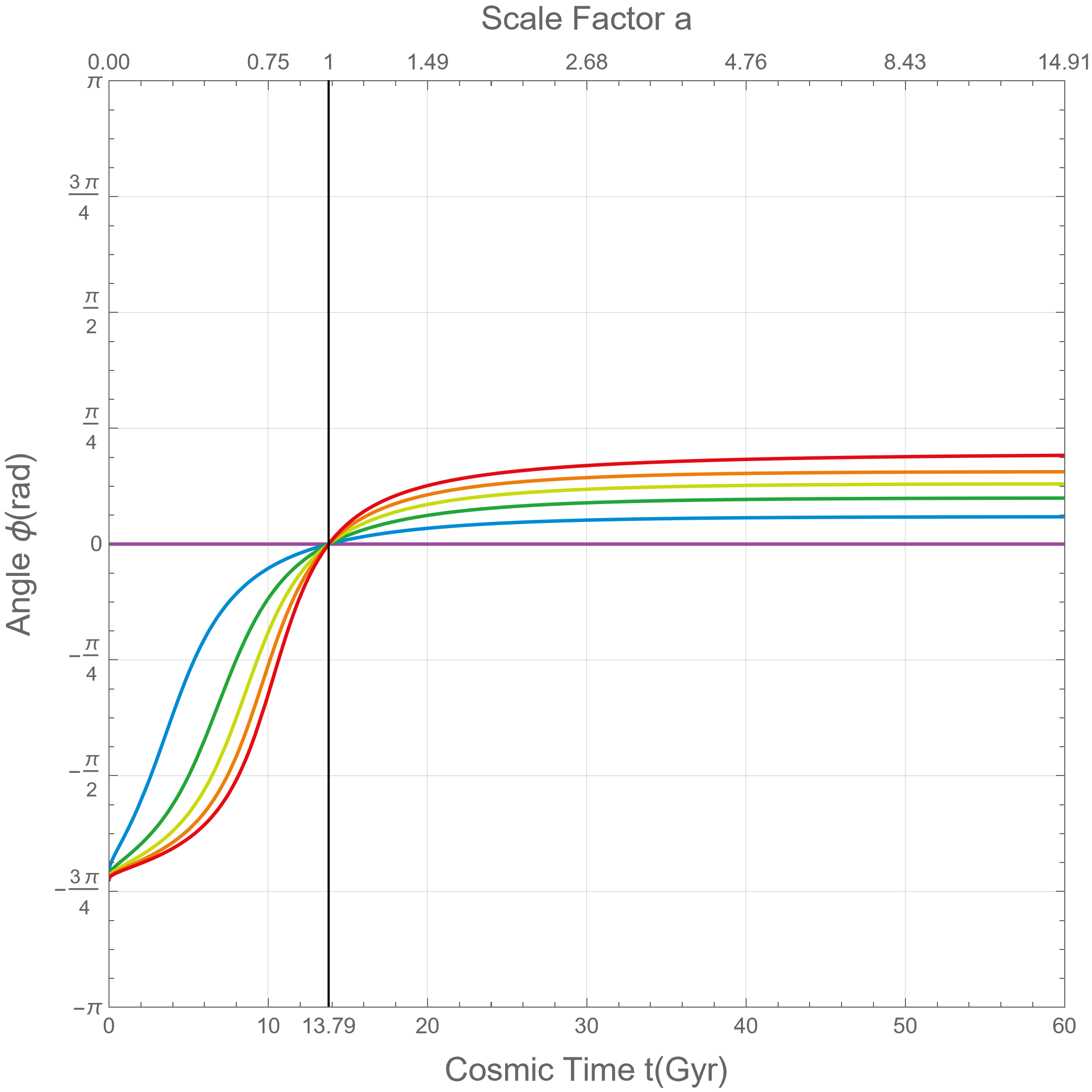}
  \caption{}
  \label{fig2e}
\end{subfigure}
\hfill
\begin{subfigure}{0.48\textwidth}
  \centering
\includegraphics[width=0.3\linewidth]{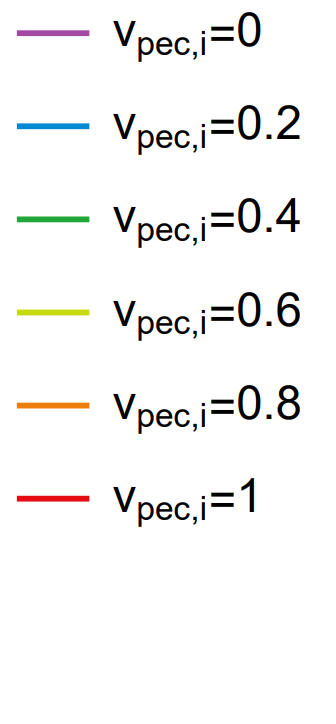}
\end{subfigure}
\caption{Dynamics of Freely-Falling Particles in FLRW Spacetime with initial conditions: $(\chi_i,\phi_i,v_{\text{vec},i},\psi_i)=(5 \ \text{Gly},0,v_{\text{vec},i},\frac{\pi}{4})$ for different peculiar velocity magnitudes $v_{\text{vec},i}$.}
\label{fig2}
\end{figure*}

\begin{figure*}[p]
\centering
\begin{subfigure}{0.48\textwidth}
  \centering
\includegraphics[width=0.8\linewidth]{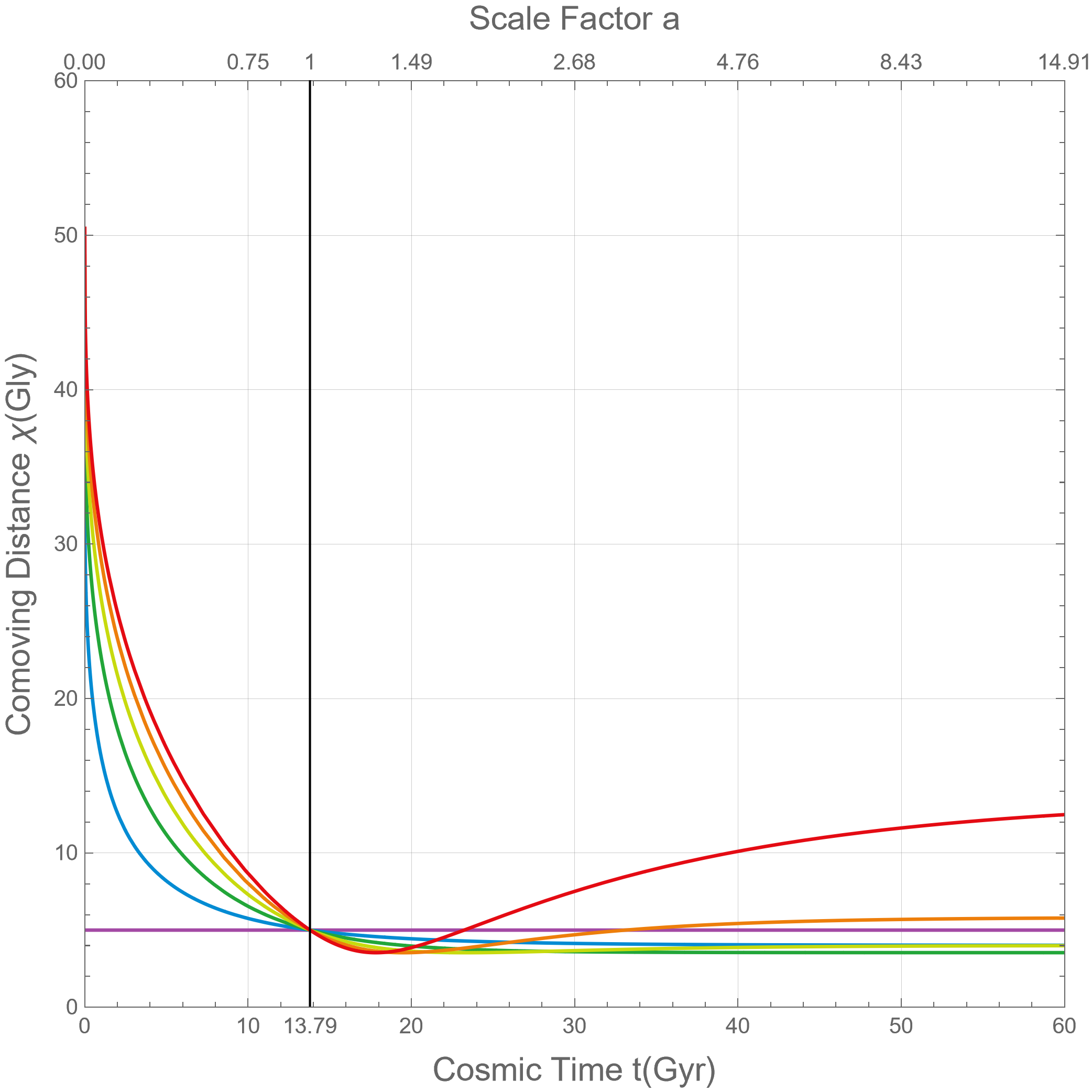}
  \caption{}
  \label{fig3a}
\end{subfigure}
\hfill
\begin{subfigure}{0.48\textwidth}
  \centering
\includegraphics[width=0.8\linewidth]{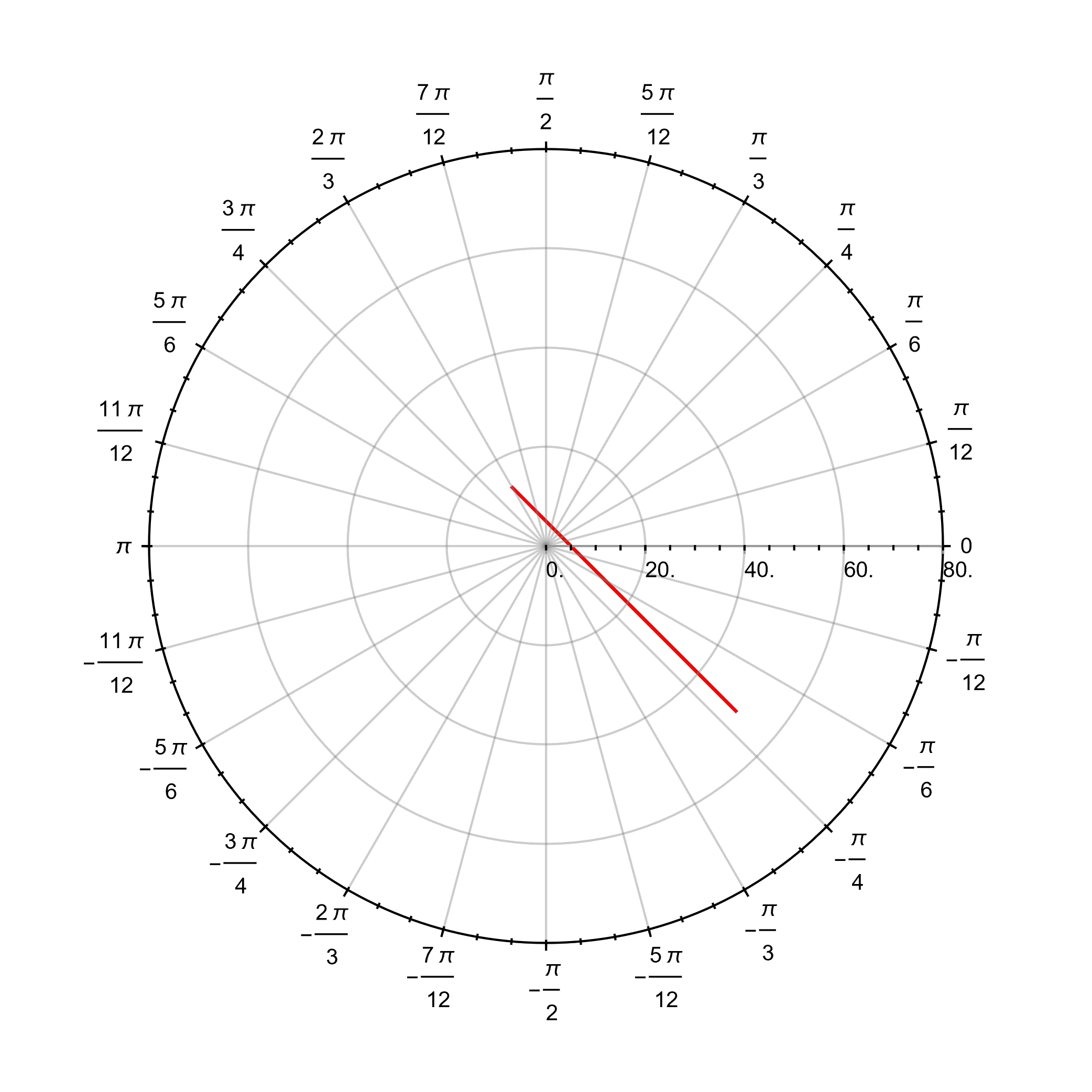}
  \caption{See Online Resource at~\cite{OnlineResource3} to view animation.}
  \label{fig3b}
\end{subfigure}
\begin{subfigure}{0.48\textwidth}
  \centering
\includegraphics[width=0.8\linewidth]{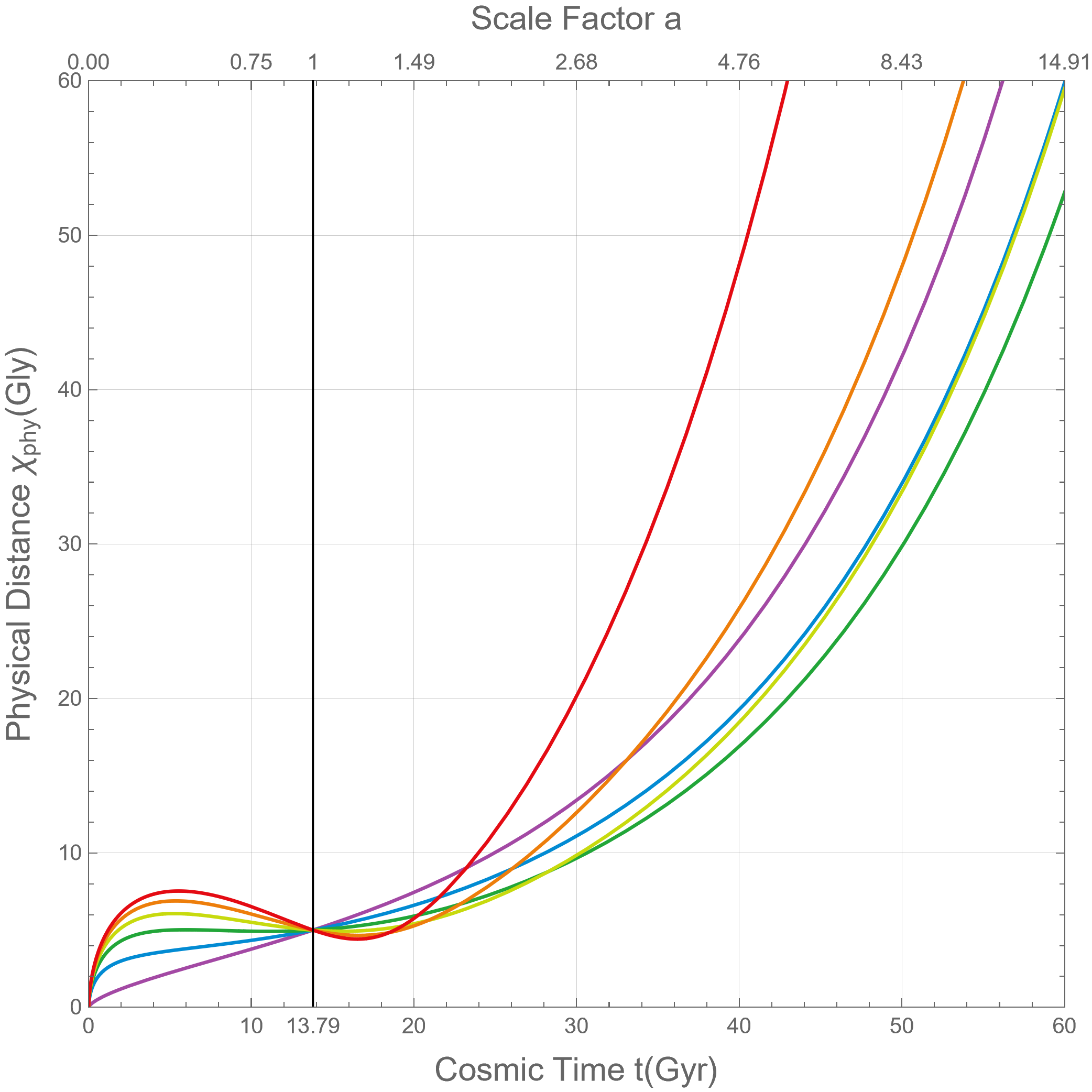}
  \caption{}
  \label{fig3c}
\end{subfigure}
\hfill
\begin{subfigure}{0.48\textwidth}
  \centering
\includegraphics[width=0.8\linewidth]{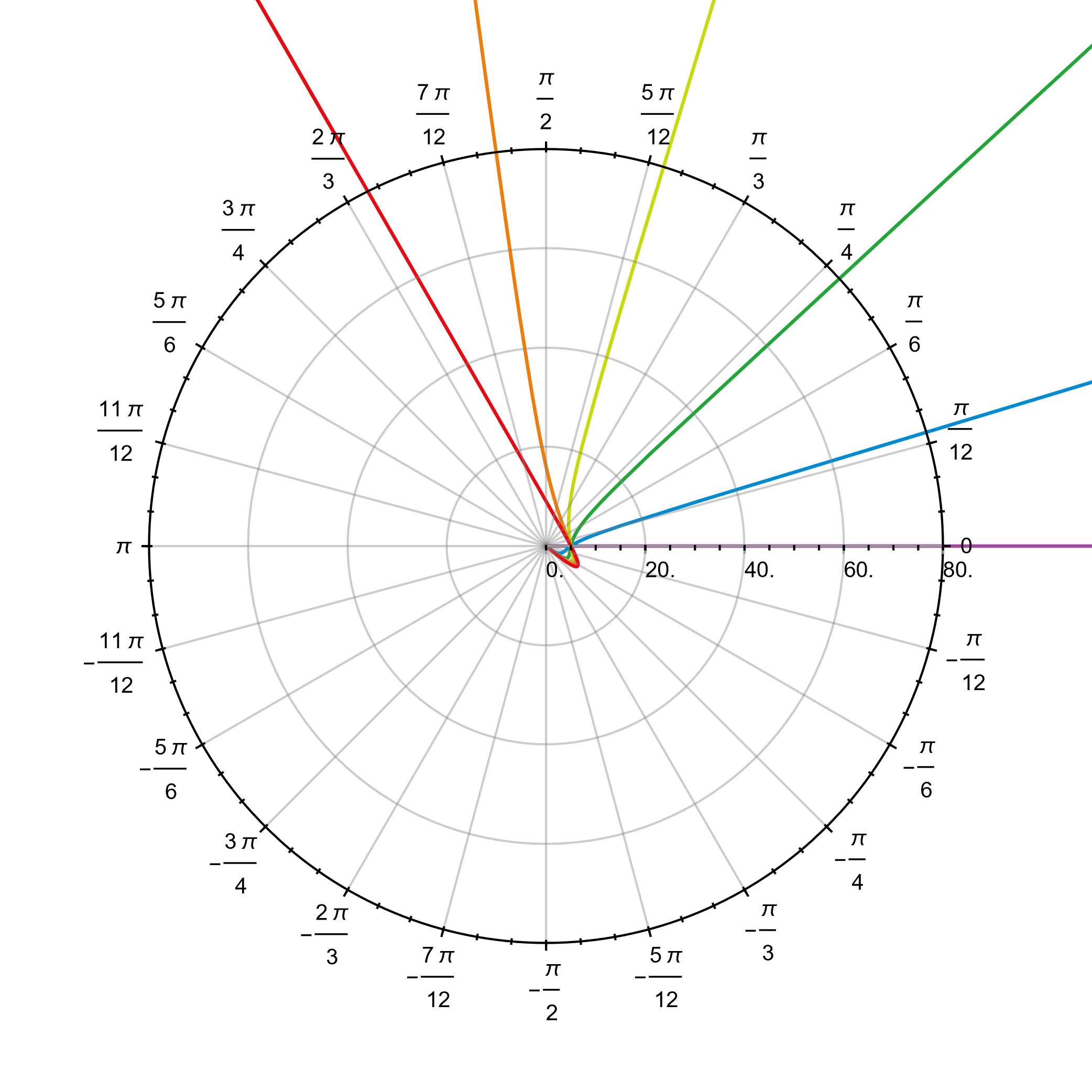}
  \caption{See Online Resource at~\cite{OnlineResource4} to view animation.}
  \label{fig3d}
\end{subfigure}
\begin{subfigure}{0.48\textwidth}
  \centering
\includegraphics[width=0.8\linewidth]{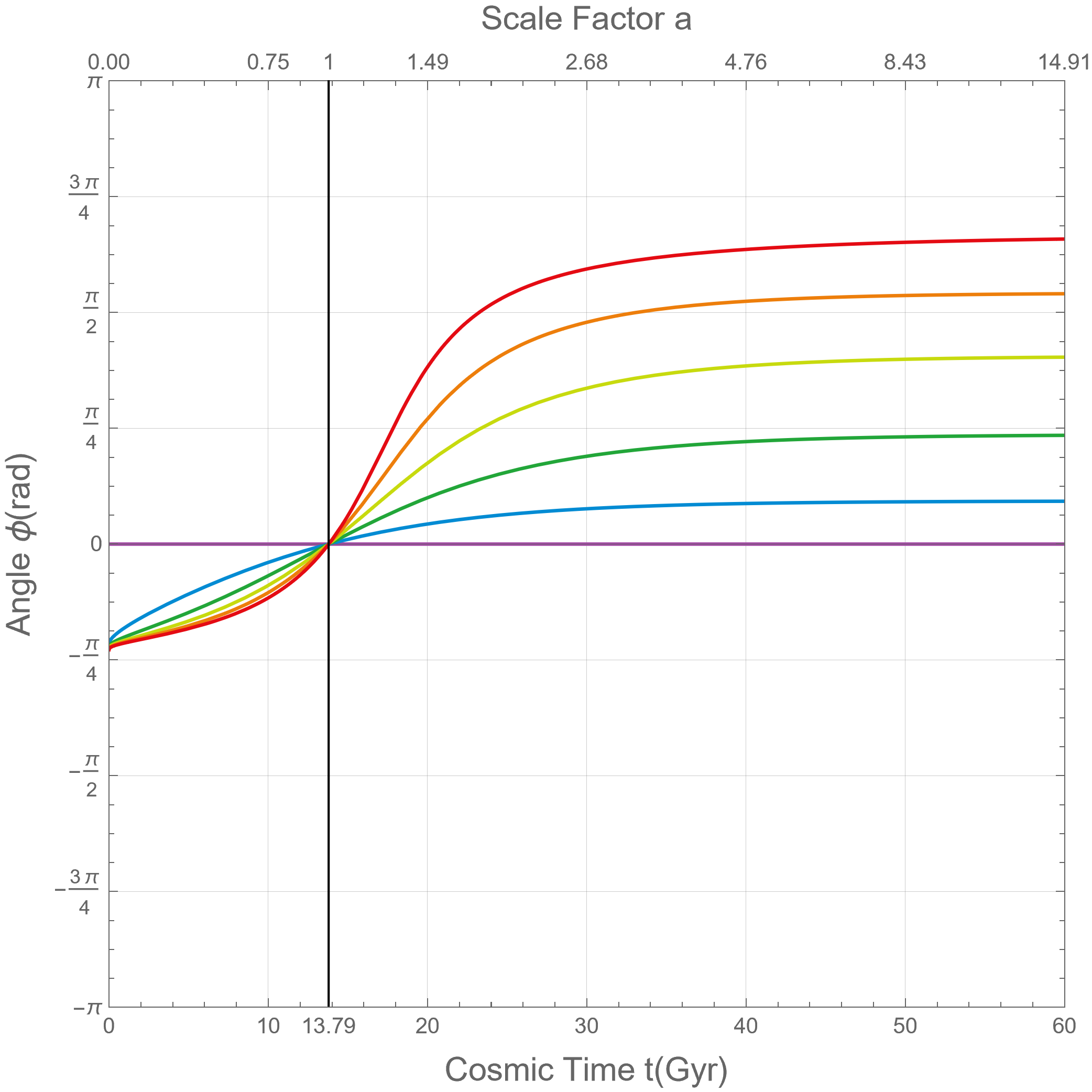}
  \caption{}
  \label{fig3e}
\end{subfigure}
\hfill
\begin{subfigure}{0.48\textwidth}
  \centering
\includegraphics[width=0.3\linewidth]{velocity.png}
\end{subfigure}
\caption{Dynamics of Freely-Falling Particles in FLRW Spacetime with initial conditions: $(\chi_i,\phi_i,v_{\text{vec},i},\psi_i)=(5 \ \text{Gly},0,v_{\text{vec},i},\frac{3\pi}{4})$ for different peculiar velocity magnitudes $v_{\text{vec},i}$.}
\label{fig3}
\end{figure*}

\begin{figure*}[p]
\centering
\begin{subfigure}{0.48\textwidth}
  \centering
\includegraphics[width=0.8\linewidth]{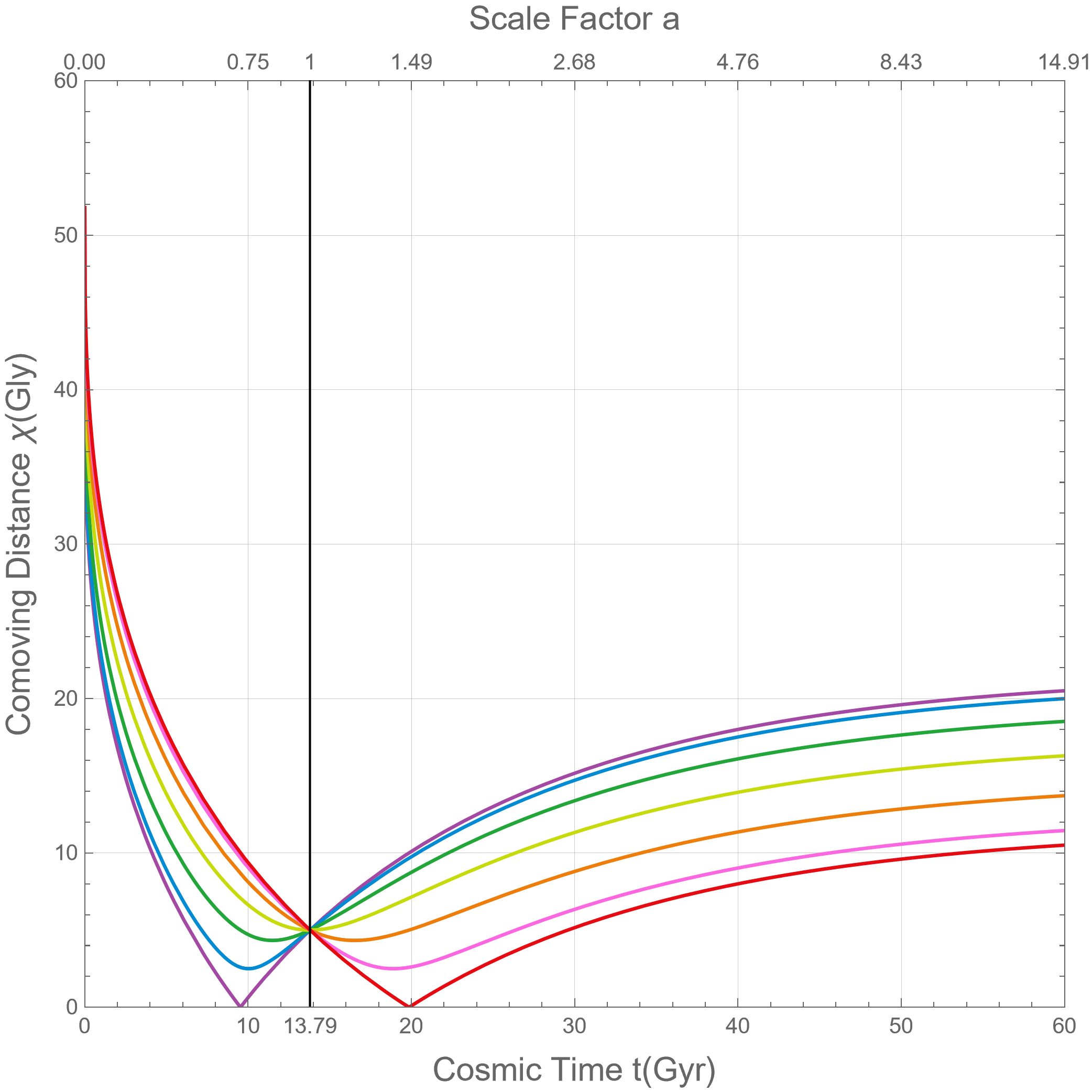}
  \caption{}
  \label{fig4a}
\end{subfigure}
\hfill
\begin{subfigure}{0.48\textwidth}
  \centering
\includegraphics[width=0.8\linewidth]{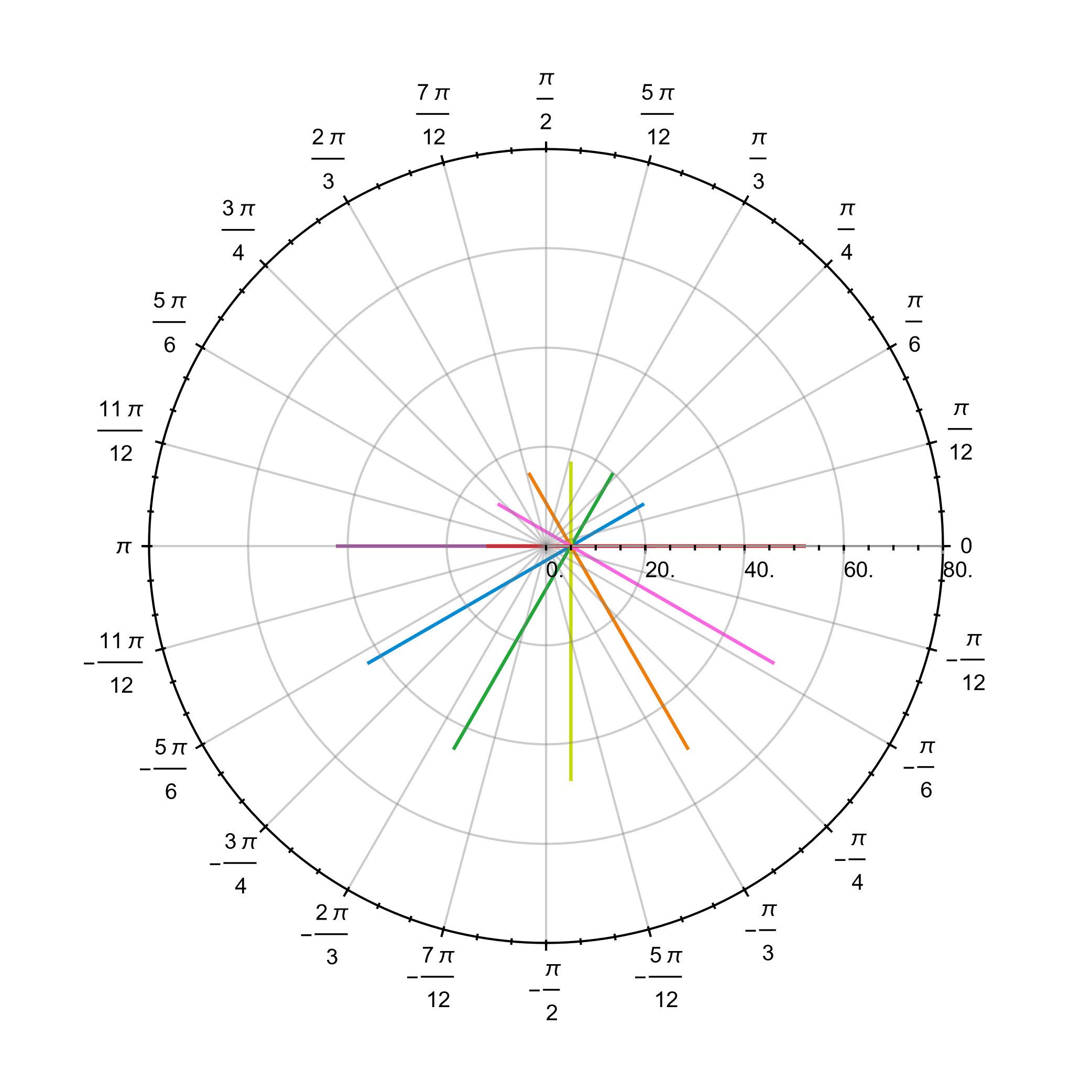}
  \caption{See Online Resource at~\cite{OnlineResource5} to view animation.}
  \label{fig4b}
\end{subfigure}
\begin{subfigure}{0.48\textwidth}
  \centering
\includegraphics[width=0.8\linewidth]{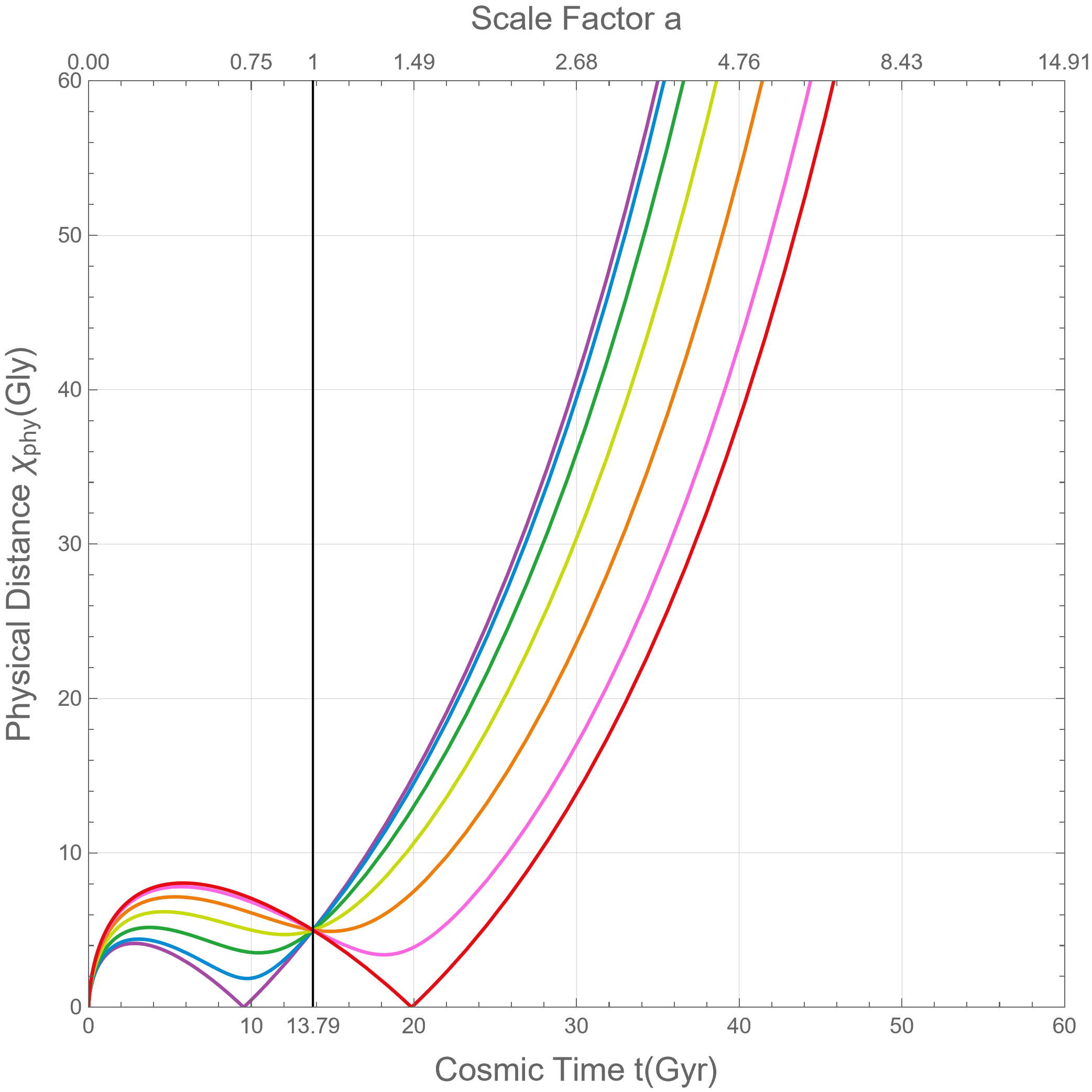}
  \caption{}
  \label{fig4c}
\end{subfigure}
\hfill
\begin{subfigure}{0.48\textwidth}
  \centering
\includegraphics[width=0.8\linewidth]{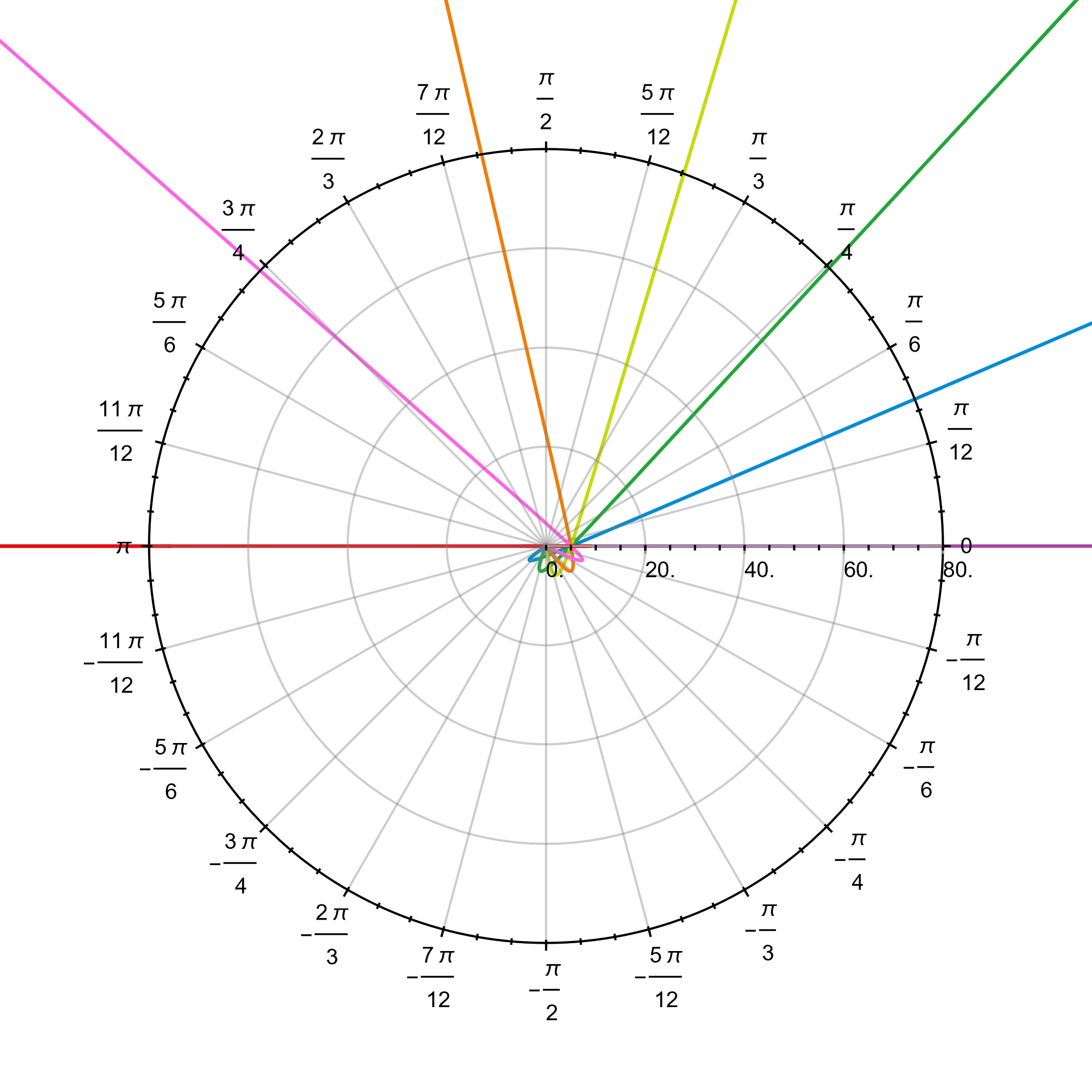}
  \caption{See Online Resource at~\cite{OnlineResource6} to view animation.}
  \label{fig4d}
\end{subfigure}
\begin{subfigure}{0.48\textwidth}
  \centering
\includegraphics[width=0.8\linewidth]{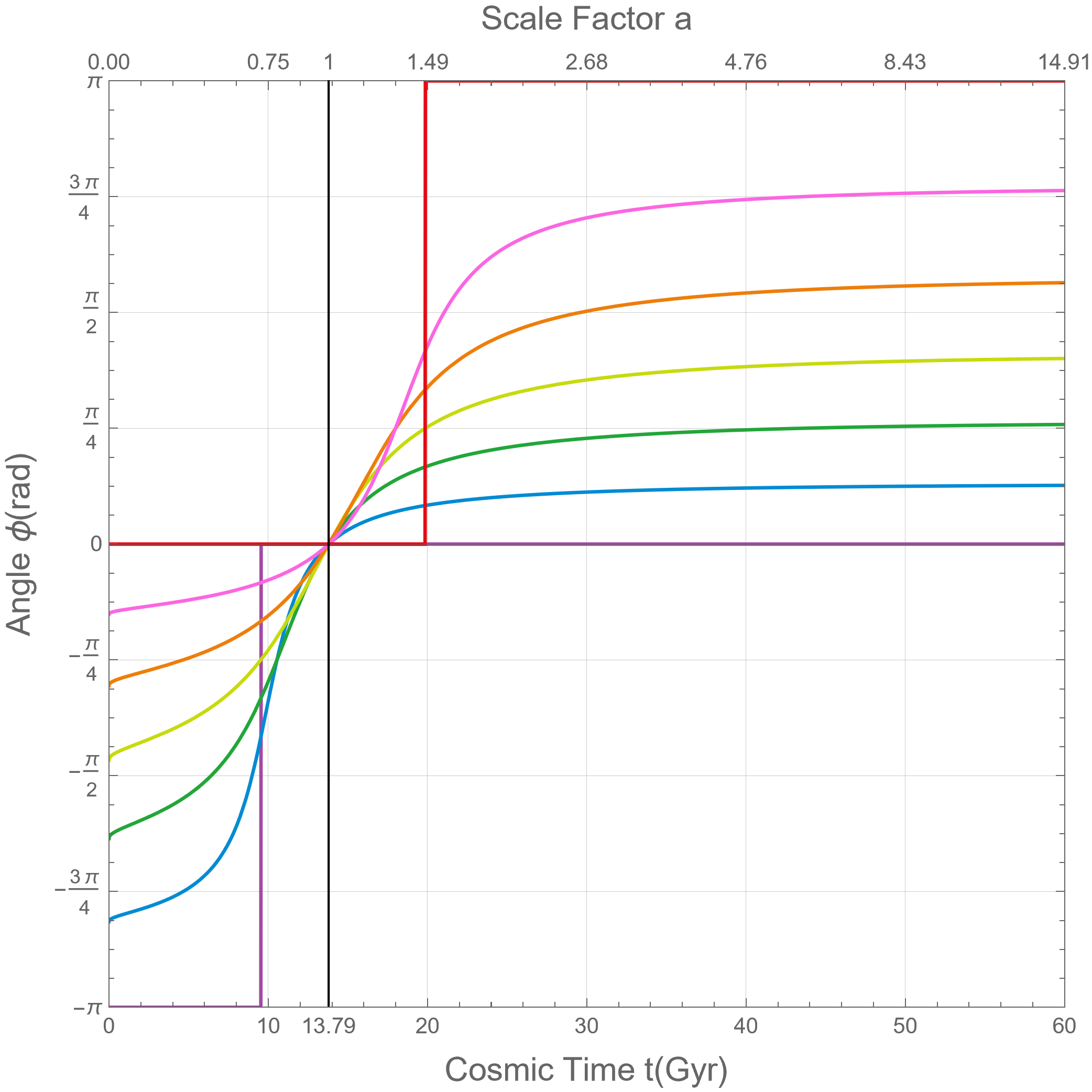}
  \caption{}
  \label{fig4e}
\end{subfigure}
\hfill
\begin{subfigure}{0.48\textwidth}
  \centering
\includegraphics[width=0.3\linewidth]{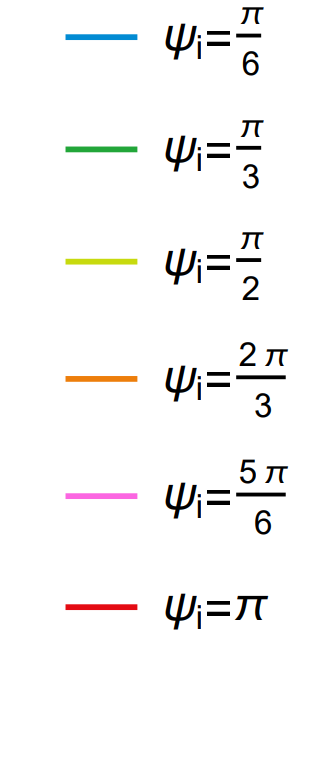}
\end{subfigure}
\caption{Dynamics of Freely-Falling Particles in FLRW Spacetime with initial conditions: $(\chi_i,\phi_i,v_{\text{vec},i},\psi_i)=(5 \ \text{Gly},0,1,\psi_i)$ for different initial peuliar velocity angles $\psi_i$.}
\label{fig4}
\end{figure*}

\begin{figure*}[p]
\centering
\begin{subfigure}{0.48\textwidth}
  \centering
\includegraphics[width=0.8\linewidth]{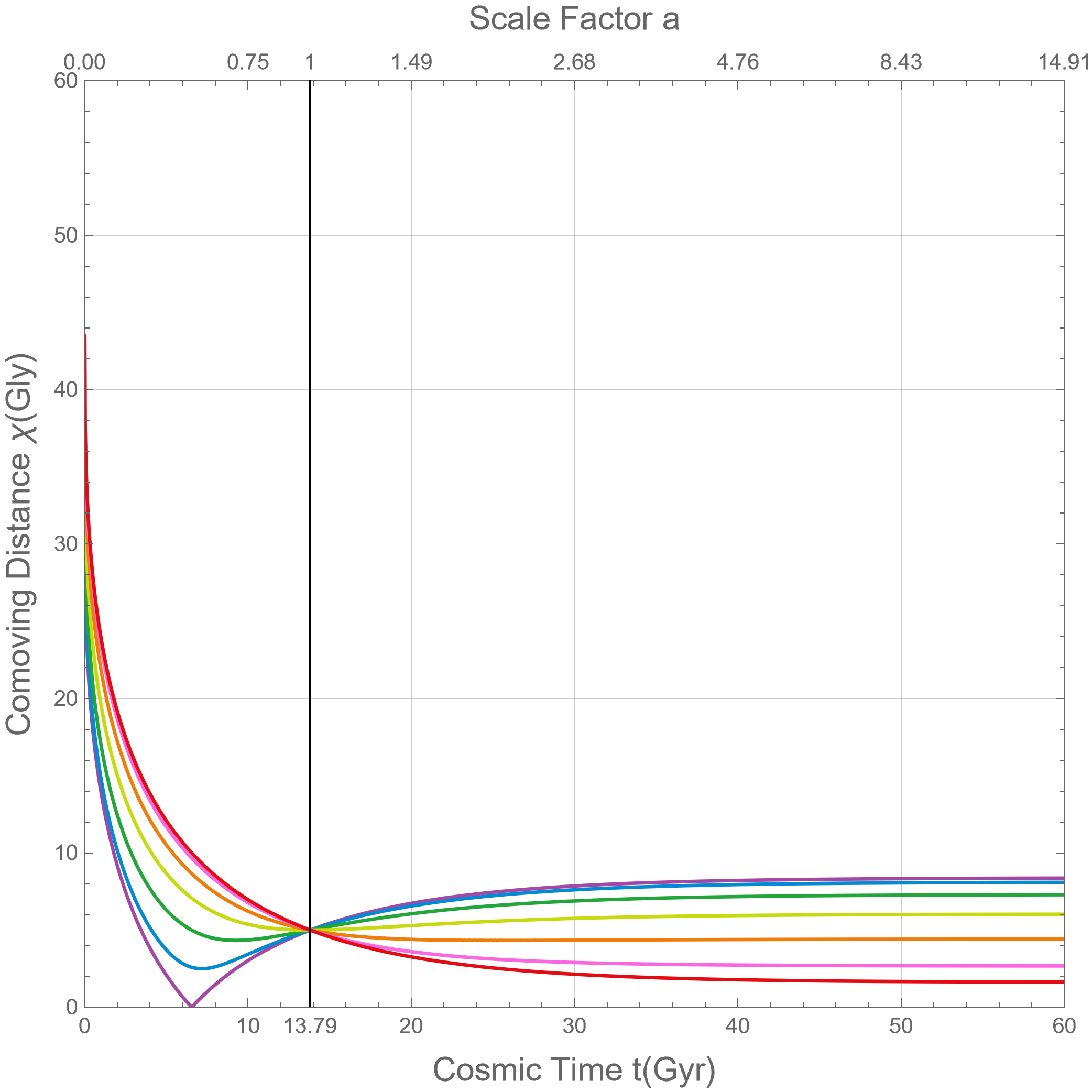}
  \caption{}
  \label{fig5a}
\end{subfigure}
\hfill
\begin{subfigure}{0.48\textwidth}
  \centering
\includegraphics[width=0.8\linewidth]{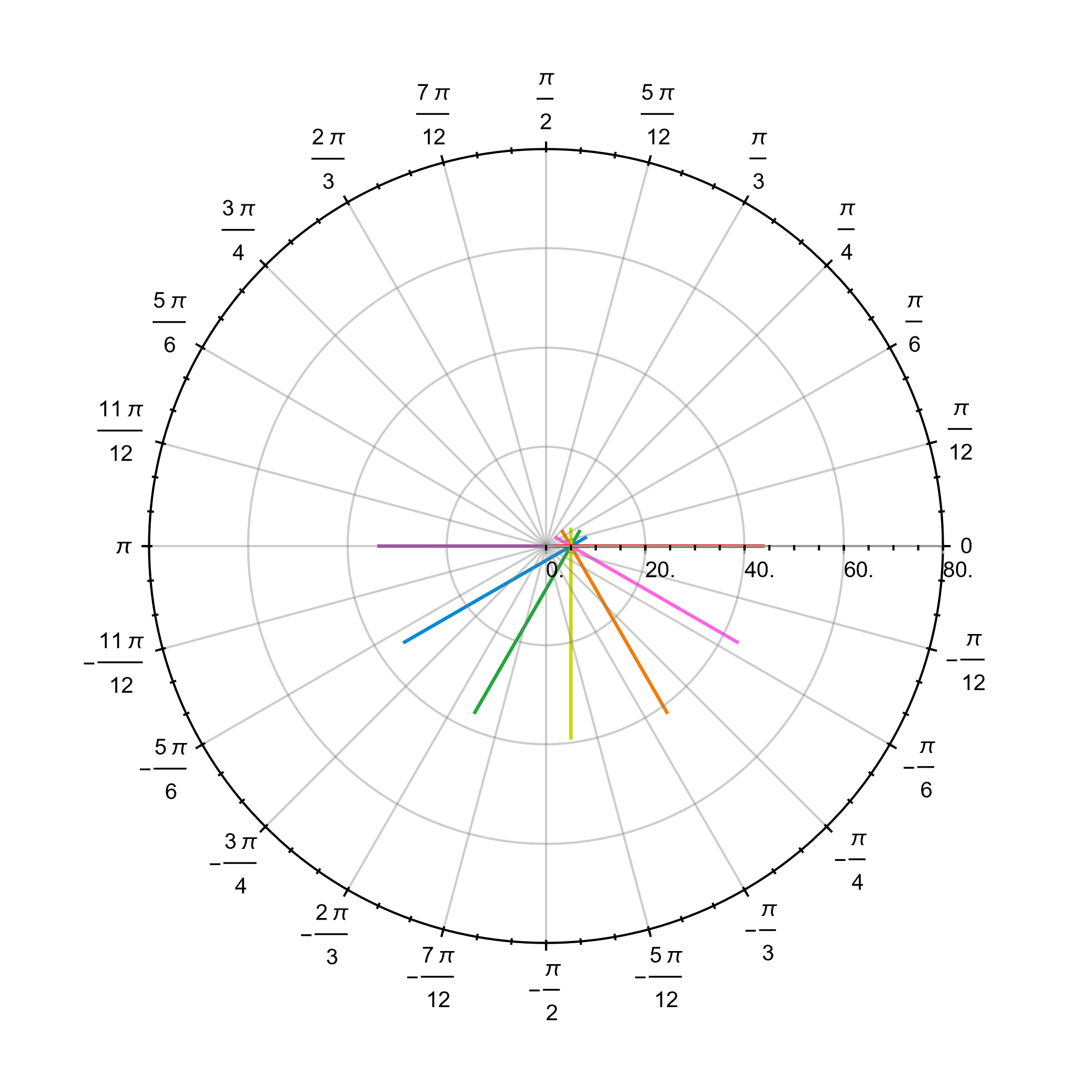}
  \caption{See Online Resource at~\cite{OnlineResource7} to view animation.}
  \label{fig5b}
\end{subfigure}
\begin{subfigure}{0.48\textwidth}
  \centering
\includegraphics[width=0.8\linewidth]{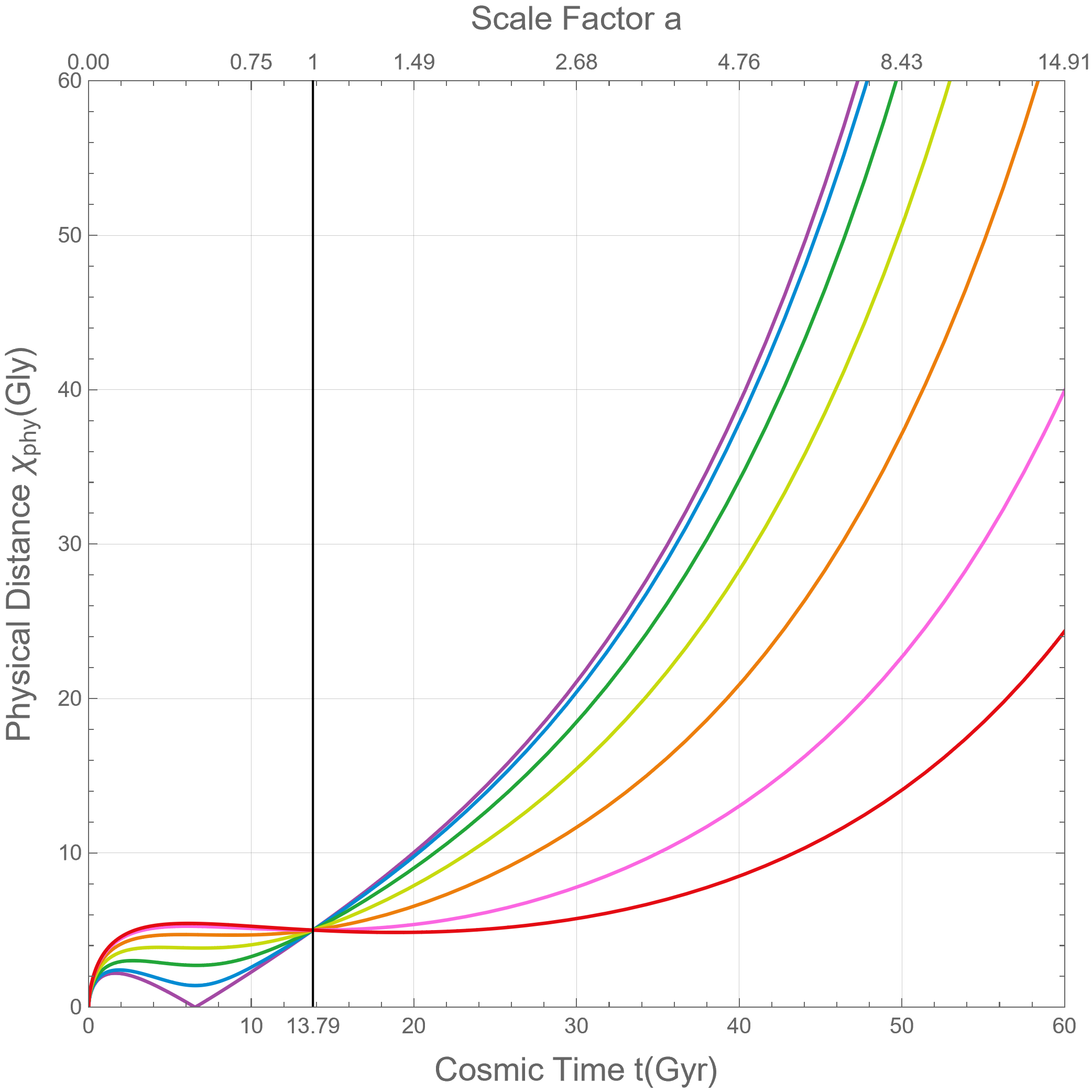}
  \caption{}
  \label{fig5c}
\end{subfigure}
\hfill
\begin{subfigure}{0.48\textwidth}
  \centering
\includegraphics[width=0.8\linewidth]{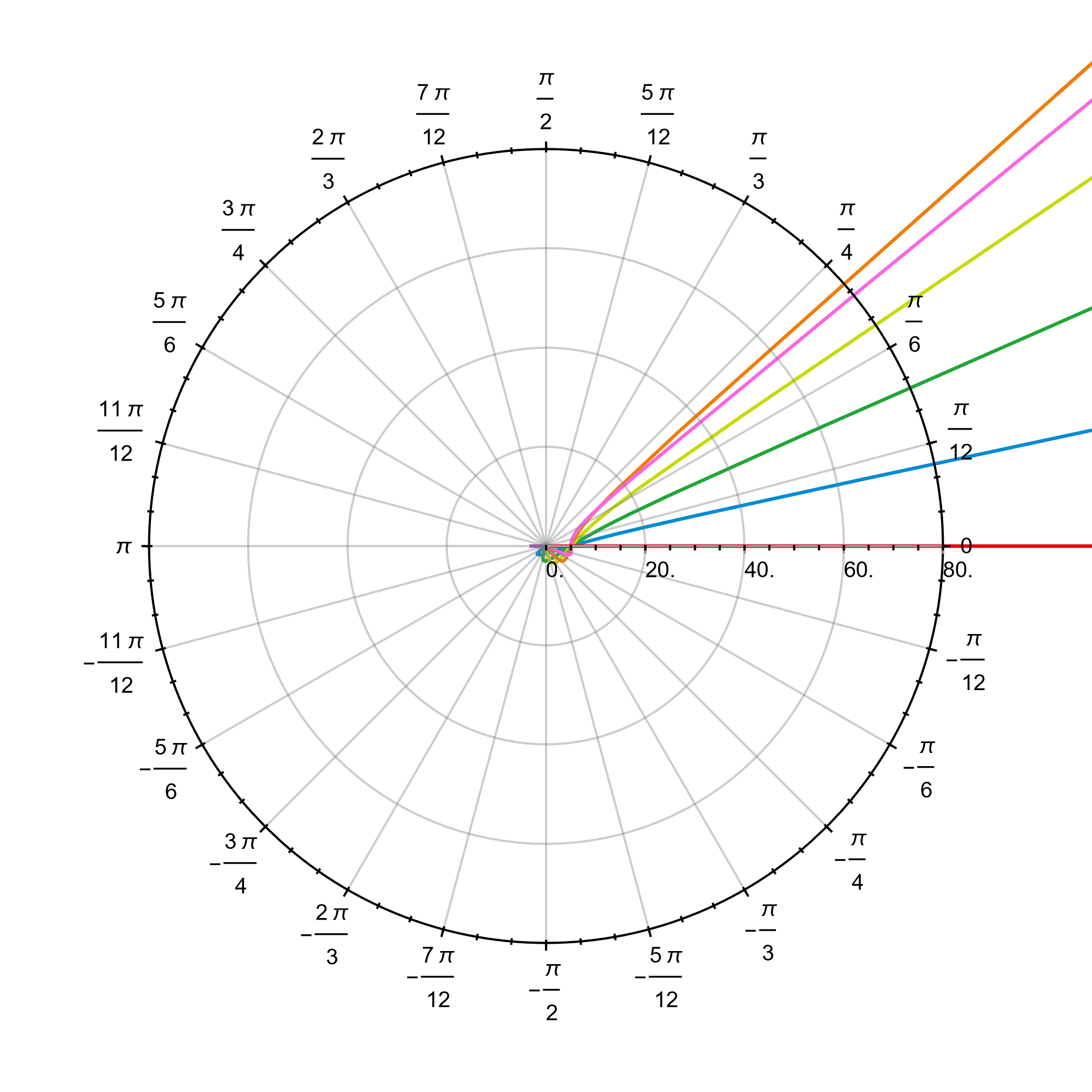}
  \caption{See Online Resource at~\cite{OnlineResource8} to view animation.}
  \label{fig5d}
\end{subfigure}
\begin{subfigure}{0.48\textwidth}
  \centering
\includegraphics[width=0.8\linewidth]{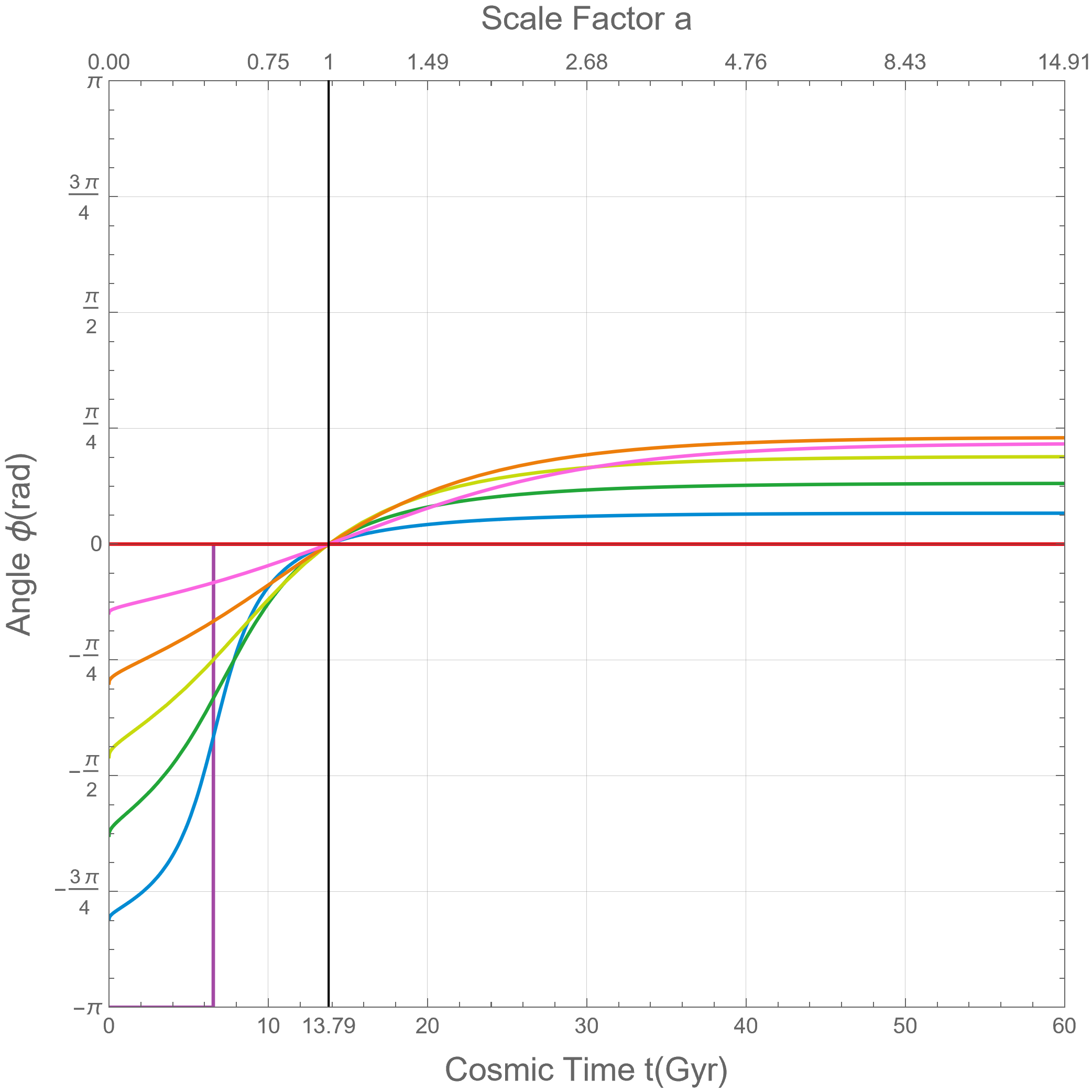}
  \caption{}
  \label{fig5e}
\end{subfigure}
\hfill
\begin{subfigure}{0.48\textwidth}
  \centering
\includegraphics[width=0.3\linewidth]{angle.png}
\end{subfigure}
\caption{Dynamics of Freely-Falling Particles in FLRW Spacetime with initial conditions: $(\chi_i,\phi_i,v_{\text{vec},i},\psi_i)=(5 \ \text{Gly},0,0.4,\psi_i)$ for different initial peuliar velocity angles $\psi_i$.}
\label{fig5}
\end{figure*}

\begin{figure*}[p]
\centering
\begin{subfigure}{0.48\textwidth}
  \centering
\includegraphics[width=0.8\linewidth]{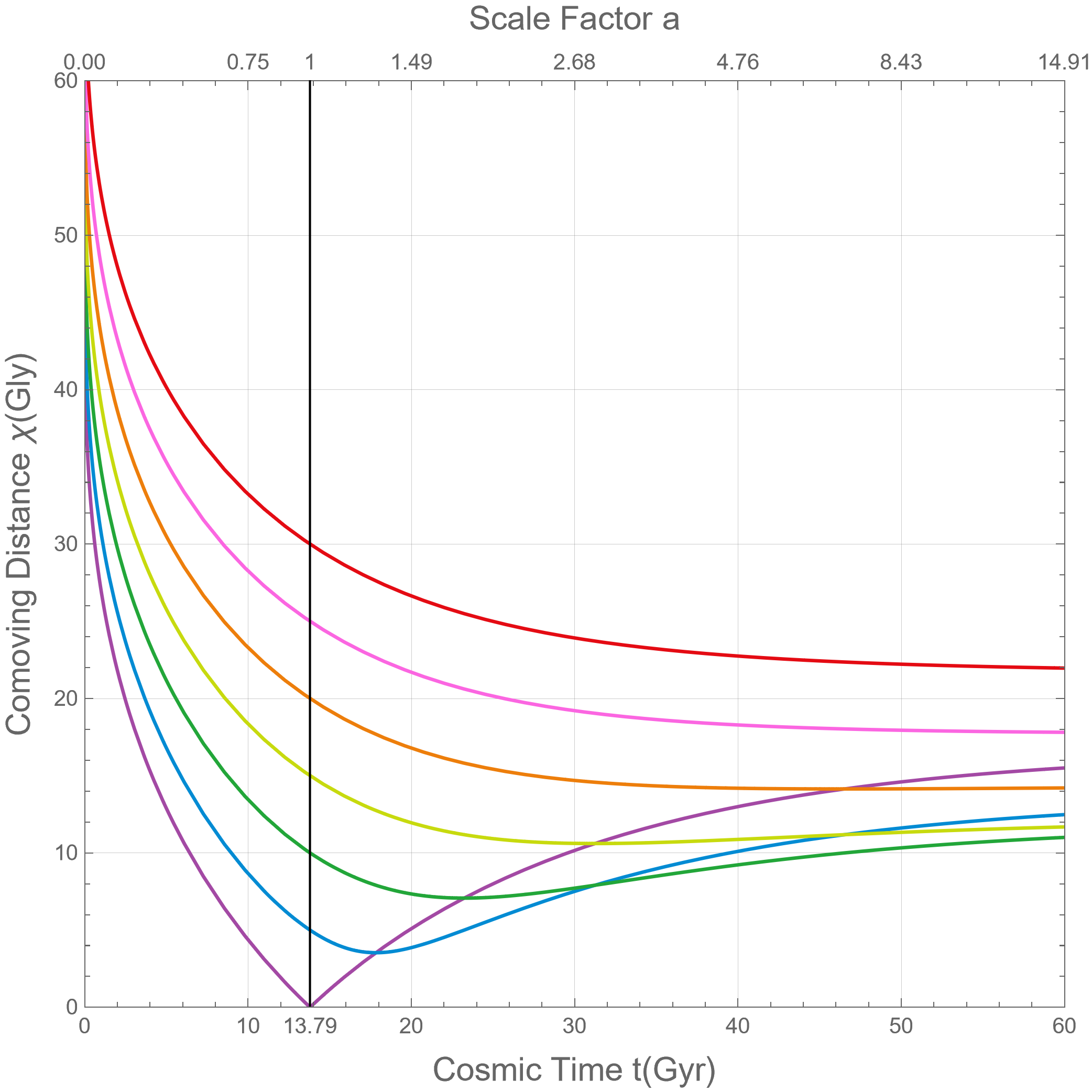}
  \caption{}
  \label{fig6a}
\end{subfigure}
\hfill
\begin{subfigure}{0.48\textwidth}
  \centering
\includegraphics[width=0.8\linewidth]{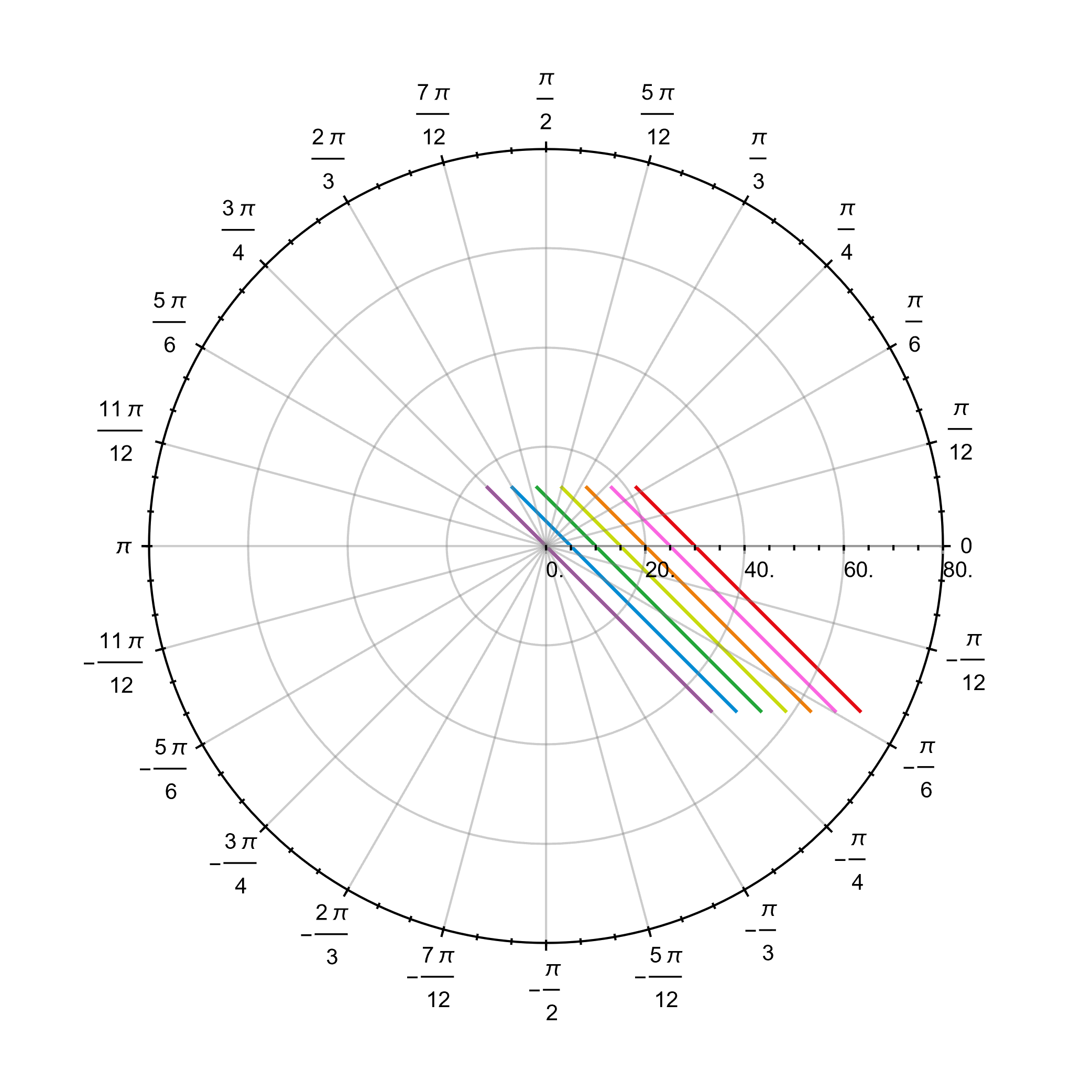}
  \caption{See Online Resource at~\cite{OnlineResource9} to view animation.}
  \label{fig6b}
\end{subfigure}
\begin{subfigure}{0.48\textwidth}
  \centering
\includegraphics[width=0.8\linewidth]{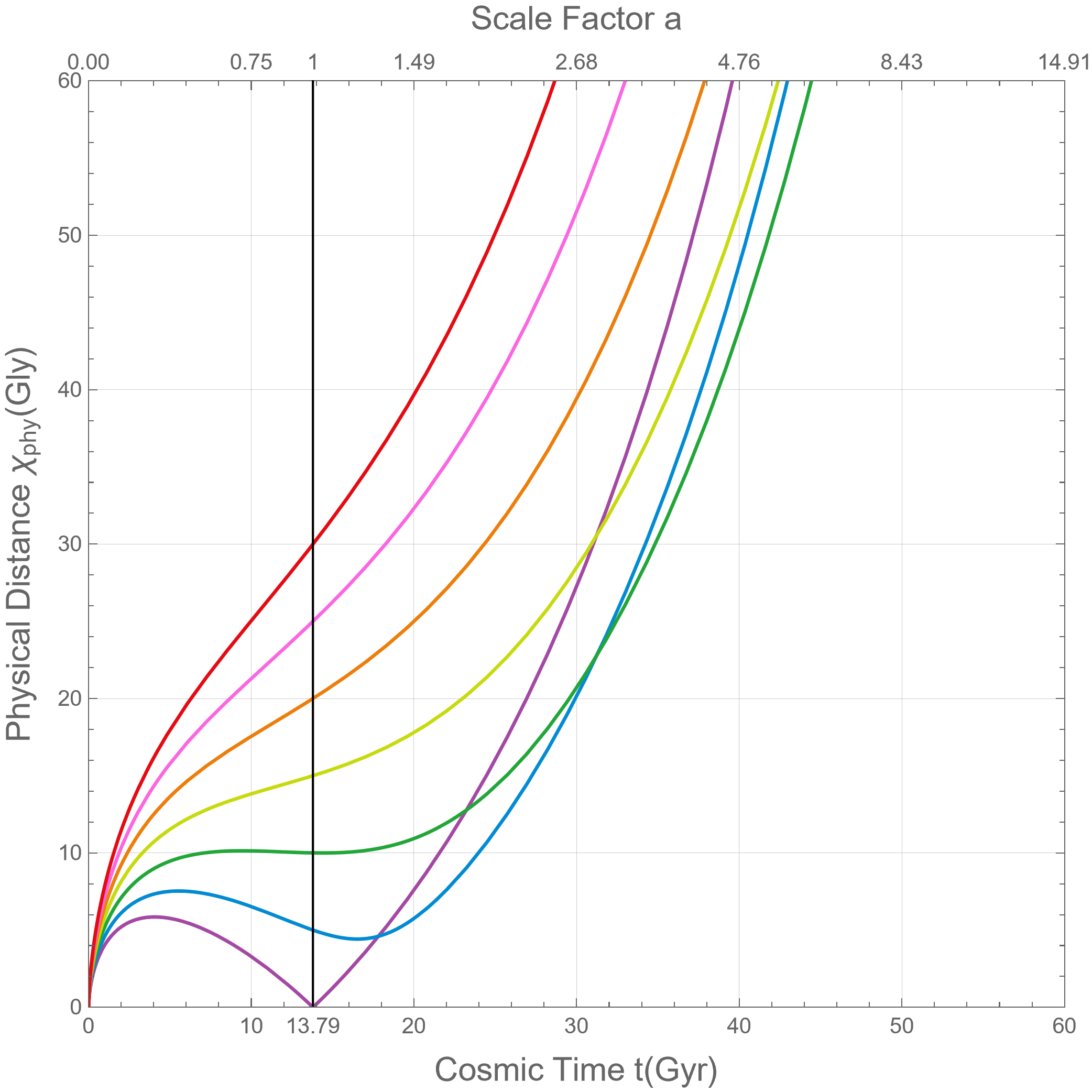}
  \caption{}
  \label{fig6c}
\end{subfigure}
\hfill
\begin{subfigure}{0.48\textwidth}
  \centering
\includegraphics[width=0.8\linewidth]{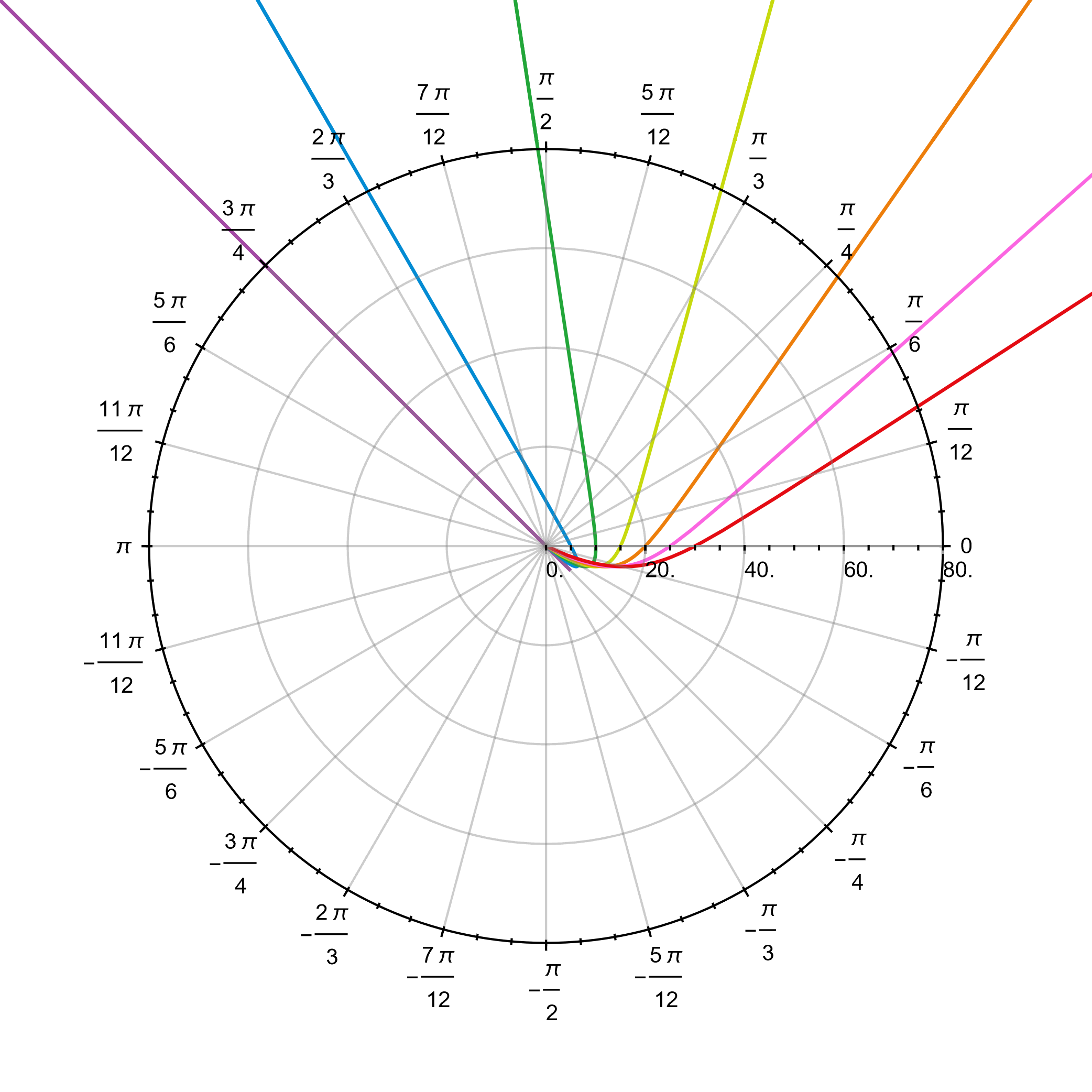}
  \caption{See Online Resource at~\cite{OnlineResource10} to view animation.}
  \label{fig6d}
\end{subfigure}
\begin{subfigure}{0.48\textwidth}
  \centering
\includegraphics[width=0.8\linewidth]{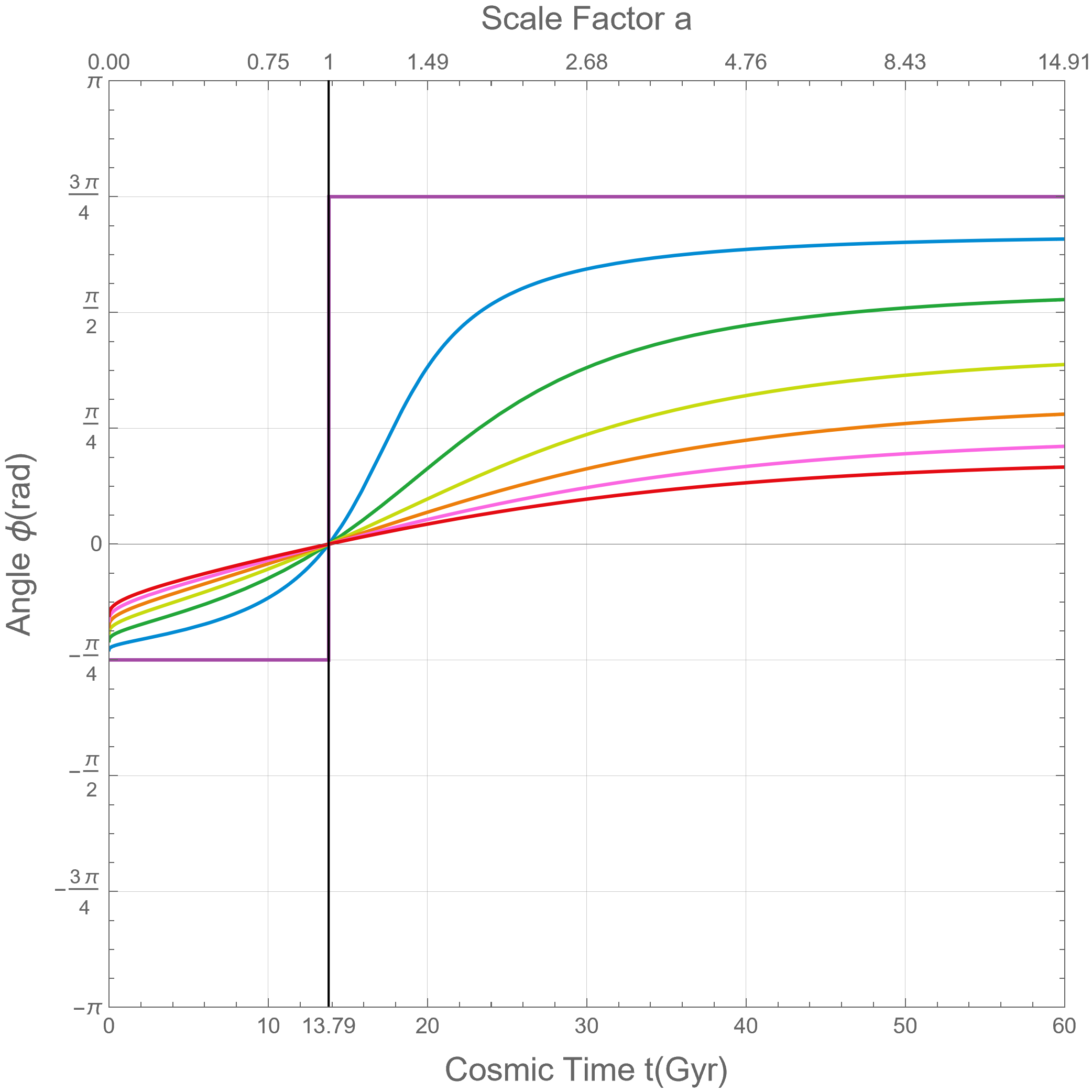}
  \caption{}
  \label{fig6e}
\end{subfigure}
\hfill
\begin{subfigure}{0.48\textwidth}
  \centering
\includegraphics[width=0.3\linewidth]{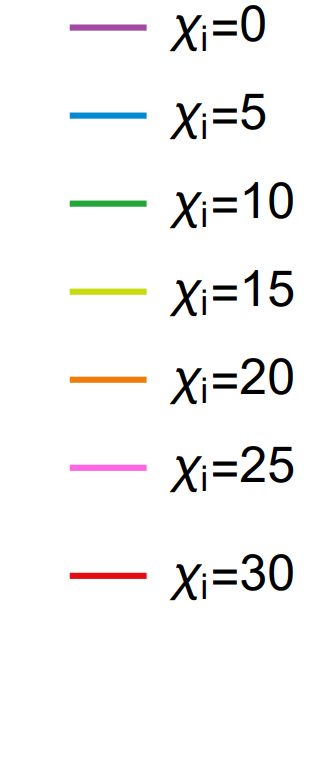}
\end{subfigure}
\caption{Dynamics of Freely-Falling Particles in FLRW Spacetime with initial conditions: $(\chi_i,\phi_i,v_{\text{vec},i},\psi_i)=(\chi_i,0,1,\frac{3\pi}{4})$ for different initial distances $\chi_i$.}
\label{fig6}
\end{figure*}

\begin{figure*}[p]
\centering
\begin{subfigure}{0.48\textwidth}
  \centering
\includegraphics[width=0.8\linewidth]{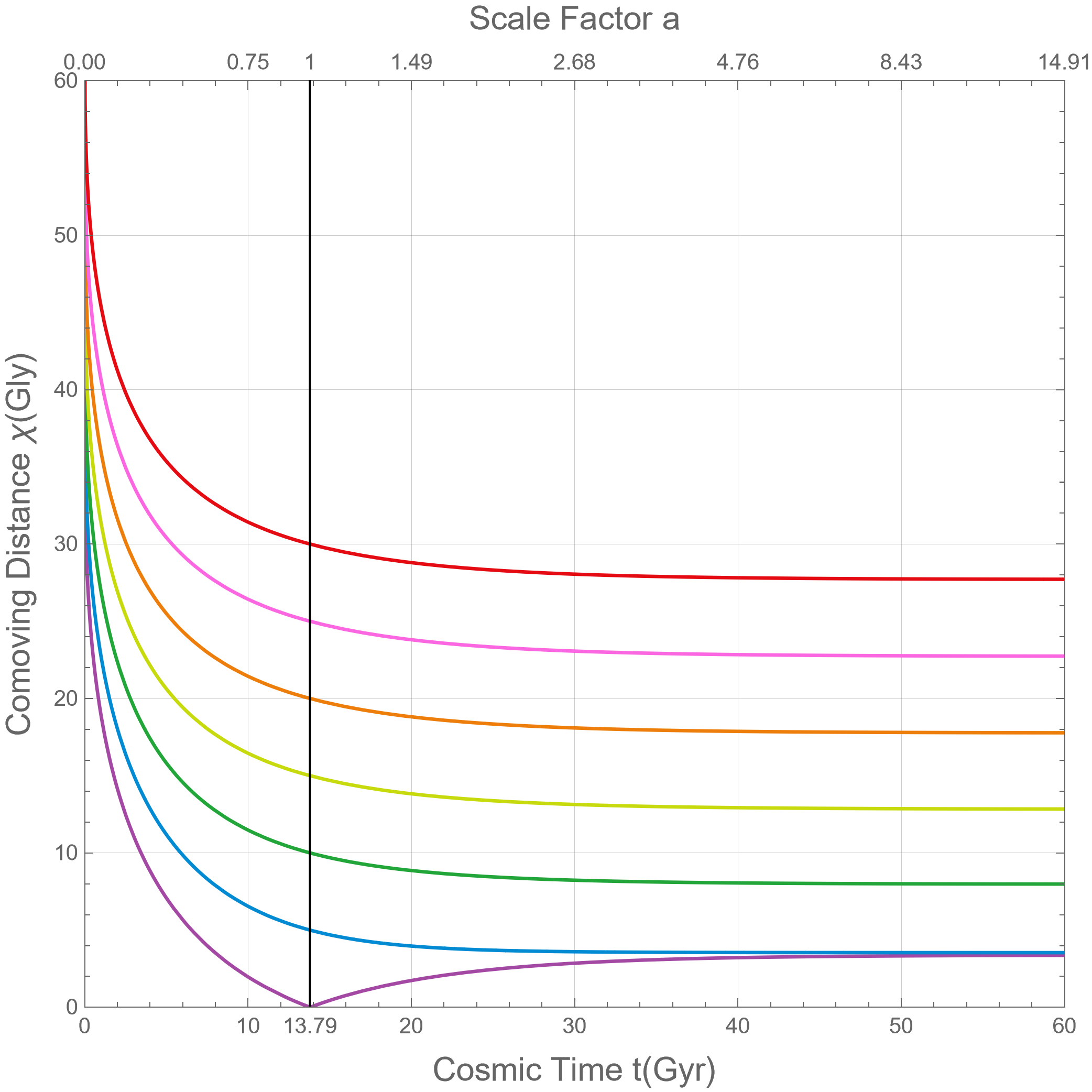}
  \caption{}
  \label{fig7a}
\end{subfigure}
\hfill
\begin{subfigure}{0.48\textwidth}
  \centering
\includegraphics[width=0.8\linewidth]{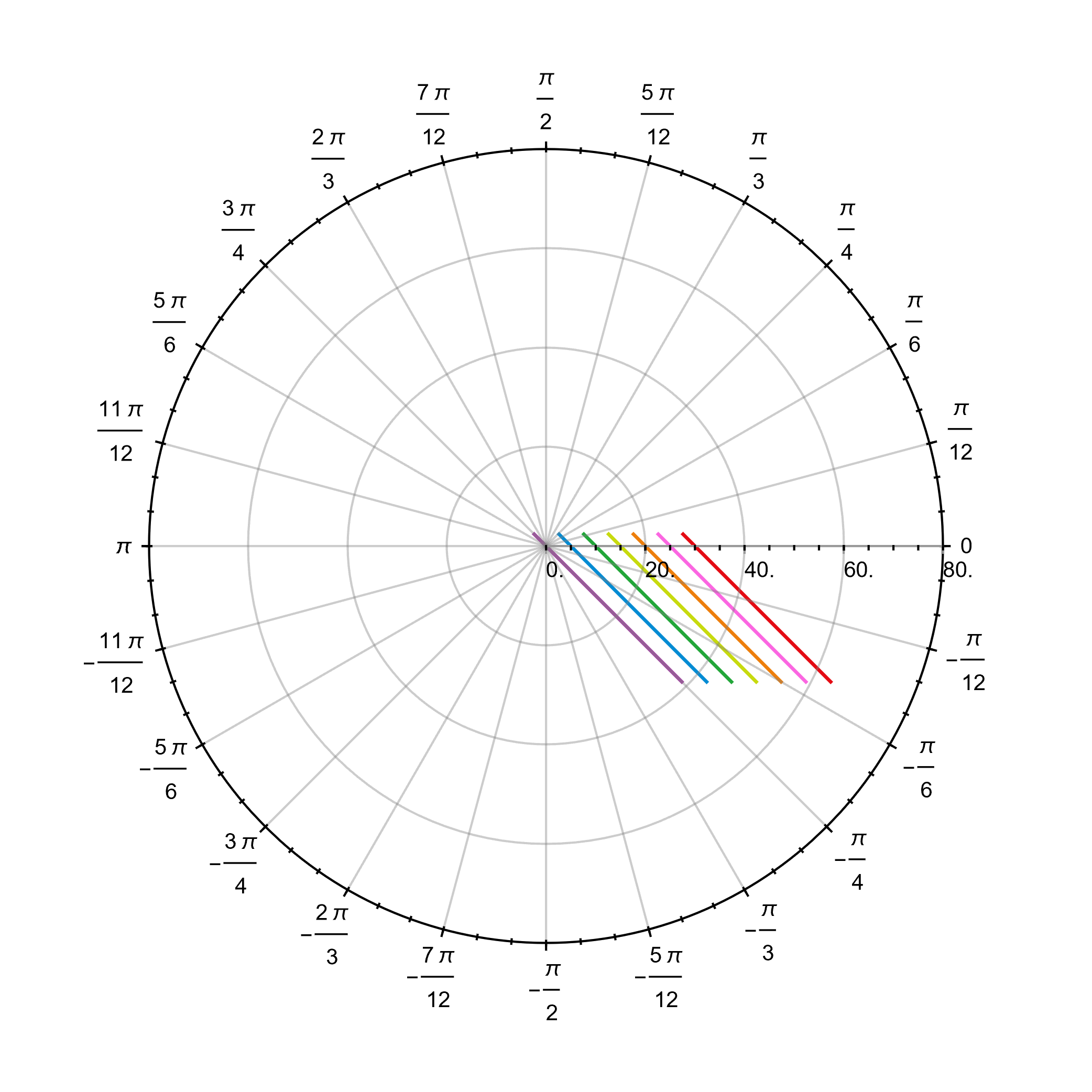}
  \caption{See Online Resource at~\cite{OnlineResource11} to view animation.}
  \label{fig7b}
\end{subfigure}
\begin{subfigure}{0.48\textwidth}
  \centering
\includegraphics[width=0.8\linewidth]{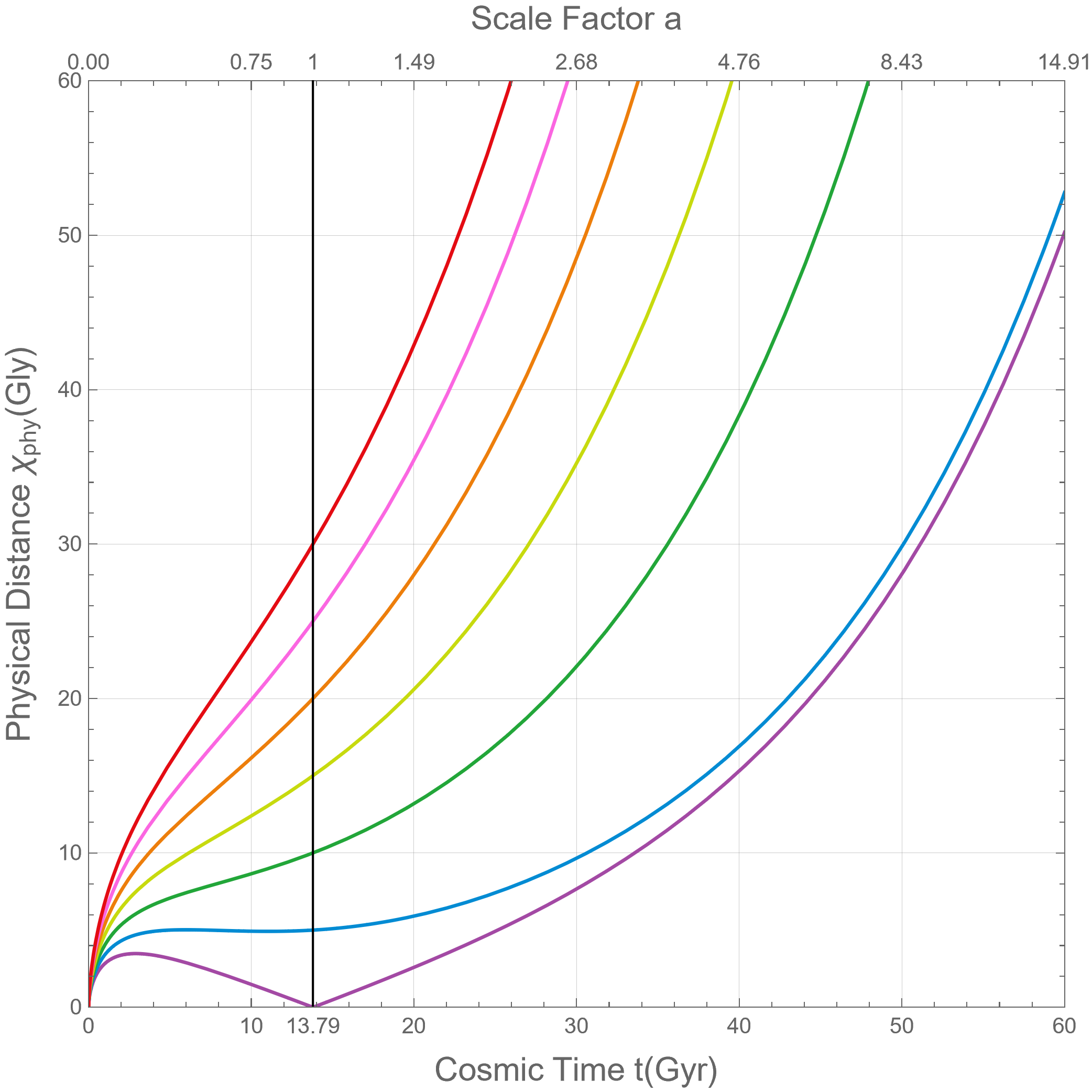}
  \caption{}
  \label{fig7c}
\end{subfigure}
\hfill
\begin{subfigure}{0.48\textwidth}
  \centering
\includegraphics[width=0.8\linewidth]{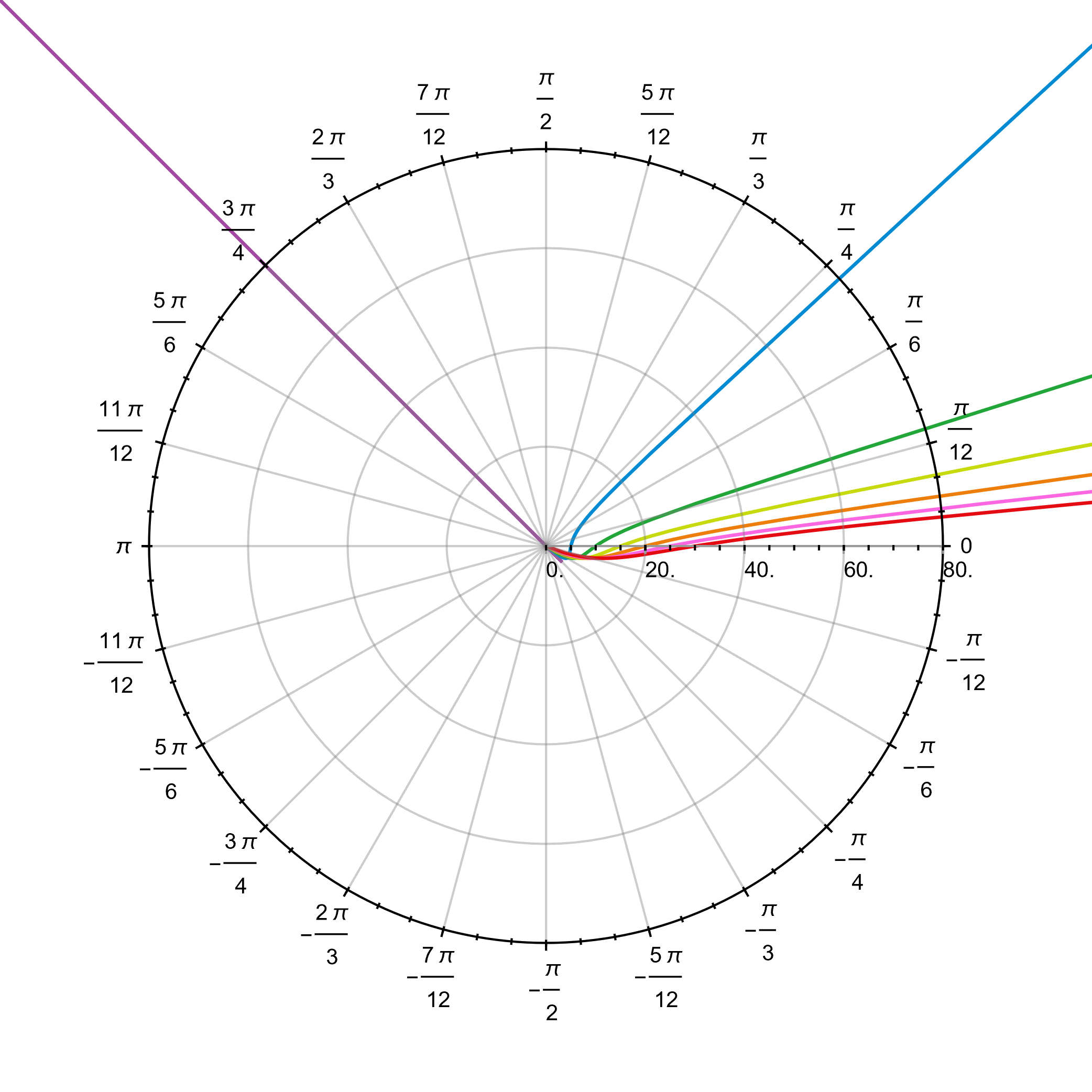}
  \caption{See Online Resource at~\cite{OnlineResource12} to view animation.}
  \label{fig7d}
\end{subfigure}
\begin{subfigure}{0.48\textwidth}
  \centering
\includegraphics[width=0.8\linewidth]{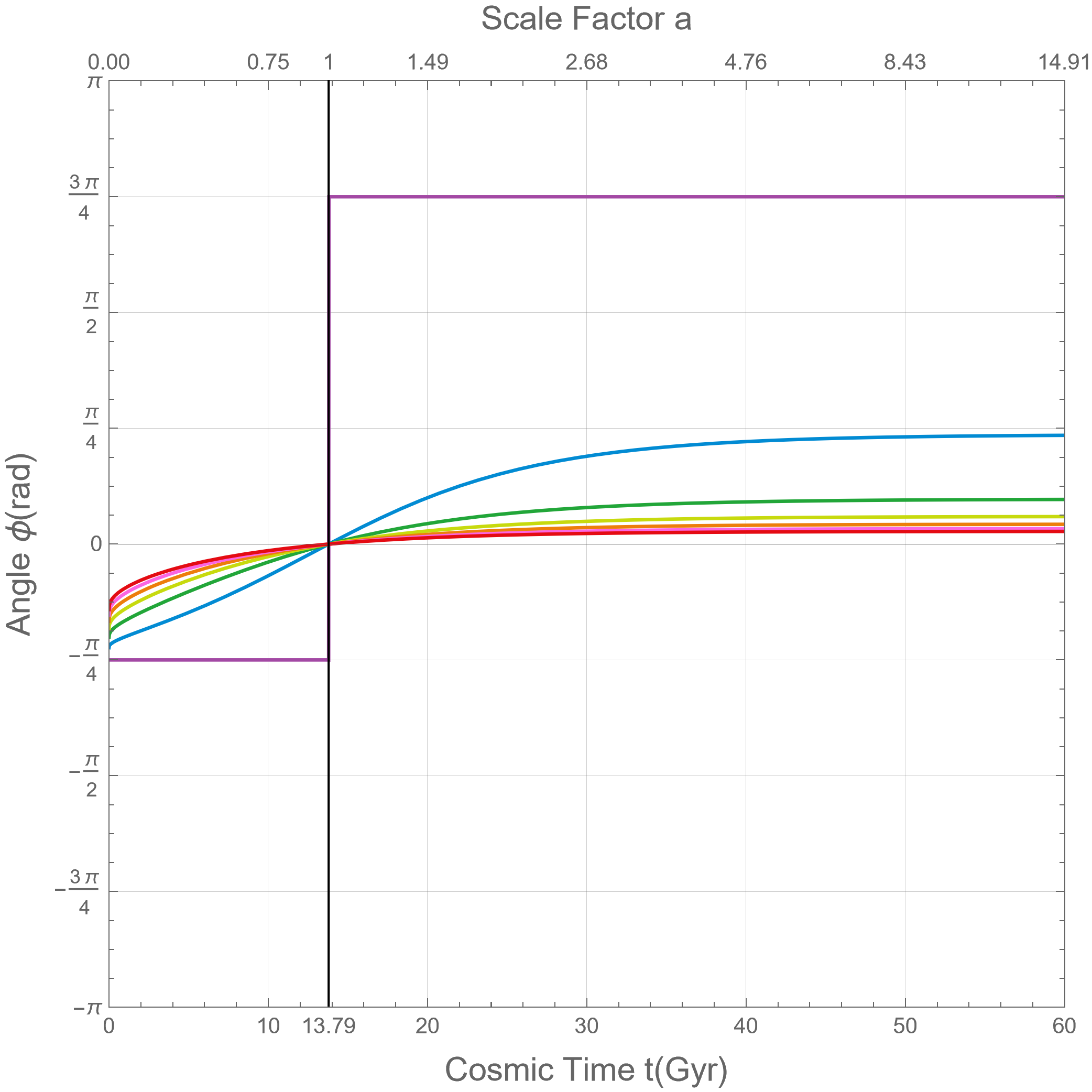}
  \caption{}
  \label{fig7e}
\end{subfigure}
\hfill
\begin{subfigure}{0.48\textwidth}
  \centering
\includegraphics[width=0.3\linewidth]{distance.png}
\end{subfigure}
\caption{Dynamics of Freely-Falling Particles in FLRW Spacetime with initial conditions: $(\chi_i,\phi_i,v_{\text{vec},i},\psi_i)=(\chi_i,0,0.4,\frac{3\pi}{4})$ for different initial distances $\chi_i$.}
\label{fig7}
\end{figure*}
\clearpage

\subsection{Results and Discussions}\label{sec5b}
With these illustrative graphs and animations in hand, we shall visualize the distinctive characteristics of our spacetime and their influence on freely-falling particles over time. Key observations from our analysis include:

\begin{itemize}[leftmargin=*]
    \item \textbf{Intersection with the x-axis at the present time}: All considered free particles are characterized by an initial angle $\phi_i = 0$ at the present time. As depicted in the animations of geodesic trajectories, it is observed that all particles intersect with the $x$-axis at the present time.
    
    \item \textbf{Inward Physical Motion despite Outward Comoving Motion}: All free particles begin with an inward physical motion ($\dot\chi_\text{phy}(0) < 0$) despite an initial outward comoving motion ($\dot\chi(0) > 0$). This phenomenon arises because, at the very beginning $(t = 0)$, the recessional velocity had infinitely positive value,  whereas the peculiar velocity consistently remains below the speed of light. This behavior indicates that the Hubble flow was the principal force driving geodesic motion in the earliest stages of the universe.
    
    \item \textbf{Discrepancy in Approaching Distances}: The closest approaching distance in the comoving frame does not necessarily align with that in the physical frame. An example of this discrepancy is illustrated by the null-geodesic curve ($v_{\text{pec},i} = 1$), represented by the red curve in Fig.~\ref{fig2}.
    
    \item \textbf{Behavior of Non-Radial Geodesics}: For non-radial geodesics, some physical trajectories continue to recede, while others return, approaching the closest physical distance before receding again.
    
    \item \textbf{Behavior of Radial Geodesics}: Radial geodesics demonstrate four potential paths (see Ref.~\cite{Omar}):
        \begin{itemize}
            \item One-Way Journey Geodesics: Some particles perpetually move away from us.
            \item Double-Reversing Geodesics: Some particles return to approach us, then start again to move away indefinitely. This unique behavior occurs only after the onset of the universe's accelerated expansion.
            \item Recrossing Geodesics: Some particles return to get closer and then cross our location to indefinitely continue their outward trajectory on the opposite side.
            \item Perpetually Approaching Geodesics: Some particles approach us indefinitely without ever crossing our location.
        \end{itemize}

    \item \textbf{Convergence in the Comoving Frame}: In the comoving frame, all geodesic trajectories converge to a point $(\chi_\infty,\phi_\infty)$, eventually aligning with the Hubble flow as time approaches infinity. This convergence occurs within the framework of the $\Lambda$CDM model, where the equation of state parameter for the dominant cosmological constant $w_\Lambda = -1$ as time tends to infinity, satisfying the `rejoin Hubble flow' condition $w_\Lambda < -\frac{1}{3}$ (see Ref.~\cite{Barnes}).
    
    \item \textbf{Correlation Between Particles Forming Straight-Line}: Preliminary observations from Animations~\ref{fig2}, ~\ref{fig3}, ~\ref{fig6}, and~\ref{fig7} suggest that freely-falling particles forming a straight line over time in the comoving frame, they definitively maintain this shape in the physical frame as well. This correlation can be established from the fact that straight lines preserve their linear characteristics under any global spatial rescaling.
    
    \item \textbf{Radial Geodesics}: In figures~\ref{fig4} and~\ref{fig5}, the purple and red curves describe radial geodesics, which are characterized by $\psi_i = 0$ for inward motion, and $\psi_i = \pi$ for outward motion.
    
    \item \textbf{Step Function Behavior of the Angle for Radial Geodesics}: As expected, the angle for radial geodesics, depicted by the purple and red curves in Figs.~\ref{fig4} and~\ref{fig5}, exhibits a step function behavior over time, jumping by $\pi$ as the particle crosses the origin to continue on the opposite side.
    
    \item \textbf{Consistency Angle Between Frames}: It is observed from the trajectories and animations that the angle made by a free particle remains consistent in both comoving and physical frames.
\end{itemize}
\section{The total angle}\label{sec6}
\begin{figure}[H]
\centering
\includegraphics[width=0.7\columnwidth]{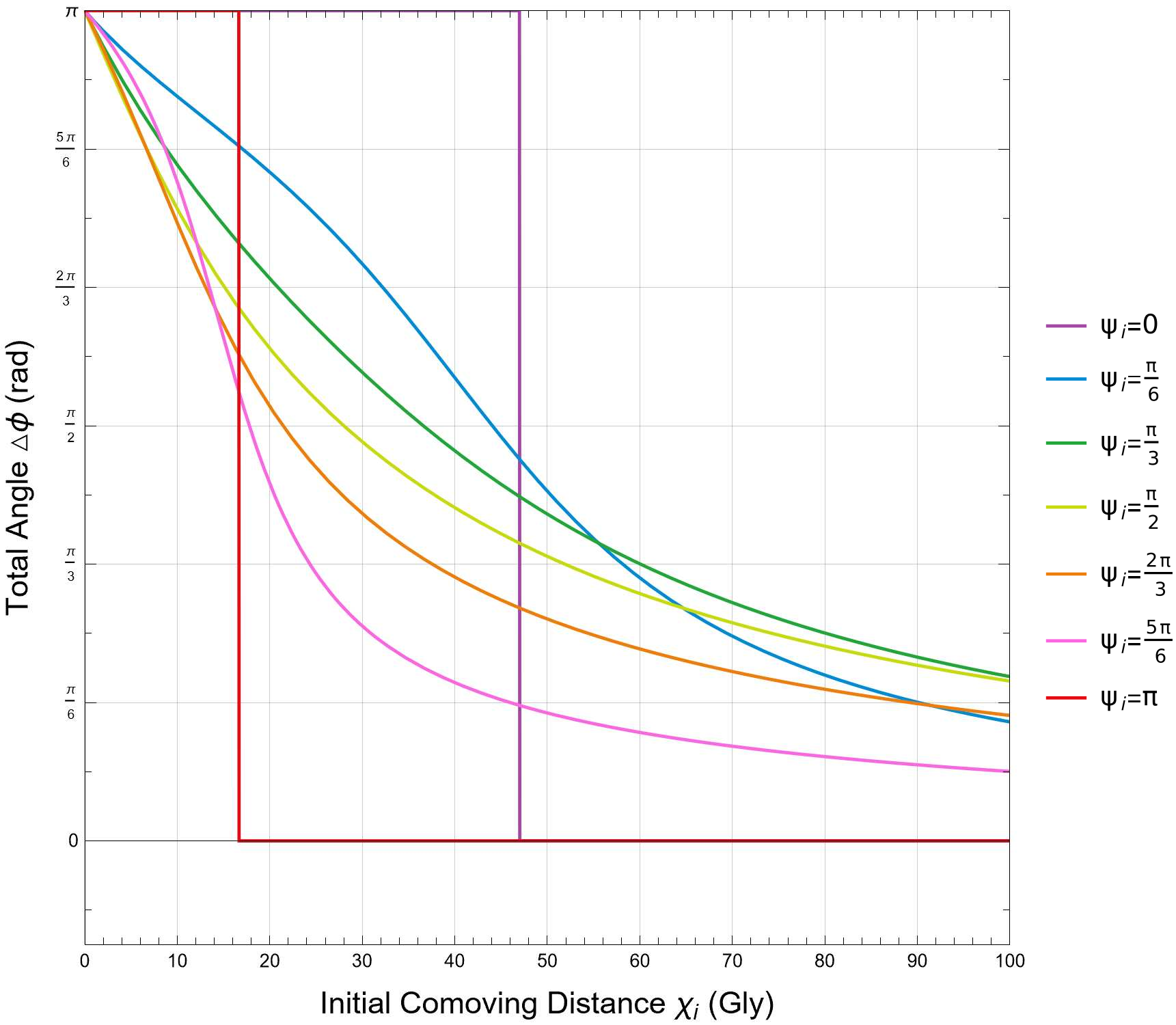}
\caption{These seven diagrams depict the total angle $\Delta\phi$ as a function of the initial comoving distance $\chi_i$ for null geodesics ($v_{\text{pec},i}=1$) with different initial deviation angles $\psi_i$.}
\label{fig8}
\end{figure}

We define the total angle, denoted by $\Delta \phi$, as the angle made by any free particle in FLRW spacetime from the Big Bang singularity to an infinite time, as observed from a comoving observer. Given that $\phi$ is a monotonic function of time, the total angle is given by
\begin{equation}
\Delta \phi = |\phi(+\infty) - \phi(0)|.
\label{eq90}
\end{equation}
It can be expressed in terms of initial conditions $(\chi_i, A, B)$ as
\begin{flalign}
&\Delta \phi(\chi_i, A, B)= \int_{0}^{+\infty}  \frac{\frac{\left|B\right|}{\chi^2(t'; \chi_i, A, B)}}{a(t') \sqrt{a^2(t') + A^2}} \, dt'\nonumber\\
&=\text{sgn}\left(\dot{\chi}(\infty)\right) \arccos\left(\frac{\left|\frac{B}{A}\right|}{\chi(\infty)} \right) - \text{sgn}\left(\dot{\chi}(0)\right) \arccos\left(\frac{\left|\frac{B}{A}\right|}{\chi(0)} \right),
\label{eq91}&&
\end{flalign}
and in terms of $(\chi_i, v_{\chi, i}, v_{\phi, i})$ as
\begin{flalign}
&\Delta \phi(\chi_i, v_{\chi, i}, v_{\phi, i}) = \int_{0}^{+\infty}  \frac{\frac{a_i \chi_i \left|v_{\phi, i}\right|}{\chi^2(t'; \chi_i, v_{\chi, i}, v_{\phi, i})}}{a(t') \sqrt{a^2(t')(1 - v_{\chi, i}^2 - v_{\phi, i}^2) + a_i^2 (v_{\chi, i}^2 + v_{\phi, i}^2)}} \, dt'\nonumber\\
&= \text{sgn}\left(\dot{\chi}(\infty)\right) \arccos\left(\frac{\frac{\chi_i}{\chi(\infty)} |v_{\phi,i}|}{\sqrt{v_{\chi,i}^2 + v_{\phi,i}^2}} \right) - \text{sgn}\left(\dot{\chi}(0)\right) \arccos\left(\frac{\frac{\chi_i}{\chi(0)} |v_{\phi,i}|}{\sqrt{v_{\chi,i}^2 + v_{\phi,i}^2}} \right),
\label{eq92}&&
\end{flalign}
and in terms of $(\chi_i, v_{\text{pec}, i}, \psi_i)$ as
\begin{flalign}
&\Delta \phi(\chi_i, v_{\text{pec}, i}, \psi_i)= \int_{0}^{+\infty}  \frac{\frac{a_i \chi_i v_{\text{pec}, i}\left|\sin{\psi_i}\right|}{\chi^2(t'; \chi_i, v_{\text{pec}, i}, \psi_i)}}{a(t') \sqrt{a^2(t')(1 - v^2_{\text{pec}, i}) + a_i^2 v^2_{\text{pec}, i}}} \, dt'\nonumber\\&=\text{sgn}\left(\dot{\chi}(\infty)\right) \arccos\left(\frac{\chi_i}{\chi(\infty)} \left|\sin(\psi_i)\right|\right) - \text{sgn}\left(\dot{\chi}(0)\right) \arccos\left(\frac{\chi_i}{\chi(0)} \left|\sin(\psi_i)\right|\right).
\label{eq93}&&
\end{flalign}
In this discussion, we present a series of diagrams depicting the total angle $\Delta\phi$ as a function of the initial radial distance $\chi_i$ at the present time. These diagrams represent free geodesics traveling at the speed of light $v_{\text{pec},i}=1$ with various choices of deviation angle $\psi_i$ at $0$, $\frac{\pi}{6}$, $\frac{\pi}{3}$, $\frac{\pi}{2}$, $\frac{2\pi}{3}$, $\frac{5\pi}{6}$, and $\pi$. Figure~\ref{fig8} displays the graphs corresponding to each specified deviation angle $\psi_i$. At first glance, it becomes evident that the total angle $\Delta\phi$ approaches zero as the free particle's initial comoving distance $\chi_i$ goes to infinity. This phenomenon is observed in our $\Lambda$CDM cosmological model, where all geodesics tend to converge with the Hubble flow (see Ref.~\cite{Barnes}). This implies that \textit{extremely distant free particles will appear stationary in the sky over time}. Notably, this is not due to a decrease in angular velocity with radial distance but rather a characteristic of the $\Lambda$CDM model, dictating a finite comoving distance journey for these particles. Additionally, $\Delta\phi=\pi$ for the radial limit $\chi_i=0$ signifies that the particle crosses the origin and transitions to the other side, resulting in an instantaneous $\pi$ shift in its angle.
Furthermore, as visualized in Fig.~\ref{fig8}, the behavior of the total angle for radial motion corresponding to $\psi_i=0$ and $\psi_i=\pi$ for outward and inward traveler is delineated by the purple and red curves, respectively. These curves exhibit a step-function behavior, with a discontinuous jump of $-\pi$ at the initial conditions $\chi_i\approx47$ Gly for $\psi_i=0$ and $\chi_i\approx16.68$ Gly for $\psi_i=\pi$. This indicates that free particles in outward motion from our position has only reached us if its radial comoving distance today $\chi_i<47$ Gly, while those beyond $\chi_i>47$ Gly will never do so. Similarly, only particles in inward motion will only reach us if $\chi_i<16.68$ Gly at the present time. These observations are consistent with cosmological predictions, where the furthest currently outgoing signal makes a maximum total angle $\pi$, was emitted from our location at $t=0$ and now spans a comoving distance of $47$ Gly, defines the current particle horizon. Likewise, the furthest incoming signal that will reach us in the infinite future is now at a comoving distance of $16.68$ Gly, which is known by the current event horizon. For more detail, see Ref.~\cite{Tamara}.

\begin{figure}[H]
\centering
\begin{subfigure}{\columnwidth}
    \centering
    \includegraphics[width=0.55\linewidth]{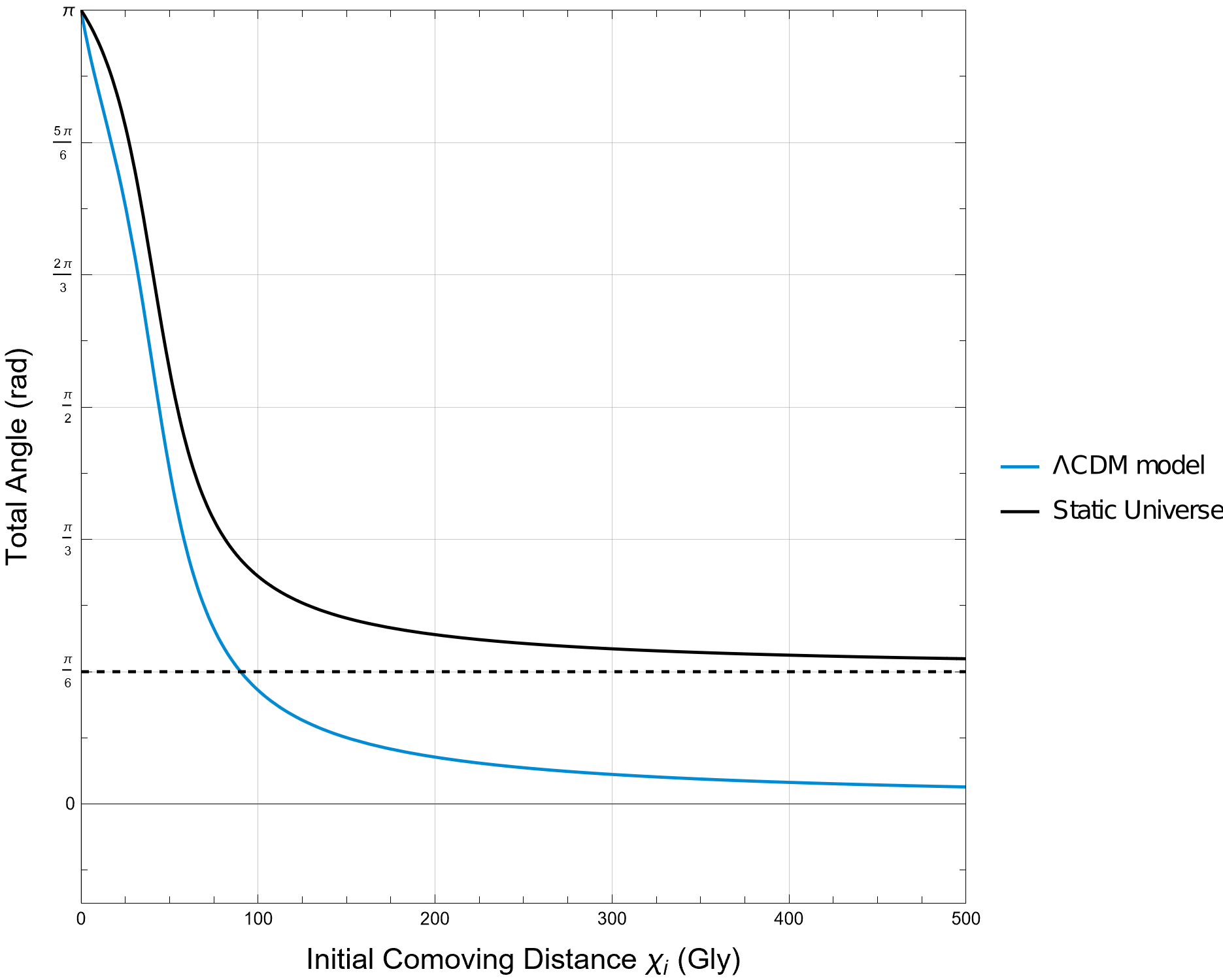}
    \caption{$(v_{\text{pec},i},\psi_i)=(1,\frac{\pi}{6})$}
    \label{fig9a}
\end{subfigure}

\begin{subfigure}{\columnwidth}
    \centering
    \includegraphics[width=0.55\linewidth]{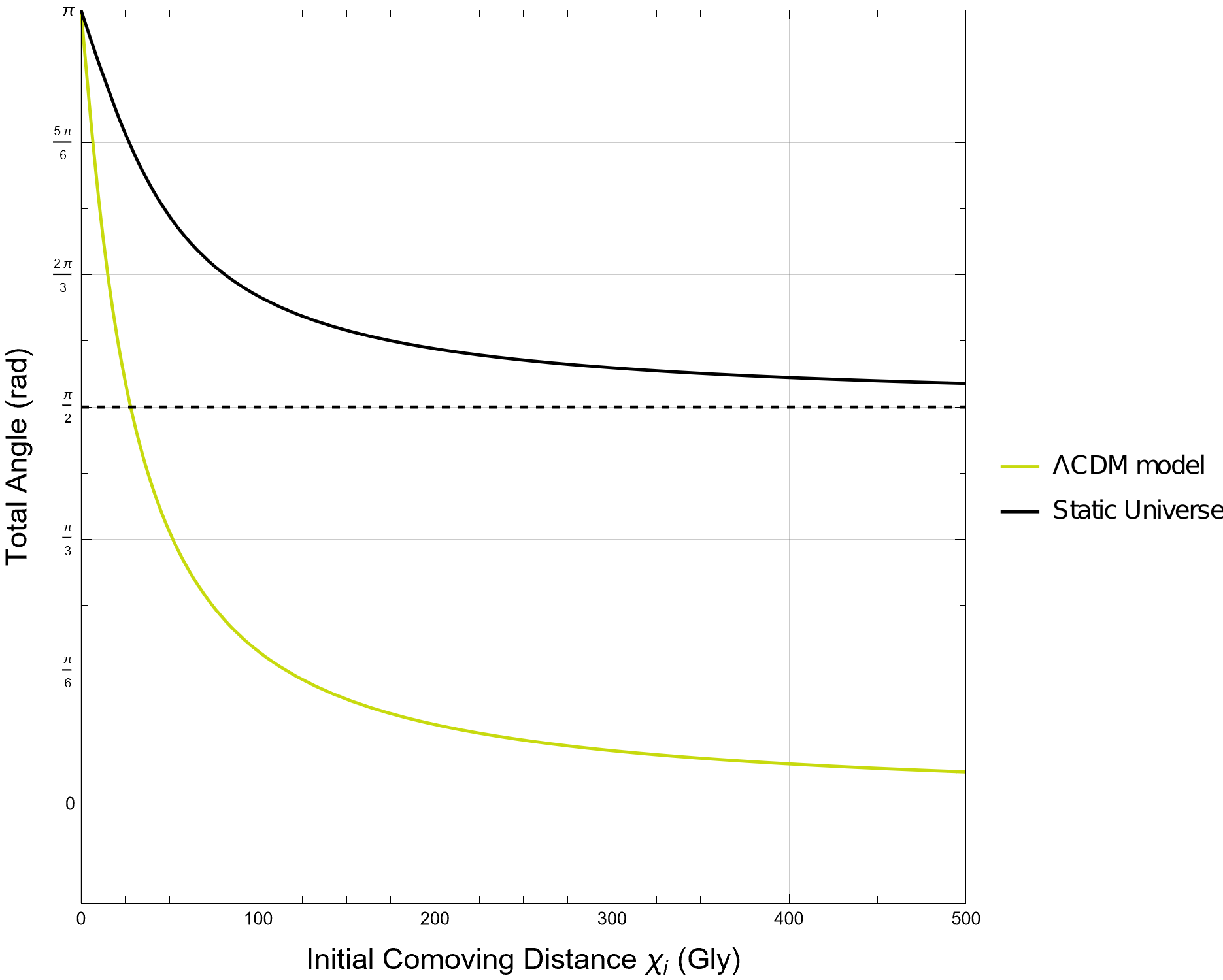}
    \caption{$(v_{\text{pec},i},\psi_i)=(1,\frac{\pi}{2})$}
    \label{fig9b}
\end{subfigure}

\begin{subfigure}{\columnwidth}
    \centering
    \includegraphics[width=0.55\linewidth]{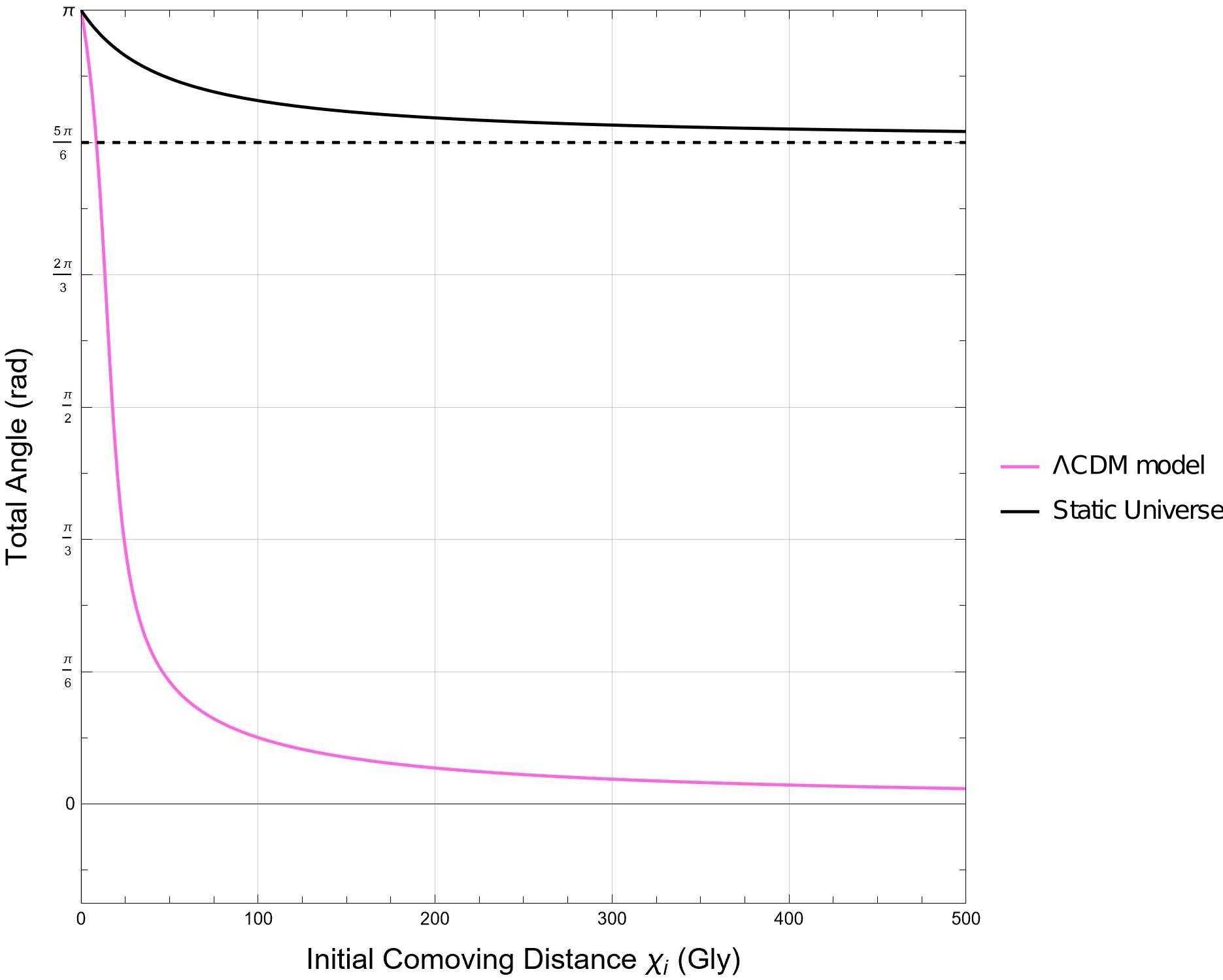}
    \caption{$(v_{\text{pec},i},\psi_i)=(1,\frac{5\pi}{6})$}
    \label{fig9c}
\end{subfigure}

\caption{This set of figures illustrates a comparative analysis of the total angle in the $\Lambda$CDM model ($\Delta\phi$) versus a static universe ($\Delta^*\phi$) in terms of the initial comoving distance $\chi_i$ under a range of deviation angles.}
\label{fig9}
\end{figure}

We define the total angle made by a free particle for a static spatially flat universe, denoted by $\Delta^*\phi$. Its expression can be shown to take the following form
\begin{align}
\Delta^*\phi=\left|\phi_i-\phi(0)+\psi_i\right|.
\label{eq94}
\end{align}
To compare the total angle of a non-radial signal within the $\Lambda$CDM Model and the static universe, we have assumed all signals start their propagation at $t=0$ for the static universe. We can express $\Delta^*\phi$ in terms of $(\chi_i,v_{\text{pec},i},\psi)$ by using Eq.~\eqref{eq68} as follows
\begin{align}
\Delta^*\phi=\left| \text{sgn}\left(\dot{\chi}(0)\right) \arccos{\left(\frac{\chi_i}{\chi(0)}\left|\sin\psi_i\right|\right)} - \frac{\pi}{2} \right|.
\label{eq95}
\end{align}
In contrast to the $\Lambda$CDM model where the total angle $\Delta\phi\rightarrow0$ as $\chi_i\rightarrow\infty$, a free particle in a static universe with infinitely distant initial comoving radial distance $\chi_i\rightarrow\infty$ at the present time would have a total angle corresponding to $\Delta^*\phi\rightarrow\psi_i$. 
To visualize this fact, we present a series of diagrams (See Figs.~\ref{fig9}) comparing the total angle for the $\Lambda$CDM model $\Delta\phi$ in Eqs.~\eqref{eq93} with that of a static universe $\Delta^*\phi$ in Eqs.~\eqref{eq95} as a function of the initial radial distance $\chi_i$ at the present time, for a variety of null geodesics with initial deviation angles $\psi_i=\frac{\pi}{6}$, $\frac{\pi}{2}$, and $\frac{5\pi}{6}$. The behavior of the total angle in the $\Lambda$CDM model can be explained by the finite nature of their geodesics in the comoving frame (joining the Hubble flow; See Ref.~\cite{Barnes}), where they reach a finite comoving distance, denoted as $\chi_\infty$.

\section{Conclusion}\label{sec7}
In summary, we have presented a comprehensive analysis of non-radial, free, timelike geodesics in spatially curved FLRW spacetimes, using the symmetry of the system to construct the corresponding two conserved quantities. This allows for a straightforward and simpler determination of the radial $\dot{\chi}(t)$ and the angular $\dot{\phi}(t)$ velocities along with the evolution of the radial distance $\chi(t)$ and the angle $\phi(t)$ in terms of the constant of motions $A$ and $B$, offering an advantageous alternative to traditional Euler-Lagrange or even the geodesic equation methods. In this framework, geodesics are fully determined by three appropriately parameterizations of initial conditions $(\chi_i,\phi_i, A,B)$ as detailed in Eqs.~\eqref{eq52}, $(\chi_i,\phi_i, v_{\chi,i},v_{\phi,i})$ in Eqs.~\eqref{eq53}, and $(\chi_i,\phi_i, v_{\text{pec},i},\psi_i)$ in Eqs.~\eqref{eq54}. Furthermore, we have elucidated the radial geodesic limit, the null geodesic limit, and the comoving geodesic limit for the established non-radial solution. Our investigation into non-radial geodesic motions has addressed the geometrical picture within the comoving frame, characterized by straight-line trajectories. This analysis has enabled us to derive an alternative formula for the angle $\phi(t)$, as shown in Eqs.~\eqref{eq66},~\eqref{eq67}, and~\eqref{eq68}, which is explicitly determined only by the comoving radial distance $\chi(t)$. These outcomes have been applied in particular to the context of our expanding universe, as evidenced by the recent results from the Planck mission~\cite{Planck}. Using this data, we have constructed time diagrams to illustrate the variations in comoving and physical radial distance, as well as the angle, effectively depicting the motion of a free particle. Additionally, we enhanced these diagrams with animations that trace the trajectories of geodesics under various initial conditions, further enriching our insights. Moreover, we analyzed the total angle $\Delta\phi$ swept by a free particle from the Big Bang singularity to an infinite time, considering its dependence on both initial comoving distance and peculiar velocity. We have further compared the characteristics of this total angle for two cosmological frameworks: the $\Lambda$CDM model and a static universe. This study potentially enables a more profound understanding of the behavioral aspects of free particles from our perspective, encompassing not only radial but also non-radial motions. It is worth mentioning that this analysis has been extended to encompass non-Euclidean spatial spaces, as elaborated in Ref.~\cite{Omar2}. However, we aim to expand this investigation to include the gravitational effects of galaxies and Earth's peculiar velocity relative to the CMB.
\appendix
\section{The non-zero Christoffel symbol components}\label{A1}
The interval element in Eq.~\eqref{eq8} can indeed be represented in matrix form for the ordinary basis $(\partial_t, \partial_\chi, \partial_\theta, \partial_\phi)$ induced by the coordinates system $(t, \chi, \theta, \phi)$ as
\begin{equation}
g_{\mu\nu} = 
\begin{pmatrix}
1 & 0 & 0 & 0 \\
0 & -a^2(t) & 0 & 0 \\
0 & 0 & -a^2(t)\chi^2 & 0 \\
0 & 0 & 0 & -a^2(t)\chi^2 \sin^2\theta
\end{pmatrix},
\label{eqA1}
\end{equation}
and its inverse is
\begin{equation}
g^{\mu\nu} = 
\begin{pmatrix}
1 & 0 & 0 & 0 \\
0 & -\frac{1}{a^2(t)} & 0 & 0 \\
0 & 0 & -\frac{1}{a^2(t)\chi^2} & 0 \\
0 & 0 & 0 & -\frac{1}{a^2(t)\chi^2 \sin^2\theta}
\end{pmatrix}.
\label{eqA2}
\end{equation}
The Christoffel symbols defined in terms of the metric and its inverse, are given by
\begin{equation}
\Gamma^\mu_{\alpha\beta} = \frac{1}{2} g^{\mu\nu} (\partial_\alpha g_{\beta\nu} + \partial_\beta g_{\alpha\nu} - \partial_\nu g_{\alpha\beta}).
\label{eqA3}
\end{equation}
The only non-zero Christoffel symbol components are
\begin{subequations}
\label{eqA4}
\begin{align}
\Gamma^t_{ij} &= -\frac{1}{2} g^{tt} \partial_t g_{ij} = a(t) \dot{a}(t) \gamma_{ij}, \label{eqA4a}\\
\Gamma^i_{tj} &= \frac{1}{2} g^{ik} \partial_t g_{jk} = \frac{\dot{a}(t)}{a(t)} \delta^i_j, \label{eqA4b}\\
\Gamma^\chi_{\theta\theta} &= -\frac{1}{2} g^{\chi\chi} \partial_\chi g_{\theta\theta} = -\chi, \label{eqA4c}\\
\Gamma^\chi_{\phi\phi} &= -\frac{1}{2} g^{\chi\chi} \partial_\chi g_{\phi\phi} = -\chi \sin^2\theta, \label{eqA4d}\\
\Gamma^\theta_{\chi\theta} &= \frac{1}{2} g^{\theta\theta} \partial_\chi g_{\theta\theta} = \frac{1}{\chi}, \label{eqA4e}\\
\Gamma^\theta_{\phi\phi} &= -\frac{1}{2} g^{\theta\theta} \partial_\theta g_{\phi\phi} = -\sin\theta \cos\theta, \label{eqA4f}\\
\Gamma^\phi_{\chi\phi} &= \frac{1}{2} g^{\phi\phi} \partial_\chi g_{\phi\phi} = \frac{1}{\chi}, \label{eqA4g}\\
\Gamma^\phi_{\theta\phi} &= \frac{1}{2} g^{\phi\phi} \partial_\theta g_{\phi\phi} = \frac{\cos\theta}{\sin\theta}.\label{eqA4h}
\end{align}
\end{subequations}
By implementing the geodesic equations~\eqref{eq7} for each spacetime index $\mu$ and using the non-zero Christoffel symbols~\eqref{eqA4}, one can ultimately derive a system of four equations~\eqref{eq14}.
\section{Derivation of the radial comoving distance formula}\label{A2}
The comoving radial distance for free non-radial geodesic motion in flat FLRW spacetime can be determined by integrating Eq.~\eqref{eq44a}. To achieve this, the coordinate variable $\chi$ must be separated from the time coordinate $t$. Once this separation is achieved, integration can be performed to obtain the desired result
\begin{equation}
\int_{\chi_i}^{\chi} \frac{d\chi}{\sqrt{A^2 - \frac{B^2}{\chi^2}}} = \int_{t_i}^{t} \frac{\text{sgn}[\dot{\chi}(t')]}{a(t') \sqrt{a^2(t') + A^2}} dt'.
\label{eqB1}
\end{equation}
The left-hand side of Eq.~\eqref{eqB1} can be simply integrated using the formula
\begin{equation}
\int \frac{d\chi}{\sqrt{A^2 - \frac{B^2}{\chi^2}}} = \frac{1}{|A|} \sqrt{\chi^2 - \frac{B^2}{A^2}}.
\label{eqB2}
\end{equation}
By performing integration on Eq.~\eqref{eqB1}, we derive the following results
\begin{equation}
\chi(t) = \sqrt{\left[ \int_{t_i}^{t} \frac{\text{sgn}(\dot{\chi}(t')) |A| dt'}{a(t') \sqrt{a^2(t') + A^2}} + \sqrt{\chi_i^2 - \frac{B^2}{A^2}} \right]^2 + \frac{B^2}{A^2}}.
\label{eqB3}
\end{equation}
The integral in Eq.~\eqref{eqB3} presents a non-trivial challenge, mainly due to the complexities of addressing $\text{sgn}(\dot{\chi}(t'))$. To effectively handle this problem, it is crucial to consider all possible geodesic scenarios in the comoving frame. In general cosmological models, free geodesics do not necessarily converge with the Hubble flow as time approaches infinity. Indeed, as shown in Ref.~\cite{Barnes}, for this convergence to occur, the equation of state parameter $ w_d $ for the dominant cosmological component, as time tends to infinity, must satisfy $w_d < -\frac{1}{3}$. Consequently, this classification enables the identification of various scenarios:
\begin{itemize}[leftmargin=*]
\item \textbf{changed $\dot{\chi}(t)$ sign}
\begin{itemize}[leftmargin=*]
\item For a general cosmological model, if a particle starts with a negative comoving radial velocity $\dot\chi(0)<0$, it will reach a minimum distance $\chi_*$ at $t=t_*$, then ultimately surpassing it, and finally attain a positive comoving radial velocity $\dot\chi(0)>0$ (See Fig.~\ref{fig10}).
\end{itemize}
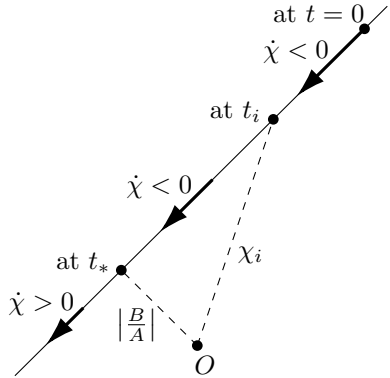
\begin{figure}[H]
\centering
\begin{tikzpicture}
\draw (0.6,0.6) -- (5.5,5.5); 
\draw[dashed] (3,1) -- (2,2); 
\draw[dashed] (3,1) -- (4,4); 
\fill (3,1) circle (2pt);
\fill (2,2) circle (2pt);
\fill (4,4) circle (2pt);
\fill (5.2,5.2) circle (2pt);
\node[] at (3.7, 2.2) {$\chi_i$};
\node[] at (2.2, 1.28) {$\left|\frac{B}{A}\right|$};
\node[] at (4.6, 5.4) {at $t=0$};
\node[] at (3.5, 4.1) {at $t_i$};
\node[] at (1.5, 2.1) {at $t_*$};
\node[] at (4.3, 4.9) {$\dot\chi<0$};
\node[] at (2.5, 3.1) {$\dot\chi<0$};
\node[] at (0.95, 1.55) {$\dot\chi>0$};
\draw[-{Latex[length=4mm, width=2.3mm]}, line width=1.2pt] (5.2,5.2) -- (4.3,4.3);
\draw[-{Latex[length=4mm, width=2.3mm]}, line width=1.2pt] (3.2,3.2) -- (2.5,2.5);
\draw[-{Latex[length=4mm, width=2.3mm]}, line width=1.2pt] (1.5,1.5) -- (1,1);
\node[] at (3.1, 0.75) {$O$};
\end{tikzpicture}
\caption{Visualization of the geodesic trajectory in the comoving
frame, depicting a geodesic that initiates with $\dot\chi(0)<0$ and crosses the closest approach point at $t=t_*$.}
\label{fig10}
\end{figure}
\item \textbf{unchanged $\dot{\chi}(t)$ sign}
\begin{itemize}[leftmargin=*]
\item For a general cosmological model, a particle with an initial positive comoving radial velocity $\dot\chi(0)>0$ will continue to move away indefinitely. In this case,$\dot\chi(t)$ is always positive (See Fig.~\ref{fig11a}).
\item For a Hubble flow convergence model ($w_d < -\frac{1}{3}$), a particle starting with a negative comoving radial velocity $\dot\chi(0)<0$ will converge to a radial distance $\chi_\infty$, without reaching the minimum distance $\chi_*$. Here, $\dot\chi(t)$ is always negative (See Fig.~\ref{fig11b}).
\item For a general cosmological model, a particle beginning with zero peculiar velocity $v_\text{pec}(0)=0$ remains comoving indefinitely. Consequently, $\dot\chi(t)=0$ at all times $t$ (See Fig.~\ref{fig11c}).
\end{itemize}
\end{itemize}
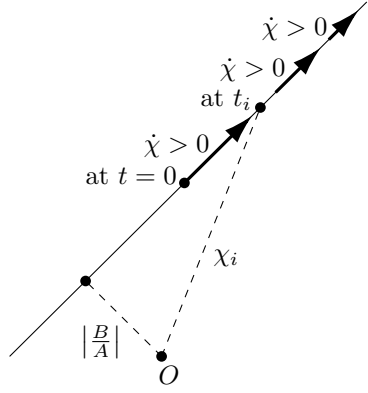
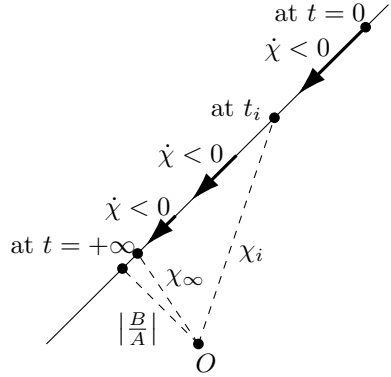
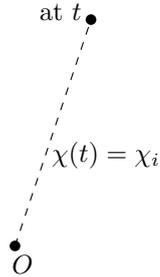
\begin{figure}[H]
\centering
\begin{subfigure}[b]{\columnwidth}
\centering
\begin{tikzpicture}
\draw (1,1) -- (5.7,5.7); 
\draw[dashed] (3,1) -- (2,2); 
\draw[dashed] (3,1) -- (4.3,4.3); 
\fill (3,1) circle (2pt);
\fill (2,2) circle (2pt);
\fill (3.3,3.3) circle (2pt);
\fill (4.3,4.3) circle (2pt);
\node[] at (3.85, 2.3) {$\chi_i$};
\node[] at (2.2, 1.2) {$\left|\frac{B}{A}\right|$};
\node[] at (3.85, 4.4) {at $t_i$};
\node[] at (2.6, 3.4) {at $t=0$};
\node[] at (3.2, 3.8) {$\dot\chi>0$};
\node[] at (4.2, 4.8) {$\dot\chi>0$};
\node[] at (4.75, 5.35) {$\dot\chi>0$};
\draw[-{Latex[length=4mm, width=2.3mm]}, line width=1.2pt] (3.3,3.3) -- (4.2,4.2);
 \draw[-{Latex[length=4mm, width=2.3mm]}, line width=1.2pt] (4.5,4.5) -- (5.1,5.1);
\draw[-{Latex[length=4mm, width=2.3mm]}, line width=1.2pt] (5.2,5.2) -- (5.6,5.6);
\node[] at (3.1, 0.75) {$O$};
\end{tikzpicture}
\caption{Visualization of the geodesic trajectory in the comoving frame, where $\dot\chi(t)>0$ at any given time $t$.}
\label{fig11a}
\end{subfigure}
\vspace{15pt}
\begin{subfigure}[b]{\columnwidth}
\centering
\begin{tikzpicture}
\draw (1,1) -- (5.5,5.5); 
\draw[dashed] (3,1) -- (2,2); 
\draw[dashed] (3,1) -- (4,4); 
\draw[dashed] (3,1) -- (2.2,2.2); 
\fill (3,1) circle (2pt);
\fill (2,2) circle (2pt);
\fill (4,4) circle (2pt);
\fill (2.2,2.2) circle (2pt);
\fill (5.2,5.2) circle (2pt);
 \node[] at (3.7, 2.2) {$\chi_i$};
\node[] at (2.2, 1.2) {$\left|\frac{B}{A}\right|$};
\node[] at (4.6, 5.4) {at $t=0$};
\node[] at (3.5, 4.1) {at $t_i$};
\node[] at (1.35, 2.3) {at $t=+\infty$};
\node[] at (4.3, 4.9) {$\dot\chi<0$};
\node[] at (2.9, 3.5) {$\dot\chi<0$};
\node[] at (2.2, 2.8) {$\dot\chi<0$};
\node[] at (2.83, 1.88) {$\chi_\infty$};
\draw[-{Latex[length=4mm, width=2.3mm]}, line width=1.2pt] (5.2,5.2) -- (4.3,4.3);
\draw[-{Latex[length=4mm, width=2.3mm]}, line width=1.2pt] (3.5,3.5) -- (2.9,2.9);
\draw[-{Latex[length=4mm, width=2.3mm]}, line width=1.2pt] (2.7,2.7) -- (2.3,2.3);
\node[] at (3.1, 0.75) {$O$};
\end{tikzpicture}
\caption{Visualization of the geodesic trajectory in the comoving frame, where $\dot\chi(t)<0$ at any given time $t$.}
\label{fig11b}
\end{subfigure}
\vspace{15pt}
\begin{subfigure}[b]{\columnwidth}
\centering
\begin{tikzpicture}
\draw[dashed] (3,1) -- (4,4); 
\fill (3,1) circle (2pt);
\fill (4,4) circle (2pt);
\node[] at (4.2, 2.2) {$\chi(t)=\chi_i$};
\node[] at (3.6, 4.1) {at $t$};
\node[] at (3.1, 0.75) {$O$};
\end{tikzpicture}
\caption{Visualization of the geodesic trajectory in the comoving frame, where $\dot\chi(t)=0$ at any given time $t$.}
\label{fig11c}
\end{subfigure}
\caption{Combined visualization of the geodesic trajectories in the comoving frame under different conditions.}
\label{fig11}
\end{figure}
In the following discussion, we address this challenge by examining $\text{sgn}(\dot{\chi}(t))$ within the integral~\eqref{eqB3} for each scenario.
\begin{itemize}[leftmargin=*]
\item \textbf{For changed $\dot{\chi}(t)$ sign}
\\
If there is a change in the sign of $\dot{\chi}(t)$, it occurs only once, transitioning from negative to positive. To mathematically prove this claim, one should first determine $\dot{\chi}(t)$ by computing the time derivative of Eq.~\eqref{eqB3}. Subsequently, by multiplying both sides of the resulting expression by $\text{sgn}(\dot{\chi}(t))$, one arrives at the following equation
\begin{align}
|\dot{\chi}(t)| =\left[ \int_{t_i}^{t} \frac{\text{sgn}(\dot{\chi}(t')) |A|dt'}{a(t') \sqrt{a^2(t') + A^2}}  + \sqrt{\chi_i^2 - \frac{B^2}{A^2}} \right]  \times\frac{|A|}{\chi(t)a(t) \sqrt{a^2(t) + A^2}}.
\label{eqB4}
\end{align}
From this equation, and considering arbitrary initial conditions $\chi_i, A,$ and $B$, it follows that $\dot{\chi}(t)$ can reach zero only once, precisely at $t_*$ if and only if
\begin{equation}
\int_{t_i}^{t_*} \frac{\text{sgn}(\dot{\chi}(t')) |A|dt'}{a(t') \sqrt{a^2(t') + A^2}} + \sqrt{\chi_i^2 - \frac{B^2}{A^2}} = 0.
\label{eqB5}
\end{equation}
Since $\dot{\chi}(t)$ can only vanish at $t_*$, its sign remains unchanged between $t_i$ and $t_*$. This allows us to extract $\text{sgn}(\dot{\chi}(t'))$ from the integral in the condition~\eqref{eqB5} to $\text{sgn}(\dot{\chi}_i)$, as illustrated below
\begin{equation}
\text{sgn}(\dot{\chi}_i) \int_{t_i}^{t_*} \frac{|A|dt'}{a(t') \sqrt{a^2(t') + A^2}}+ \sqrt{\chi_i^2 - \frac{B^2}{A^2}} = 0.
\label{eqB6}
\end{equation}
This condition can also be expressed by moving the $\text{sgn}(\dot{\chi}_i)$ to the second term, as follows
\begin{equation}
\int_{t_i}^{t_*} \frac{|A|dt'}{a(t') \sqrt{a^2(t') + A^2}}+\text{sgn}(\dot{\chi}_i)  \sqrt{\chi_i^2 - \frac{B^2}{A^2}} = 0.
\label{eqB7}
\end{equation}
This expression remains valid for all values of $\dot\chi_i$, including when it equals zero. This is because $\dot\chi_i=0$ corresponds to $t_i=t_*$ leading to both the first and second terms in Eq.~\eqref{eqB7} vanishing, thereby ensuring the equation's validity in this specific case. The sign of the initial comoving velocity $\dot{\chi}_i$ can be expressed from Eq.~\eqref{eqB6} as
\begin{equation}
\text{sgn}(\dot{\chi}_i) = \frac{-\sqrt{\chi_i^2 - \frac{B^2}{A^2}}}{\int_{t_i}^{t_*} \frac{|A|dt'}{a(t') \sqrt{a^2(t') + A^2}}}.
\label{eqB8}
\end{equation}
Consequently, the validity of condition.~\eqref{eqB6} depends on the following requirement
\begin{equation}
\text{sgn}(\dot{\chi}_i) = \text{sgn}(t_i - t_*).
\label{eqB9}
\end{equation}
This proves our claim that $\dot{\chi}(t)$ can change its sign just once at $t_*$, transitioning from negative to positive value. As highlighted earlier, we show that this specific time $t_*$ corresponds to the minimum distance at which the particle approaches the original point
\begin{equation}
t = t_* \Longleftrightarrow \dot{\chi}(t_*) = 0 \Longleftrightarrow \chi(t_*) = \chi_* = \left|\frac{B}{A}\right|.
\label{eqB10}
\end{equation}
Now, to tackle the integral term in Eq.~\eqref{eqB3}, we divide the integration interval into two segments, within each $\text{sgn}(\dot{\chi}(t))$ remains constant, allowing us to extract $\text{sgn}(\dot{\chi}(t))$ from the integral. Using $t_*$ as the division point for the integral in Eq.~\eqref{eqB3}, the expression transforms as follows
{\small
\begin{flalign}
\chi(t)&=\sqrt{
\biggl[ \int_{t_*}^t \frac{\text{sgn}(\dot{\chi}(t'))|A|}{a(t') \sqrt{a^2(t') + A^2}} \, dt' 
+\int_{t_i}^{t_*} \frac{\text{sgn}(\dot{\chi}(t'))|A|}{a(t') \sqrt{a^2(t') + A^2}}\, dt' + \sqrt{\chi_i^2 - \frac{B^2}{A^2}} \biggr] ^2 + \frac{B^2}{A^2}}\nonumber\\
&=\sqrt{\biggl[\text{sgn}(\dot{\chi}(t))\int_{t_*}^t \frac{|A|}{a(t') \sqrt{a^2(t') + A^2}} \, dt' \biggr] ^2 + \frac{B^2}{A^2}}\nonumber\\
&=\sqrt{\biggl[\int_{t_*}^t \frac{|A|}{a(t') \sqrt{a^2(t') + A^2}} \, dt' \biggr] ^2 + \frac{B^2}{A^2}}\nonumber\\
&=\sqrt{\biggl[\int_{t_*}^t \frac{|A|}{a(t') \sqrt{a^2(t') + A^2}} \, dt' +\int_{t_i}^{t_*} \frac{|A|dt'}{a(t') \sqrt{a^2(t') + A^2}}+\text{sgn}(\dot{\chi}_i)  \sqrt{\chi_i^2 - \frac{B^2}{A^2}}\biggr] ^2 + \frac{B^2}{A^2}}\nonumber\\
&=\sqrt{\biggl[\int_{t_i}^t \frac{|A|}{a(t') \sqrt{a^2(t') + A^2}} \, dt' +\text{sgn}(\dot{\chi}_i)  \sqrt{\chi_i^2 - \frac{B^2}{A^2}}\biggr] ^2 + \frac{B^2}{A^2}}.
\label{eqB11}&&
\end{flalign}
}
In the second step, the vanishing condition from Eq.~\eqref{eqB5} was applied, noting that $\text{sgn}(\dot{\chi}(t'))$ stays constant over the interval $t_* \rightarrow t$. In the third step, we removed the term $\text{sgn}(\dot{\chi}(t))$ as its squared value always results in a positive sign. Finally, in the fourth step, the zero condition term from Eq.~\eqref{eqB7} was inserted into the squared term, ultimately leading to the derivation of Eq.~\eqref{eq52a}.
\\
In scenarios where $t_*$ does not exist ($\dot{\chi}(t)\neq0$), $\text{sgn}(\dot{\chi}(t'))$ in Eq.~\eqref{eqB3} remains constant and can be extracted from the integral as $\text{sgn}(\dot{\chi}_i)$, subsequently moving it to the second term to arrive at the same expression~\eqref{eqB11}.
\end{itemize}
Thus, the comoving radial solution, detailed in Eq.~\eqref{eq52a}, is convincingly proven.
\section{Derivation of the geometric formula for angular coordinates}\label{A3}
Now, let us demonstrate that the geodesic solution for the angle $\phi$, as shown in Eq.~\eqref{eq52b}, is identical to the geometrical formula presented in Eq.~\eqref{eq66}. This formula was derived by considering the geodesic trajectories of free motion as straight lines. Initially, it appears that the geodesic angular solution~\eqref{eq52b} explicitly depends on the scale factor $a(t)$, unlike the geometric formula~\eqref{eq66}. However, we aim to prove that $\phi(t)$ does not depend explicitly on $a(t)$.
Consider Eqs.~\eqref{eq44a} and~\eqref{eq44b}, by dividing them we can eliminate the explicit relation with the scale factor, resulting in the following expression
\begin{equation}
\frac{\dot{\chi}(t)}{\dot{\phi}(t)} = \frac{\text{sgn}(\dot{\chi}(t))}{B}\chi^2(t)\sqrt{A^2 - \frac{B^2}{\chi^2(t)}}.
\label{eqC1}
\end{equation}
Upon separating variables $\phi$ and $t$, we arrive at
\begin{equation}
\int_{\phi_i}^{\phi} d\phi = B \int_{t_i}^{t} \frac{\text{sgn}(\dot{\chi}(t)) \dot{\chi}(t)}{\chi^2(t) \sqrt{A^2 - \frac{B^2}{\chi^2(t)}}} dt.
\label{eqC2}
\end{equation}
\begin{itemize}[leftmargin=*]
\item \textbf{For changed $\dot{\chi}(t)$ sign}
\\
If $\dot{\chi}(t)$ changes its sign at time $t_*$, we divide the integral into two intervals: $t_i$ to $t_*$ and $t_*$ to $t$, where $\text{sgn}(\dot{\chi}(t))$ remains constant within each interval. This leads us to
\begin{flalign}
\phi(t)&= B \ \text{sgn}(\dot{\chi}_i) \int_{t_i}^{t_*} \frac{\dot{\chi}(t)}{\chi^2 \sqrt{A^2 - \frac{B^2}{\chi^2(t)}}} dt + B \ \text{sgn}(\dot{\chi}(t)) \int_{t_*}^{t} \frac{\dot{\chi}(t)}{\chi^2 \sqrt{A^2 - \frac{B^2}{\chi^2(t)}}} dt + \phi_i\nonumber&\\
&= B \ \text{sgn}(\dot{\chi}_i) \int_{\chi_i}^{\chi_*} \frac{d\chi}{\chi^2 \sqrt{A^2 - \frac{B^2}{\chi^2}}}  + B \ \text{sgn}(\dot{\chi}(t)) \int_{\chi_*}^{\chi} \frac{d\chi}{\chi^2 \sqrt{A^2 - \frac{B^2}{\chi^2}}} + \phi_i.
\label{eqC3}&
\end{flalign}
Using the integration formula
\begin{equation}
\int \frac{d\chi}{\chi^2 \sqrt{A^2 - \frac{B^2}{\chi^2}}} = \frac{1}{|B|} \arctan \left( \sqrt{\frac{A^2}{B^2} \chi^2 - 1} \right),
\label{eqC4}
\end{equation}
\\
we find the following
\begin{align}
\phi(t) = &B \ \text{sgn}(\dot{\chi}_i) \frac{1}{|B|} \left[ \arctan\left( \sqrt{\frac{A^2}{B^2} \chi^2(t_*) - 1} \right) - \arctan\left( \sqrt{\frac{A^2}{B^2} \chi_i^2 - 1} \right) \right] \nonumber\\
+ &B \ \text{sgn}(\dot{\chi}(t)) \frac{1}{|B|} \left[ \arctan\left( \sqrt{\frac{A^2}{B^2} \chi^2(t) - 1} \right) - \arctan\left( \sqrt{\frac{A^2}{B^2} \chi^2(t_*) - 1} \right) \right] + \phi_i.
\label{eqC5}
\end{align}
Applying the relation~\eqref{eq55a} for the radial distance at $t_*$, we simplify the expression to
\begin{align}
\phi(t) = \text{sgn}(B) \left[ \text{sgn}(\dot{\chi}(t)) \arctan \left( \sqrt{\frac{A^2}{B^2} \chi^2(t) - 1} \right) - \text{sgn}(\dot{\chi}_i) \arctan \left( \sqrt{\frac{A^2}{B^2} \chi_i^2 - 1} \right) \right].
\label{eqC6}
\end{align}
The integral in Eq.~\eqref{eq52b} simplifies and leads to the same geometrical relation as Eq.~\eqref{eq66}.
\item \textbf{For unchanged $\dot{\chi}(t)$ sign}
\\
 In this case, $\text{sgn}(\dot{\chi}(t'))$ in Eq.~\eqref{eqC2} is a constant and can be extracted from the integral to $\text{sgn}(\dot{\chi}_i)$ and by using the integral formula~\eqref{eqC4}, we arrive at the same expressions in Eq.~\eqref{eqC6}. \\ \\
\end{itemize} 
Thus, we conclude that the geodesic solution for the angle $\phi$ is indeed identical to the geometrical formula derived under the assumption that the geodesic trajectories of free motion are straight lines. This substantiates our conclusion that the general geodesic trajectory in our spacetime is effectively a straight line in the comoving frame.
\section{Checking that the solutions satisfy the Euleur-Lagrange Equations}\label{A4}
We now confirm that our solution for the comoving radial-angular velocity in Eqs.~\eqref{eq44a} and~\eqref{eq44b} satisfies the Euler-Lagrange equations~\eqref{eq37a} and~\eqref{eq37b}. Let's start by computing the left-hand and right-hand sides of the first equation~\eqref{eq37a}, resulting in the following
\begin{subequations}
\begin{flalign}
\text{LHS}&=\frac{d}{dt} \left[ \frac{a^2 \dot{\chi}}{\sqrt{1 - a^2 \dot{\chi}^2 - a^2 \chi^2 \dot{\phi}^2}} \right]=\text{sgn}(\dot{\chi}(t)) \frac{d}{dt} \left[ \sqrt{A^2 - \frac{B^2}{\chi^2}} \right]\nonumber\\
&= \text{sgn}(\dot{\chi}(t)) \frac{\frac{B^2 \dot{\chi}}{\chi^3}}{\sqrt{A^2 - \frac{B^2}{\chi^2}}}=\frac{\frac{B^2}{\chi^3}}{a \sqrt{a^2 + A^2}},\\
\text{RHS}&=\frac{\chi a^2 \dot{\phi}^2}{\sqrt{1 - a^2 \dot{\chi}^2 - a^2 \chi^2 \dot{\phi}^2}}=\frac{\frac{B^2}{\chi^3}}{a \sqrt{a^2 + A^2}}.&&
\end{flalign}
\end{subequations}
This confirms that the comoving velocities in Eqs.~\eqref{eq44a} and~\eqref{eq44b} indeed satisfy the Euler-Lagrange equation~\eqref{eq37a}. Similarly, for the Euler-Lagrange equation~\eqref{eq37b}, its satisfaction can be easily verified using the constant of motion defined in Eq.~\eqref{eq40}. It yields
\begin{equation}
\frac{d}{dt} \left[ \frac{a^2 \chi^2 \dot{\phi}}{\sqrt{1 - a^2 \dot{\chi}^2 - a^2 \chi^2 \dot{\phi}^2}} \right]=\frac{d}{dt}B=0,
\label{eqD3}
\end{equation}
which confirms that our solution in Eqs.~\eqref{eq44a} and~\eqref{eq44b} also satisfies the Euler-Lagrange equation~\eqref{eq37b}. Therefore, it is verified that our solutions in Eqs.~\eqref{eq44a} and~\eqref{eq44b} satisfy both Euler-Lagrange equations~\eqref{eq37a} and~\eqref{eq37b}.
\section{Checking that the solutions satisfy the Geodesic Equations}\label{A5}
Now, let's verify that our solution for the comoving radial-angular velocity in Eqs.~\eqref{eq44a} and~\eqref{eq44b} satisfy the geodesic equations~\eqref{eq38a},~\eqref{eq38b}, and~\eqref{eq38c}. This verification will be discussed for each equation separately. To simplify the expressions, we will use a prime ($'$) in the upper right to indicate the derivative with respect to the proper time $s$. Therefore, the geodesic equations (38) can be reformulated as
\begin{subequations}
\begin{align}
t'' + a(t) \dot{a}(t)(\chi'^2 + \chi^2 \phi'^2) &= 0, \label{eqE1a}\\
\chi'' + 2 \frac{\dot{a}(t)}{a(t)} t' \chi' - \chi \phi'^2 &= 0, \label{eqE1b}\\
\phi'' + 2 \frac{\dot{a}(t)}{a(t)} t' \phi' + \frac{2}{\chi} \chi' \phi' &= 0.\label{eqE1c}
\end{align}
\end{subequations}
In what follows, we will substitute $\dot{\chi}$ and $\dot{\phi}$ by their expressions in Eqs.~\eqref{eq44a} and~\eqref{eq44b} in terms of $A$ and $B$.
\begin{itemize}[leftmargin=*]
\item \textbf{For the first equation~\eqref{eqE1a}:} one can write its left-hand side as
\end{itemize}
\begin{flalign}
t'' + a \dot{a}(\chi'^2 + \chi^2 \phi'^2) &= t'' + a \dot{a} t'^2 (\dot{\chi}^2 + \chi^2 \dot{\phi}^2)\nonumber\\
&=T_1 +T_2,
\label{eqE2}&&
\end{flalign}
where
\begin{subequations}
\begin{align}
&T_1=t'',\label{eqE3a}\\
&T_2= a \dot{a} t'^2 (\dot{\chi}^2 + \chi^2 \dot{\phi}^2).\label{eqE3b}
\end{align}
\end{subequations}
Firstly, computing the differentiating of the cosmological time $t$ with respect to the proper time $s$ 
\begin{align}
\frac{dt}{ds} = \frac{1}{\sqrt{1 - a^2\dot{\chi}^2 - a^2 \chi^2 \dot{\phi}^2}} = \frac{\sqrt{a^2(t) + A^2}}{a(t)}.
\label{eqE4}
\end{align}
Now, computing each term of Eq.~\eqref{eqE2} separately gives
\begin{subequations}
\begin{flalign}
&T_1=\frac{d^2 t}{ds^2}= \frac{d}{ds} \left[ \frac{\sqrt{a^2(t) + A^2}}{a(t)} \right]
= \frac{dt}{ds} \frac{d}{dt} \left[ \frac{\sqrt{a^2(t) + A^2}}{a(t)} \right]
= -\frac{A^2 \dot{a}(t)}{a(t)^3},\label{eqE5b}&\\
&T_2=a \dot{a} t'^2 (\dot{\chi}^2 + \chi^2 \dot{\phi}^2)= \frac{A^2 \dot{a}(t)}{a(t)^3}.
\label{eqE5a}&
\end{flalign}
\end{subequations}
Therefore, $T_1$ is the opposite of $T_2$, ensuring they cancel each other out. Consequently, the comoving velocities in Eqs.~\eqref{eq44a} and~\eqref{eq44b} satisfy the geodesic equation~\eqref{eqE1a}.
\begin{itemize}[leftmargin=*]
\item \textbf{For the second equation~\eqref{eqE1b}:} one can write its left-hand side as
\end{itemize}
\begin{align}
\chi'' + 2 \frac{\dot{a}}{a} t' \chi' - \chi \phi'^2 
= t'^2 [U_1 + U_2 + U_3]
\label{eqE6},
\end{align}
where
\begin{subequations}
\begin{align}
&U_1 = \ddot\chi,\label{eqE7a}\\
&U_2 = -a \dot{a} \dot\chi(\dot\chi^2 + \chi^2\dot\phi^2),\label{eqE7b}\\
&U_3 = 2 \frac{\dot{a}}{a}\dot\chi - \chi \dot\phi^2.\label{eqE7c}
\end{align}
\end{subequations}
We will proceed by computing each term in the equation~\eqref{eqE6} separately by using the solution in Eqs.~\eqref{eq44a} and~\eqref{eq44b}. We obtain the following
\begin{subequations}
\begin{flalign}
U_1& =\text{sgn}(\dot{\chi}(t)) \left[ \frac{-2\dot{a}(t) A^2 + 2\dot{a}(t) \frac{B^2}{\chi^2(t)} - \frac{\dot{a}(t)}{a^2(t)} A^4 + \frac{\dot{a}(t)}{a^2(t)} \frac{A^2 B^2}{\chi^2(t)}}{(a^2(t) + A^2)^{3/2} \sqrt{A^2 - \frac{B^2}{\chi^2(t)}}} \right] + \frac{\frac{B^2}{a^2(t) \chi^3(t)}}{a^2(t) + A^2},\label{eqE8a}&\\
U_2 &= \text{sgn}(\dot{\chi}(t))\left[ \frac{\frac{-\dot{a}(t)}{a^2(t)} A^4 + \frac{\dot{a}(t)}{a^2(t)} \frac{A^2 B^2}{\chi^2(t)}}{(a^2(t) + A^2)^{3/2} \sqrt{A^2 - \frac{B^2}{\chi^2(t)}}}\right],\label{eqE8b}&\\
U_3 &=\text{sgn}(\dot{\chi}(t)) \left[ \frac{2\dot{a}(t) A^2 - 2\dot{a}(t) \frac{B^2}{\chi^2(t)} + 2 \frac{\dot{a}(t)}{a^2(t)} A^4 - 2 \frac{\dot{a}(t)}{a^2(t)} \frac{A^2 B^2}{\chi^2(t)}}{(a^2(t) + A^2)^{3/2} \sqrt{A^2 - \frac{B^2}{\chi^2(t)}}} \right] - \frac{ \frac{B^2}{a^2(t) \chi^3(t)}}{a^2(t) + A^2}.\label{eqE8c}&
\end{flalign}
\end{subequations}
One can check that $U_1 + U_2 + U_3 = 0$, confirming that the comoving velocities in Eqs.~\eqref{eq44a} and~\eqref{eq44b} satisfy the geodesic equation~\eqref{eqE1b}.
\begin{itemize}[leftmargin=*]
\item \textbf{For the third equation~\eqref{eqE1c}:} one can write its left-hand side as
\end{itemize}
\begin{align}
\phi'' + 2 \frac{\dot{a}}{a} t' \phi' + \frac{2}{\chi} \chi' \phi' 
= t'^2 [V_1 + V_2 + V_3],
\label{eqE9}
\end{align}
where
\begin{subequations}
\begin{align}
&V_1 = \ddot\phi,\label{eqE10a}\\
&V_2 = -a \dot{a} \dot\phi(\dot\chi^2 + \chi^2 \dot\phi^2),\label{eqE10b}\\
&V_3 = 2 \frac{\dot{a}}{a}\dot\phi + \frac{2}{\chi} \dot\chi\dot\phi.\label{eqE10c}
\end{align}
\end{subequations}
Each term in equation~\eqref{eqE9} will be calculated separately by using the solution in Eqs.~\eqref{eq44a} and~\eqref{eq44b}. We obtain the following
\begin{subequations}
\begin{flalign}
V_1 &= -\text{sgn}(\dot{\chi}(t)) \frac{\frac{2B}{a^2(t) \chi^3(t)}}{a^2(t) + A^2} \sqrt{A^2 - \frac{B^2}{\chi^2(t)}}-  \frac{\frac{2\dot{a}(t) B}{\chi^2(t)}+\frac{\dot{a}(t) A^2B}{a^2(t)\chi^2(t)}}{(a^2(t) + A^2)^{3/2}},
\label{eqE11a}&\\
V_2 &= - \frac{\frac{\dot{a}(t)A^2 B}{a^2(t)\chi^2(t)}}{(a^2(t) + A^2)^{3/2}},
\label{eqE11b}&\\
V_3 &= \frac{\frac{2\dot{a}(t) B}{\chi^2(t)}+\frac{2\dot{a}(t) A^2B}{a^2(t)\chi^2(t)}}{(a^2(t) + A^2)^{3/2}} + \text{sgn}(\dot{\chi}(t)) \frac{\frac{2B}{a^2(t) \chi^3(t)}}{a^2(t) + A^2} \sqrt{A^2 - \frac{B^2}{\chi^2(t)}}.\label{eqE11c}&
\end{flalign}
\end{subequations}
It can be verified that $V_1 + V_2 + V_3 = 0$, thereby proving that the comoving velocities in Eqs.~\eqref{eq44a} and~\eqref{eq44b} satisfy the geodesic equation~\eqref{eqE1c}.
In conclusion, these calculations thoroughly confirm that the comoving radial-angular velocities presented in Eqs.~\eqref{eq44a} and~\eqref{eq44b} indeed satisfy the geodesic equations~\eqref{eq38a},~\eqref{eq38b}, and~\eqref{eq38c}.

\bibliography{sn-bibliography}

\end{document}